\documentclass[11pt,letterpaper]{article}
\usepackage{graphicx} 

\usepackage[ruled]{algorithm2e}
\SetArgSty{textnormal}
\usepackage{amsmath}
\usepackage{amssymb}
\usepackage{mathtools}
\usepackage{pifont}
\usepackage{amsthm}
\usepackage{csquotes}
\usepackage{thm-restate}
\usepackage{bbm}
\usepackage{hyperref}
\usepackage[nameinlink]{cleveref}

\newtheorem{theorem}{Theorem}

\newtheorem{lemma}{Lemma}

\newtheorem{fact}{Fact}
\usepackage[dvipsnames]{xcolor}
\usepackage{paralist}
\usepackage{nicefrac}
\usepackage[margin=0.95in]{geometry}
\usepackage{tikz}
\usepackage{forest}
\usepackage{setspace}
\usepackage{orcidlink}
\usetikzlibrary{decorations.pathreplacing,shapes.geometric,arrows.meta,calc}

\usepackage{thm-restate}

\hypersetup{
	colorlinks,
	linkcolor={red!50!black},
	citecolor={blue!50!black},
	urlcolor={blue!80!black}
}

\allowdisplaybreaks

\newcount\Comments  
\usepackage{comment}
\usepackage{color}
\definecolor{darkgreen}{rgb}{0,0.5,0}
\definecolor{darkgreen2}{rgb}{0,0.7,0}
\definecolor{purple}{rgb}{1,0,1}
\newcommand{\kibitz}[2]{\ifnum\Comments=1{\color{#1} #2}\fi}

\newcommand{\email}[1]{\protect\href{mailto:#1}{#1}}

\newcommand{\prunings}{\mathrm{randPrunings}}
\newcommand{\fM}{f^{\mathcal{M}}}

\newcommand{\bpleft}{\mathopen{}\mathclose\bgroup\left}
\newcommand{\bpright}{\aftergroup\egroup\right}

\newcommand{\RB}[1]{\ensuremath{\bpleft( {#1} \bpright)}}
\newcommand{\SB}[1]{\ensuremath{\bpleft[ {#1} \bpright]}}
\newcommand{\CB}[1]{\ensuremath{\{ {#1} \}}}
\newcommand{\eps}{\ensuremath{\varepsilon}}

\newcommand{\Ex}[1]{\ensuremath{\mathbb{E} \SB{#1}}}
\newcommand{\ExIl}[1]{\ensuremath{\mathbb{E} [#1]}}
\newcommand{\Exc}[2]{\Ex{\bpleft. #1 \;\middle|\; #2 \bpright.}}
\newcommand{\Pty}[1]{\ensuremath{\mathrm{Pr}\SB{{#1}}}}

\newcommand{\Cost}[1]{\ensuremath{\cost\RB{#1}}}
\newcommand{\cost}{\ensuremath{\mathrm{cost}}}
\newcommand{\acost}{\ensuremath{\mathrm{a}\cost}}
\newcommand{\Acost}[1]{\ensuremath{\acost\RB{#1}}}
\newcommand{\util}{\ensuremath{U}}
\newcommand{\Util}[1]{\ensuremath{\util\RB{#1}}}
\newcommand{\Utilk}[2]{\ensuremath{\util_{#1}\RB{#2}}}

\newcommand{\kk}{\ensuremath{\bar{k}}}
\newcommand{\qq}[1][]{\ensuremath{\bar{p}_{#1}}}
\newcommand{\PP}{\ensuremath{\overline{P}}}

\tikzset{
    stopnodebase/.style={regular polygon, minimum width=1.82cm,regular polygon sides=8, draw, line width = 1.5pt,fill=red!20}
    }
\forestset{
    leaftree/.style={isosceles triangle,draw, shape border rotate=90, thick,child anchor=north,minimum width=2cm,fill=white,line width = 1.5pt,anchor=south},
    innernode/.style={circle, draw, fill=gray!30, minimum width=1.7cm,line width = 1.5pt,anchor=south,font=\scriptsize},
    textbox/.style={rectangle,align=left,l=2cm,very thick,fill=gray!30,inner sep=6pt,outer sep=3pt,child anchor=north,label distance=3pt},
    stopnode/.style={stopnodebase},
    endnodef/.style={stopnode,child anchor=north,anchor=south},
    endnodes/.style={rectangle, draw, minimum width = 1.7cm, minimum height = 1.7cm,child anchor=north, anchor=south,line width=1.5pt,fill=OliveGreen!20},
    nonode/.style={parent anchor=center, child anchor=center}
}

\newcommand{\OPTone}{\text{OPT}${}_1$}
\newcommand{\KS}{\text{SteepestAscent}}

\title{Approximating Matroid Basis Testing for Partition Matroids\\ using Budget-In-Expectation}
\author{Lisa Hellerstein \orcidlink{0000-0002-3743-7965} \thanks{Department of Computer Science and Engineering, New York University, New York, United States (\email{lisa.hellerstein@nyu.edu}).} \and Benedikt M. Plank \orcidlink{0000-0002-7949-3738} \thanks{Berlin, Germany (\email{b.plank@mail.de}).} \and Kevin Schewior \orcidlink{0000-0003-2236-0210} \thanks{Department of Mathematics and Computer Science, University of Cologne, Köln, Germany; Department of Mathematics and Computer Science, University of Southern Denmark, Odense, Denmark (\email{k.schewior@uni-koeln.de}).}}

\date{\today}

\begin{document}

\maketitle
\begin{abstract}
We consider the following Stochastic Boolean Function Evaluation problem, which is closely related to several problems from the literature. A matroid $\mathcal{M}$ (in compact representation) on ground set $E$ is given, and each element $i\in E$ is active independently with known probability $p_i\in(0,1)$. The elements can be queried, upon which it is revealed whether the respective element is active or not. The goal is to find an adaptive querying strategy for determining whether there is a basis of $\mathcal{M}$ in which all elements are active, with the objective of minimizing the expected number of queries.

When $\mathcal{M}$ is a uniform matroid, this is the problem of evaluating a $k$-of-$n$ function, first studied in the 1970s. This problem is well-understood, and has an optimal adaptive strategy that can be computed in polynomial time.

Taking $\mathcal{M}$ to instead be a partition matroid, we show that previous approaches fail to give a constant-factor approximation. Our main result is a polynomial-time constant-factor approximation algorithm producing a randomized strategy for this partition matroid problem.  We obtain this result by combining a new technique with several well-established techniques.  Our algorithm adaptively interleaves solutions to several instances of a novel type of stochastic querying problem, with a constraint on the \emph{expected} cost. We believe that this type of problem is of independent interest, will spark follow-up work, and has the potential for additional applications.
\end{abstract}

\vspace{\fill}
\newpage

\tableofcontents

\newpage

\section{Introduction}
\label{sec:Intro}

Stochastic Boolean Function Evaluation (SBFE) problems are a class of information-acquisition problems in stochastic environments. In this paper, we focus on the fundamental unit-cost version. Here, one is given a (compactly represented) function $f:\{0,1\}^n\rightarrow\{0,1\}$ and Boolean random variables $x_1,\dots,x_n$, where for all $i\in[n]$ the variable $x_i$ is $1$ (we also say element $i$ is \emph{active}) independently with known probability $p_i\in(0,1)$. The function $f(x_1,\dots,x_n)$ must be evaluated by testing variables sequentially for their values. In other words, testing must continue until a certificate for the value of the function has been found.  

The problem is to (adaptively) determine the order in which to perform the tests so as to minimize the expected number of tests performed.
An algorithm for such a problem is not required to output the explicit testing strategy (decision tree); it is merely required to output, in any situation, the next test to perform. For a recent survey of results in the area, see~\cite{Unluyurt25}.

Recent papers on SBFE problems are concerned with (usually polynomial-time) approximation algorithms~\cite{deshpande2016approximation,gkenosis2018stochastic,ghuge2022non,plank2024simple,hellerstein2024quickly,NielsenRS25}: An algorithm is called a $c$-approximation algorithm if the expected number of tests it performs is within a factor of $c$ of the expected number of tests of the optimal strategy. Notably, for several of these problems it is open whether the problem is NP-hard \emph{and} whether a constant-factor approximation algorithm exists. A general difficulty in designing approximation algorithms for SBFE problems is the absence of natural strong lower bounds for the costs of the optimal testing strategy.

One main direction in the SBFE literature focuses on ``counting functions'' in which active variables increase a counter, and the function value depends only on the value of this counter: $k$-of-$n$ functions, which output $1$ iff there are at least $k$ active elements~\cite{halpern1974,Salloum1997FastOD,SalloumBreuer84,BenDov81,NielsenRS25}; linear-threshold functions, the weighted counterparts of $k$-of-$n$ functions~\cite{heuristicLeastCostCox,deshpande2016approximation,jiang2020algorithms}; general symmetric functions~\cite{gkenosis22stochastic,grammel2022,ghuge2022non,plank2024simple,liu20226approximation}; or, going beyond the Boolean case and a single counter, voting functions~\cite{hellerstein2024quickly}.
Work of Ghuge et al., 
on a problem they called Explainable Stochastic $d$-Halfspace Evaluation,
considered a variant of the SBFE problem in which an explanation (a certain type of certificate) has to be found for functions of the form $h(g_1,\ldots,g_d)$, where each $g_i$ is a linear threshold function, and $h$ is an arbitrary Boolean function~\cite{ghuge2022non}.  Exact or approximation algorithms have been developed for all these problems.

Another direction is that of functions related to graphs. Here, the variables correspond to the edges $E$ of a given graph $G=(V,E)$, and the value of $f$ depends on the subgraph $G'$ of $G$ consisting of all edges whose corresponding variable is active.  
An interesting problem of this type concerns $s$-$t$ connectivity: $f(x)=1$ iff there exists an $s$-$t$ path in $G'$ (for some given $s,t\in V$). This problem is NP-hard~\cite{fu2017determining,Guo0N0N24}.
The special case in which $G$ is a series--parallel graph is equivalent to the SBFE problem for Boolean read-once formulas (cf.~\cite{golumbic2011read-once}), a problem studied
since the 1970s whose NP-hardness is still open~\cite{unluyurt2004sequential,Greiner06,deshpande2016approximation}. Even for this special case, it is open whether a polynomial-time non-trivial approximation algorithm exists.

In this paper, we introduce the following framework, which is closely related to both directions. 
Let  $\mathcal{M}=(E,\mathcal{I})$, be a matroid, where $E=\{1,\dots,n\}$ is the ground set and $\mathcal{I}$ is the independence system. We define $f^{\mathcal{M}}$ to be a function that is $1$ if and only if there is a basis of $\mathcal{M}$ that consists only of active elements (variables).\footnote{Equivalently, in matroid terms, $f^{\mathcal{M}}(x)=0$ iff there is a cocircuit of $\mathcal{M}$ that consists only of inactive elements. For a definition of cocircuits and other matroid concepts, see e.g., \cite{Oxley}. We assume only very basic knowledge of matroids in this paper.}
We refer to the SBFE problem for this function, with the goal of minimizing the expected {\em number} of queries (equivalently, assuming unit testing costs), 
as the \emph{Matroid Basis Testing} problem (MBT).

The case when $\mathcal{M}$ is a uniform matroid (i.e., $E'\in\mathcal{I}$ iff $|I|\leq k$ for some given $k$)
is identical to the SBFE problem for $k$-of-$n$ functions. The case when $\mathcal{M}$ is a graphical matroid (i.e., $E'\in\mathcal{I}$ iff $(V,E')$ is an acyclic subgraph of a given graph $G=(V,E)$) is the global-connectivity version of the aforementioned $s$-$t$ connectivity function.  

While some of our results concern more general problems, the bulk of this paper focuses on MBT when $\mathcal{M}$ is a partition matroid.  In this case, the matroid $\mathcal{M}$ is represented by a partition $\mathcal{P}=\{P_1,\dots,P_d\}$ of $E$ as well as upper bounds $k_1,\dots,k_d$, such that $E'\in\mathcal{I}$ iff $|E'\cap P_i|\leq k_i$ for all $i\in\{1,\dots,d\}$. \textbf{Our main result is an $O(1)$-approximation algorithm for MBT on partition matroids.}

The MBT problem on partition matroids is equivalent to the SBFE problem for Boolean functions $f=g_1 \wedge g_2 \wedge \ldots \wedge g_d$, where the $g_i$ are defined on disjoint sets of variables, and each $g_i$ is a $k_i$-of-$n_i$ function, for some $1 \leq k_i \leq n_i$.  For example, $f(x_1,x_2,x_3,x_4,x_5) = g_1(x_1,x_2) \wedge g_2(x_3,x_4,x_5)$ where $g_1(x_1,x_2)$ is the 1-of-2 function (Boolean OR), and $g_2(x_3,x_4,x_5)$ is the 2-of-3 function (majority).
Because the $k_i$-of-$n_i$ functions are linear threshold functions, and the $g_i$ are defined on disjoint sets of variables, the MBT problem on partition matroids can be shown to be a special case of 
Explainable Stochastic $d$-Halfspace Evaluation.  Ghuge et al.\ presented an approximation algorithm for that more general problem, with an approximation factor of $O(d^2 \log d)$~\cite{ghuge2022non}.

A special case of MBT on partition matroids is where each $k_i=1$, equivalently, where $f$ is representable by a \emph{read-once CNF formula}.
Along with $k$-of-$n$ functions, functions represented by read-once CNF (dually, DNF) formulas are among the very few Boolean function classes for which the SBFE problem is known to be exactly solvable by a polynomial-time algorithm~\cite{Boros2000}.\footnote{By Boolean duality, the MBT problem on partition matroids is effectively equivalent to the above SBFE problem for the disjunction, rather than conjunction, of the $g_i$'s. Similarly, the SBFE problem for read-once DNF formulas is effectively equivalent to the SBFE problem for CNF formulas.}

Although partition matroids are arguably the next step up from uniform matroids, and seem to be simple, the MBT problem on partition matroids poses a significant new challenge.
In terms of the SBFE formulation above, we know how to optimally evaluate each of the $g_i$ on its own because each is a $k_i$-of-$n_i$ function. But because $f$ is the conjunction of the $g_i$, we need to decide how and when to switch between testing variables in each of the $g_i$.  A natural approach is to perform the tests in a depth-first order, where once we begin testing the variables of one $g_i$, we complete the evaluation of that $g_i$ before performing tests on the variables in another $g_i$.  In fact, a depth-first approach is optimal for the special case of read-once CNF functions. However, when the $g_i$ are arbitrary $k_i$-of-$n_i$ functions, the approximation ratio of a depth-first approach is not bounded by any function of $d$ (the number of $g_i$). On the other hand, the round-robin approach that Ghuge at al.\ employ to get their $O(d^2 \log d)$-approximation for Explainable Stochastic $d$-Halfspace Evaluation inevitably loses a factor of $d$. \textbf{How then, should one decide when to switch from performing tests on variables in one $g_i$ to performing tests in another $g_i$?} 
Our approach to answering this question is based on a reduction to a novel type of maximization problem, with a constraint on the \emph{expected} cost. 
We refer to this type of problem as a ``budget-in-expectation problem''.

Our particular budget-in-expectation problem, which we call MP0, is as follows. Given a single $g_i$, 
the probabilities $p_i$, and a budget $B$, find a strategy that maximizes the probability of obtaining a certificate that $f(x)=0$, with the constraint that the \emph{expected} cost of the strategy is at most $B$. \textbf{Here, the term ``in expectation'' is essential and causes an arbitrarily large factor (``gap'') in the attainable probability} relative to a hard, per-realization constraint on cost.
This is obvious for $B<1$, but, more importantly, it also holds for arbitrarily large budgets. 
Algorithmically, with a constraint on expected cost,
the challenge is to figure out when test results are promising enough to justify performing further tests, and when one should just abort. Interestingly, for MP0, conducting the tests in the same (adaptive) order as the optimal strategy for the SBFE problem on $k$-of-$n$ functions does not lead to a constant-factor approximation, even if one aborts this strategy in an optimal way. 
In this paper, we show how to achieve a constant-factor approximation for MP0.

In the literature, there are several problems where one also has to adaptively choose between $d$ alternatives, e.g., Markov chains. A deep solution concept that often leads to optimality is based on the celebrated Gittins index and related indices~\cite{Gittins,Weitzman79OptimalSearch,DumitriuTW03,Singla18,GuptaJSS19}.  
Our approach of reducing to the aforementioned budget-in-expectation problem is very different and has the potential to have applications not only to other SBFE problems, but to other stochastic combinatorial optimization problems. 
While we only obtain an $O(1)$-approximation, the problem we are solving poses a significant challenge not present in problems typically addressed by the Gittins index, namely not only choosing between the $d$ different alternative $g_i$, but also choosing which test to perform to further evaluate that $g_i$. In other words, in each step, we are not merely selecting a Markov chain to advance but a Markov decision process along with an action within that process, a task that tends to be much harder (see, e.g.,~\cite{FuLL23,ScullyT25}).

We summarize our contributions. Our main results are an $O(1)$-approximation algorithm for MBT on partition matroids and an $O(1)$-approximation algorithm for the budget-in-expectation problem on uniform matroids (in fact, on partition matroids as well). On a more conceptual level, we define MBT, a new class of SBFE problems, and the budget-in-expectation problem. We believe that the latter contributions, in particular, open up intriguing directions for future research.

\subsection{Technical Overview}
\label{sec:techoverview}

\begin{figure}
\centering
    \scalebox{.8}{
        \begin{forest}
            for tree ={edge={very thick,Stealth-,color=black!80},s sep = .6cm}
            [{\Cref{thm:maintheorem}: $O(1)$-approximation for MBT on partition matroids},textbox,
                [{\Cref{lem:latency}: latency inequality for MBT on partition matroids},textbox,
                [{\Cref{OPT1BOptimal}: optimal algorithm for\\ finding $1$-certificates on matroids},textbox
                ]
                [{},textbox,color=white,no edge,edge label={node[midway,anchor=center] {
                }}
                ]
                [{\Cref{TableKeyLemma}: $O(1)$-approximation for\\ MP0 on partition matroids with budget $O(B)$},textbox, 
                    [{\Cref{lem:prune}: efficient optimal pruning\\of compactly represented strategies},textbox,edge label={node[midway,right,outer sep=6pt] {
                    }}
                    ]
                    [{\Cref{lem:IOapprox}: pruning achieving opt.\ probability for\\ MP0 on uniform matroids with budget $O(B)$},textbox,edge label={node[midway,right,outer sep=6pt] {
                    }}
                    ]
                ]
                ]
            ]
        \end{forest}
    }

\caption{Overview of the results and how they interact with each other.}
\label{fig:overview}
\end{figure}
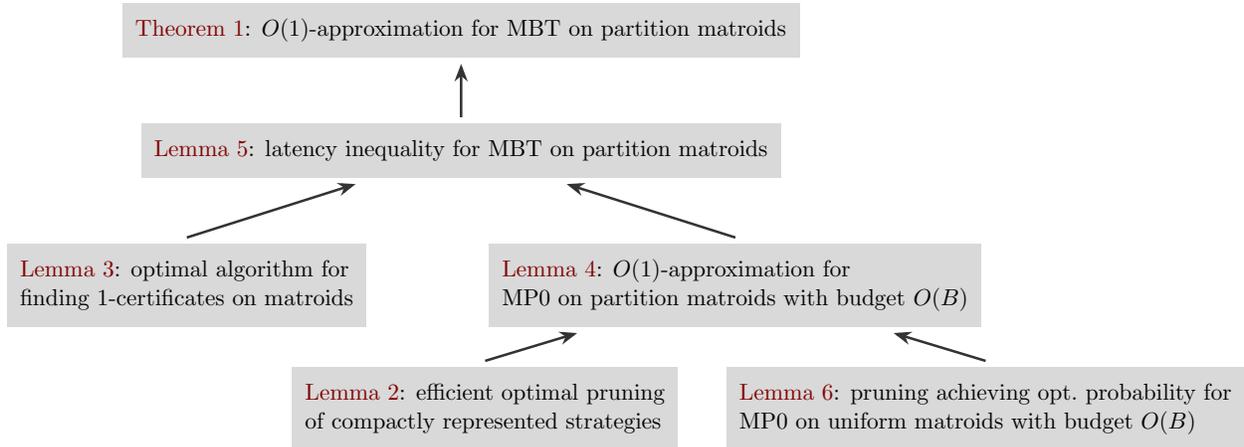

Before discussing further related work,
we start by giving an informal overview of our approach to obtaining an $O(1)$-approximation for the MBT problem on partition matroids. In \Cref{fig:overview}, we present a diagram illustrating the relationship between the main results in the paper.
Readers who prefer to skip this overview and proceed immediately to the detailed presentation of the results can do so without loss of continuity.

\paragraph{Known results on uniform matroids.} 
Let us first recall the elegant optimal strategy for MBT on uniform matroids~\cite{halpern1974} (see also e.g., ~\cite{unluyurtBorosDoubleRegular,BenDov81}).  
As is intuitively clear, testing in decreasing order of probabilities maximizes the probability of certifying $\fM(x)=1$ ($1$-certificate in the following), i.e., finding $k$ active elements, after any number of steps (tests).  
Similarly, testing in the opposite order maximizes the probability of certifying $\fM(x)=0$ ($0$-certificate), i.e, finding $n-k+1$ inactive elements, after any number of steps.
The crucial observation is that the first strategy has to perform at least the first $k$ of its tests before it can terminate, and the second has to perform at least the first $n-k+1$ of its tests before it can terminate. Thus both need to perform the test with the $k$-th largest probability before terminating.  Therefore, there is a strategy that maximizes the probability of finding a certificate (i.e., either a $0$-certificate or a $1$-certificate), again after any number of steps, that performs the test with the $k$-th largest probability first.
Such a strategy can be computed by performing that test and then recursing.
That strategy also minimizes the expected number of queries until a certificate is found.

In fact, as we focus on approximation algorithms in this paper, we may also consider the following simple approximation algorithm for MBT on uniform matroids: Alternate in round-robin fashion between performing the tests in increasing probability order and performing the tests in decreasing probability order.  For any $\ell > 0$, after $2\ell$ steps, the tests with the $\ell$ highest probability and the tests with the $\ell$ lowest probability have all been performed.  Therefore this strategy matches (or surpasses) the probability that the optimal strategy finds a certificate after $\ell$ steps, and this strategy is a $2$-approximation. 

\paragraph{Separating out the value-$1$ case.}

Some of the above can be generalized to \emph{any} matroid $\mathcal{M}$: 
Consider the strategy that tests in decreasing order of probabilities, but modify it to skip any element that would not contribute to a basis, i.e., that would form a dependent set with the elements that have been found to be active already. We call this strategy \OPTone{}.
By a straightforward generalization of the proof for uniform matroids,
we show in \Cref{OPT1BOptimal} (\Cref{sec:OPT1}) that \OPTone{} maximizes the probability of having found a $1$-certificate after any number of steps. Indeed, in the following we focus on finding $0$-certificates, and we eventually combine a strategy handling this task with the strategy \OPTone{}, again using a form of round robin.

\paragraph{Some intuition about the value-$0$ case.} 

Clearly, on partition matroids, finding $0$-certificates is very different from finding $1$-certificates---one only needs to show that there are less than $k_i$ active elements in \emph{some} partition class $i$. One may therefore be tempted to use a ``depth-first'' strategy: Choose a partition class $P_i$ according to some criterion and use the optimal evaluation strategy for uniform matroids to determine whether or not $P_i$ has $k_i$ active elements. Stop testing if that is not the case (having found a $0$-certificate); otherwise proceed to another $P_i$.

\begin{figure}[t!]
    \centering
    \includegraphics[scale=.75]{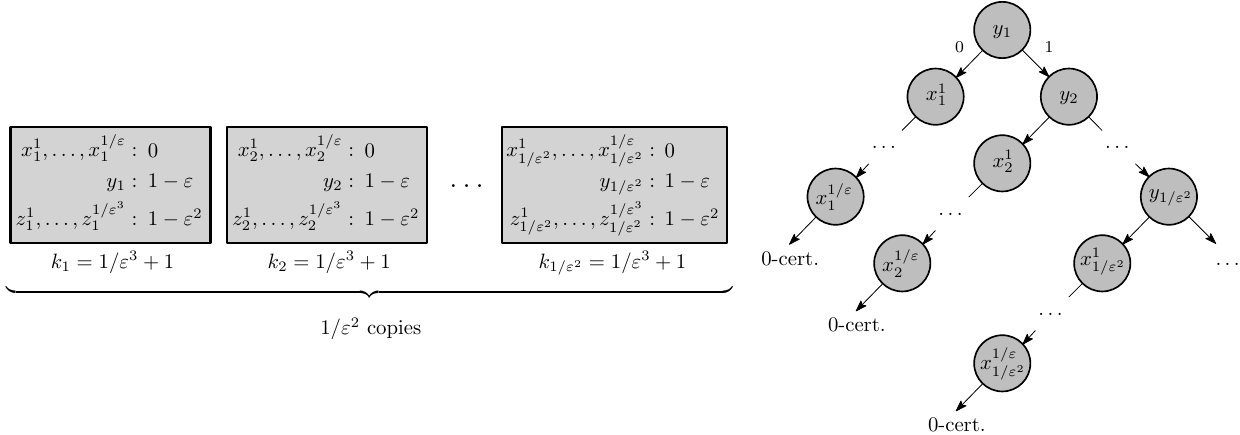}
    \caption{Left: the instance on which depth-first strategies and non-adaptive strategies perform poorly (notation: ``variable name: probability''). Right: the optimal strategy for that instance.}
    \label{fig:depth-first-counter}
\end{figure}

We give a running example in which $f(x)=0$ with very high probability (so that we can ignore the case $f(x)=1$) on which no depth-first strategy is a constant-factor approximation. In our family of instances\footnote{We allow probabilities $0$ and $1$ in some of our counterexamples for the sake of presentation. Such variables still have to be tested in order to include them in a certificate, even if their value is known. It is not difficult to see that such probabilities can be simulated by probabilities very close to $0$ and $1$, respectively. We also name variables differently than $x_1,\dots,x_n$, again for the sake of presentation.}, $d=1/\eps^2$, and for each $i\in[d]$ partition class $i$ contains the following $n_i$ variables:
\begin{compactitem}
    \item $1/\eps$ variables $x_i^1,\dots,x_i^{1/\eps}$ that take value $1$ with probability $0$,
    \item a variable $y_i$ that takes value $1$ with probability $1-\eps$, and
    \item $1/\eps^3$ variables $z_i^1,\dots,z_i^{1/\eps^3}$  that take value $1$ with probability $1-\eps^2$,
\end{compactitem}
for some small $\eps>0$ (see \Cref{fig:depth-first-counter}), and we also let $k_i=1/\eps^3+1=n_i-1/\eps$. Note that any partition class has a $0$-certificate with probability approaching $1$ (as $\eps\to 0$). Intuitively, the challenge is to identify a partition class in which this certificate can be obtained cheaply. We consider the optimal strategy and a depth-first strategy:
\begin{compactitem}
    \item The optimal strategy tests the $y_i$ in distinct partition classes until it finds a partition class $i$ in which $y_i$ has value $0$ (which exists with probability $1-(1-\eps)^{1/\eps^2}$, a probability large enough to ignore the other case), within expected cost less than $1/\eps$, and then tests all $x_i^j$ in this partition class. The expected total cost is $O(1/\eps)$ (as $\eps\to0$).
    \item In contrast, consider a depth-first strategy. In the first partition class $i$ it considers, $y_i$ will have value $1$ with probability $1-\eps$, in which case, taking into account the $0$s from the $x_i^j$, there is a single $0$ missing toward a $0$-certificate. The depth-first strategy will likely (with probability $1-(1-\eps^2)^{1/\eps^3}$) find another $0$ among the $z_i^j$, but it has has to test approximately $1/\eps^2$ of them in expectation to find such a $0$. The expected total cost of a depth-first strategy is therefore $\Omega(1/\eps^2)$.
\end{compactitem}

Observe that the same instance shows that non-adaptive strategies (strategies that test in a fixed order, independently of the outcomes) are not constant-factor approximations. Indeed, a non-adaptive strategy cannot couple the investment into the $x_i^j$ of some partition class $i$ to the value $0$ of $y_i$. In any partition class, it can thus only find $0$-certificates at rate $O(\eps^2)$ (cost per probability of finding a certificate), resulting in a total expected cost of again $\Omega(1/\eps^2)$.

In \Cref{sec:related}
we will discuss the other relevant approaches from the SBFE literature, and will also rule them out as ways to achieve a
 constant-factor approximation for MBT on partition matroids.

\paragraph{Overview of our approach.} As mentioned earlier, at the heart of our new approach, which does lead to a constant-factor approximation, we approximate the following novel problem. Given budget $B$, Boolean function $f$, and probabilities $p_i$, find a testing strategy that uses at most budget (cost) $B$ \emph{in expectation} and maximizes the probability of finding a $0$-certificate for $f$.

Crucially, the strategy is allowed to stop testing without having found a certificate, and it is allowed to use internal randomization. We call this problem the Maximum-Probability 0-certificate problem (MP0). In particular, we show how to approximate MP0 (in a sense that we will describe later) for functions $f^\mathcal{M}$ where $\mathcal{M}$ is a uniform matroid, corresponding to one of the partition classes. Along the way, we also approximate MP0 for $\fM$ where $\mathcal{M}$ is a partition matroid. Indeed, we believe that having to meet a budget constraint in expectation is natural and interesting in its own right. Quite surprisingly, we are not aware of any other work in stochastic combinatorial optimization which considers problems with this feature.

To get some intuition for why approximating MP0 on uniform matroids might be useful toward approximating MBT on partition matroids, consider MP0 for
a partition class $i$ from the counterexample for depth-first strategies depicted in \Cref{fig:depth-first-counter}. Given a budget of $B=2$, the optimal solution to MP0 here is as follows: Test $y_i$ and, if its value is $0$, continue by testing all $x_i^j$ variables, otherwise stop. The resulting strategy has expected cost at most $B=2$, and the probability it finds a $0$-certificate is $\eps$.
Note that this strategy is precisely what we would get if we took the optimal strategy we described for the entire counterexample, considered the substrategy it used in performing tests in partition class $i$, and pruned that substrategy at the time it would stop testing in class $i$.
Hence, we may hope that, by combining (approximately) optimal solutions to MP0 for the different partition classes, with appropriate budgets, one can obtain an (approximately) optimal solution of MBT on partition matroids.

Our strategy for finding $0$-certificates for MBT on partition matroids operates in phases, starting with phase $0$, of exponentially increasing lengths. The key to proving the approximation is then to use a latency-based argument. More precisely, we show a recurrence that relates three quantities:
\begin{compactitem}
    \item[(i)] the probability of not having found a certificate after the current phase,
    \item[(ii)] the probability of not having done so after the previous phase, and 
    \item[(iii)] the probability that the optimal strategy has not found a certificate in a comparable (slightly shorter) period. 
\end{compactitem}
Such recurrences have previously been used~\cite{ImNZ16,EneNS17,ghuge2022non}, but the crucial difference to previous work is that the cost of any of \emph{our} phases may greatly vary depending on the values of the variables (and internal randomization of the algorithm); its cost is only upper-bounded (by said exponential) \emph{in expectation}. (Note that the periods that we are considering in our analysis of the optimal strategy \emph{do} have a deterministic length.)

To achieve this recurrence, it turns out to be sufficient to approximate MP0 on partition matroids (with the budget being the bound on the expected cost for the current phase). We show that this suffices in our proof of the main result, \Cref{thm:maintheorem} (\Cref{sec:top-level}). An optimal strategy for MP0 on partition matroids may well be complicated; we do not exclude that it switches back and forth between partition classes arbitrarily often adaptively, i.e., depending on its observations. As we show in \Cref{TableKeyLemma} (\Cref{sec:KS}), however, for an approximation, it is sufficient to allocate the budget for the phase to the different partition classes (in an approximately optimal way) and then to run (approximately optimal) solutions to the corresponding MP0 instances (on uniform matroids) in any order, without switching back and forth. 
In allocating the budgets, we seek to maximize the \emph{sum}, over all the classes of the partition matroid, of the individual probabilities of finding a $0$-certificate for each of the classes.  This sum is easily shown to be within a $1-1/e$ factor of the probability of finding a $0$-certificate for the partition matroid itself.

\begin{figure}[t!]
    \centering
    \includegraphics[scale=.75]{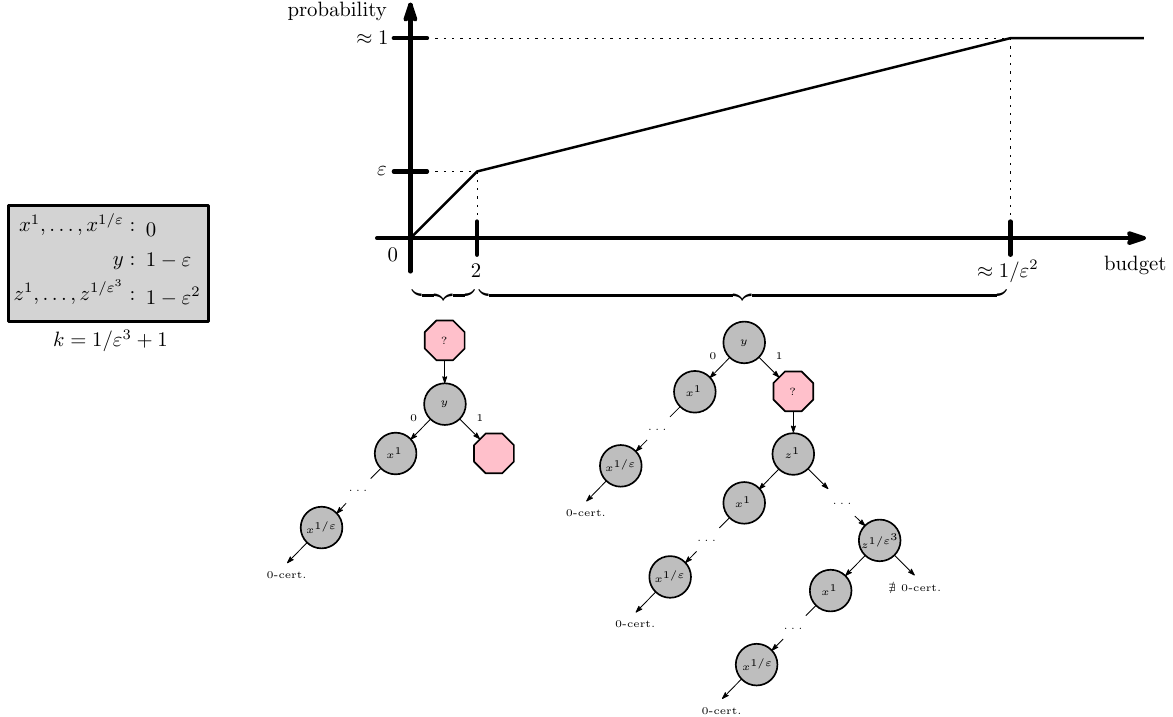}
    \caption{Left: A uniform matroid (a partition class from the counterexample for depth-first and non-adaptive strategies). Right: A (not-to-scale) graph of the function relating budget and attainable probability of finding a $0$-certificate for that uniform matroid, along with the corresponding strategies. The red octagon corresponds to stopping; when a ``?'' is depicted, it only corresponds to stopping with a certain probability. The (randomized) strategy on the left is for budgets between 0 and 2, and the strategy on the right is for budgets between 2 and approximately $\frac{1}{\eps^2}$.}
    \label{fig:function-graph}
\end{figure}

To maximize said sum of probabilities of finding a $0$-certificate, we consider the subinstance induced by all previous phases. We approximately solve MP0 on uniform matroids \emph{simultaneously for all budgets}. We show in \Cref{sec:IO} (in combination with \Cref{sec:sub:Prun}) that we can efficiently compute (a compact representation of) a piece-wise linear, monotone, and concave function $h:\mathbb{R}_+\rightarrow[0,1]$ such that, for any budget $B$, $h(B)$ is at least the probability that can be achieved with budget $c\cdot B$ for some constant $c$. \Cref{fig:function-graph} illustrates what this function looks like in the previous example (depicted in \Cref{fig:depth-first-counter}) when $c=1$. We also show how to compute, for all budgets $B$, a strategy achieving a probability of $h(B)$ with expected cost $B$. The problem of approximately maximizing the sum of probabilities can then be seen to be a continuous knapsack problem, for which the Greedy algorithm (according to density) is optimal.

\paragraph{Solving MP0 on uniform matroids.} The problem that remains to be approximated is MP0 on uniform matroids.\footnote{For uniform matroids, the corresponding problem for finding 1-certificates is effectively an equivalent problems, since the complement of a $k$-of-$n$ function is an $(n-k+1)$-of-$n$ function.}

Note that, when the budget is equal to the minimum expected evaluation cost for the corresponding instance of MBT on the same uniform matroid, the evaluation strategy discussed in the beginning of this subsection for MBT on uniform matroids is optimal. 
It achieves a probability of finding a $0$-certificate equal to the probability that $\fM(x)=0$. Note that, in the example depicted in \Cref{fig:function-graph}, \emph{pruning} the same strategy (i.e., stopping early, possibly in a randomized way) results in an optimal strategy as well. So one could (boldly) conjecture that this is true generally, or that one can at least always obtain a constant-factor approximation by pruning the optimal evaluation strategy.
To see that the conjecture is false, consider an instance of MP0 on a uniform matroid with
\begin{compactitem}
    \item a single variable that takes value $1$ with probability $1-\eps$ and 
    \item $1/\eps$ variables that take value $1$ with probability $1-\eps^2$,
\end{compactitem}
where $n=1+1/\eps$, and $k=n-1$.
 Any pruning of the optimal evaluation strategy must begin by testing 
 the probability-$(1-\eps^2)$ variables, and can only switch to testing the probability-$(1-\eps)$ variable once it has found a $0$ (or has tested all the probability-$(1-\eps^2)$ variables).  
  With a budget of $2$, the resulting probability
of finding a $0$-certificate is at most $O(\eps^3)$.  In contrast,
an optimal strategy for MP0 with a budget of $2$ first tests the probability-$(1-\eps)$
variable to see if it yields 0, and only 
makes the investment into the probability-$(1-\eps^2)$ variables in that event, yielding a probability of $\Omega(\eps^2)$ of finding a 0-certificate.
We give a rigorous argument in \Cref{app:counterexamples}. This example shows that MP0 is quite different from the evaluation problem.

One may still hope that a different such ``universal'' strategy exists---a strategy that can be pruned to be a constant-factor approximation for every budget. Indeed, showing such a structural result about some strategy would give us an algorithm already: As we show in \Cref{lem:prune} (\Cref{sec:sub:Prun}), the optimal pruning of a tree down to a certain budget $B$ can be computed by repeatedly pruning the subtree with the lowest \emph{density}, which is the ratio between two quantities, both conditioned on having reached the root of the subtree: (1) the probability of finding a $0$-certificate in $\mathcal{M}$, and (2) the expected cost incurred in the subtree.
This Greedy algorithm prunes in a continuous fashion (each chosen subtree is pruned with probability that increases continuously from 0 to 1) and stops when the budget constraint is fulfilled with equality. Conveniently, a straightforward modification of this algorithm also computes a piece-wise linear function mapping any budget to the objective achieved when pruning to that budget, with the breakpoints corresponding to the points at which the Greedy algorithm switches between subtrees. This function fulfills the aforementioned approximation guarantee if and only if the original strategy has said universality property.

A slight complication is that a na\"ive implementation of the Greedy algorithm requires a full representation of the strategy. Fortunately, although representing the strategy as a tree takes exponential space, nodes at which an equal number of $0$s has been found \emph{and} an equal number of $1$s has been found can be merged, such that the emerging directed acyclic graph (DAG) has polynomial size. This DAG can be computed directly (without explicitly merging), and the Greedy algorithm can be adapted to work on DAGs in a straightforward manner, resulting in a polynomial-time algorithm.

But what could such a universal strategy be? Another natural candidate would be to test in increasing order of probabilities---after all, we are looking for a $0$-certificate. Consider, however, an instance with $k=1$, $1/\eps$ tests with probability $0$ of being 1 and one test with probability $1-\eps$ of being 1. Given a budget of $2$, one can find a 0-certificate with probability $\eps$ by first testing the probability-$(1-\eps)$ variable (and only continuing if the test has outcome $0$). 
Testing the probability-$0$ variables first, on the other hand, means that in order to get to the probability-$(1-\eps)$ variable and find a 0-certificate, one must first perform all other tests.
The cost of performing all tests is $1/\eps+1$.
Therefore, with a budget of $2$, the best one can do is to perform all tests with probability $2/(1/\eps+1)$, and otherwise to forgo testing entirely.  The resulting probability of finding a 0-certificate is $O(\eps^2)$.

Testing variables in \emph{de}creasing order of probabilities is clearly is not universal in the above sense either; e.g., take any instance and add a lot of variables with probability very close to $1$---testing these variables just does not make enough progress toward finding a $0$-certificate. (Note that the example we gave for the optimal evaluation strategy is also a counterexample, albeit more complicated.)

Considering the previous counterexamples, we want to design a strategy that
\begin{compactitem}
    \item[(i)] ``learns'' whether it will be expensive to find a $0$-certificate, so that we have the possibility of pruning it in that case, and at the same time
    \item[(ii)] makes fast-enough progress toward finding a $0$-certificate.
\end{compactitem} This is achieved by a strategy that we call InsideOut. It is a round robin between two sequences:
\begin{compactitem}
    \item that of variables $i$ with $p_i \leq 1/2$, in decreasing order, and
    \item that of variables $i$ with $p_i > 1/2$, in increasing order.
\end{compactitem}
This intuitively takes care of both goals: It takes care of goal (i) because high-entropy tests are included relatively early on in the sequence. It also takes care of goal (ii) because, if there are tests that are active with high probability, we include some (possibly different) tests that are active with comparable probability (i.e., with probability at least $1/2$) early on in the sequence. 

In~\Cref{sec:IO}, we show that InsideOut is a universal strategy in the aforementioned sense. To this end, we modify an arbitrary solution of MP0 to obtain a pruning of InsideOut, in a way that increases the expected cost by at most a constant factor and does not decrease the probability of finding a $0$-certificate. Our proof is by induction on the length of the sequences and uses a careful case analysis and charging arguments. This is the most technical part of the paper. We give a more detailed overview of the proof in the beginning of~\Cref{sec:IO}.

\subsection{Further Discussion of Related Work}

\label{sec:related}

\paragraph{Alternative approaches.} 
There are previous approaches one could apply to the MBT problem on partition matroids which do not yield a constant-factor approximation bound.  
Several works on SBFE problems partition the distribution into, say, $\ell$ parts and, for any of these parts, devise algorithms that have approximately minimum cost conditioned on the respective part. By a round robin between these algorithms, one then obtains an algorithm that has an approximation guarantee for the entire problem that can be bounded in $\ell$ and the maximum individual approximation guarantee~\cite{allen2017evaluation,grammel2022,ghuge2022non,plank2024simple}. For our problem, a straightforward approach would be a round robin between \OPTone{} and the $d$ strategies finding $0$-certificates for each of the partition classes, but this approach inevitably loses a factor linear in $d$ in the approximation guarantee.

As mentioned above, the MBT problem on partition matroids is a special case of the
Explainable Stochastic $d$-Halfspace Evaluation problem studied by Ghuge et al.~\cite{ghuge2022non}, which involves the evaluation of a function $h(g_1,\ldots,g_d)$, where the $g_i$ are linear threshold functions.  Notably, their $g_i$ can be defined on overlapping sets of variables.
Their algorithm, like ours, proceeds in phases with exponentially increasing budgets.
However, for each phase $j$, they allocate the same budget $B_j$ for tests to each of the $g_i$ (where budget $B_j$ is a hard constraint on the maximum cost).
Within the phase, they iterate through the $g_i$, performing the tests on one $g_i$ before proceeding to the next.  This iteration leads to a dependence on $d$ in their approximation guarantee.

Deshpande et al.\ introduced a very different approach to developing approximation algorithms for SBFE problems.  The approach is to reduce the SBFE problem for a function $f$ to a Stochastic Submodular Cover problem for a submodular utility function $g$, which is constructed from $f$.  A greedy algorithm is then used to solve the Stochastic Submodular Cover problem for $g$, resulting in an approximation factor for the original SBFE problem that depends on certain properties of the constructed $g$.   For some types of functions $f$, if the utility function $g$ is constructed carefully, a small approximation factor can be achieved.
However, we show in \Cref{app:submodulargoal} that when $f$ is 
the conjunction (or disjunction) of $k_i$-of-$n_i$ functions on disjoint sets of variables, it is impossible to construct $g$ such that the resulting approximation factor is $o(n^{\beta})$, for any constant $0 < \beta < 1$.  

\paragraph{Arbitrary-Cost Tests.} 
Several of the aforementioned SBFE problems have also been considered in the more general setting where each variable $x_i$ has an associated non-negative testing cost $c_i$, and the goal is to minimize total expected testing cost, e.g.,~\cite{gkenosis22stochastic,ghuge2022non,hellerstein2024quickly}. 
The optimal strategy  
for MBT on uniform matroids (i.e., $k$-of-$n$ functions) in the arbitrary-cost case is closely related to the optimal strategy 
for the unit-cost case~\cite{SalloumBreuer84,BenDov81}. 
For the MBT problem on partition matroids, we leave as an open question whether our main result, the existence of a polynomial-time constant-factor approximation algorithm, also holds for arbitrary costs.  We note that our reduction from the MBT problem on partition matroids to MP0 on uniform matroids still works for arbitrary costs, with a straightforward modification of \OPTone{} using probability--cost ratios.  However, 
the most difficult part of our algorithm is the solution of MP0, and 
this problem appears to be significantly more challenging for arbitrary costs.   

\paragraph{Other benchmarks.} 

We remark that our benchmark is \emph{not} that of the (expected) certificate size (see, e.g.,~\cite{charikar2000query,MegowS23,kaplanMansour-Stoc05,allen2017evaluation}) or, equivalently, the performance of an all-knowing ``prophet''~\cite{Lucier17,CorreaFHOV18}. In fact, for the MBT problem, it is easy to see that it is impossible to achieve a constant-factor approximation (competitive ratio) to such a benchmark, even for uniform matroids. 
Instead of this ``online-algorithms perspective'', we take the ``approximation-algorithms perspective'', and use  the optimal \emph{strategy} as our benchmark.  
This is the benchmark used in most work on SBFE problems, and with it, we can in fact achieve a constant-factor approximation ratio.

\section{Preliminaries}
\label{sec:prelim}

In this section, we define terms related to Boolean functions, strategies, and prunings.  We exclude formal definitions already given in \Cref{sec:Intro}. We use $[n]$ to denote $\{1,\dots,n\}$ throughout.

Let $f:\{0,1\}^n \rightarrow \{0,1\}$ be a Boolean function. We treat $f$ as a function of the $n$ input variables $x_1,\ldots,x_n$, i.e., as $f(x_1,\ldots,x_n)$. An \emph{assignment} $\pi$ (to these variables) is an element of $\{0,1\}^n$. Then $\pi_i$ is the assignment to input variable $x_i$.
A {\em partial assignment} $\sigma$ is an element of $\{0,1,*\}^n$, where $\sigma_i=*$ means $\sigma$ does not assign a value to $x_i$.
 Because we consider the $x_i$ to be random variables, we will often refer to an assignment to the $x_i$ as a {\em realization} (of the $x_i$).
An {\em extension} of a partial assignment $\sigma$ to the input variables of $f$ is a partial assignment $\sigma'$ to those variables such that $\sigma'_i=\sigma_i$ for all $\sigma_i \neq *$.  
The partial assignment $\sigma$ induces a function $f_{\sigma}$ from $f$, produced by
taking all inputs $x_i$ to $f$ such that $\sigma_i \neq *$, and fixing the value of each such $x_i$ to $\sigma_i$, so that the function value depends only on the $x_i$ not assigned values by $\sigma$.
For example, if $f(x_1,x_2,x_3,x_4) = (x_1 \vee x_2) \wedge (x_3 \vee x_4)$ and $\sigma = (1,*,0,*)$, then 
$f_{\sigma}(x_2,x_4) = f(1,x_2,0,x_4) = x_4$.
For $\ell \in \{0,1\}$, an {\em $\ell$-certificate} of $f$ is a partial assignment $\sigma$ $\in \{0,1,*\}^n$ such that $f(\sigma')=\ell$ holds for all extensions $\sigma'\in\{0,1\}^n$ of $\sigma$. A {\em certificate} of $f$ is a 1-certificate or a 0-certificate of $f$.

We define a \emph{(deterministic) testing strategy} 
$S$ for $f$ through a binary decision tree, one of its possible representations. Each of the internal nodes of the decision tree is labeled with a variable $x_i$, corresponding to performing a ``test'' or ``query'' to determine the value of $x_i$.  We refer to performing this test as \emph{testing $x_i$}. 
Leaves correspond to ``stopping'' or ``terminating''. 
One child of each internal node corresponds to $x_i=1$ and the other to $x_i=0$.  On each root-leaf path, a variable $x_i$ can only appear once.
Given an assignment $\pi$ $\in \{0,1\}^n$, the strategy induces a sequence of tests to perform on $\pi$, namely the $x_i$ in the nodes on the root-leaf path traced by obtaining outcome $\pi_i$ for each test $x_i$.  
This root-leaf path corresponds to a partial assignment where $x_i=\pi_i$ for all $x_i$ tested on the path, and $x_i=*$ otherwise. If for all $\pi$ $\in \{0,1\}^n$, this partial assignment is a certificate for $f$ and the leaf at the end of the path is labeled with the value of $f(\pi)$, 
then we say that the strategy {\em computes} or {\em evaluates}~$f$.

A {\em randomized testing strategy} $S$ for $f$ can be represented by a binary decision tree with two types of internal nodes: nodes labeled $x_i$ for some $i \in [n]$, that correspond to performing the test on $x_i$, and nodes $v$ labeled with some probability $
\alpha_v$, where $0 < \alpha_v < 1$. These latter nodes correspond to flipping a coin with probability $\alpha$ of being heads, where one child of the node corresponds to heads and the other to tails.
Still, each variable may only appear once on a path from the root. 
Equivalently, we can define a randomized testing strategy to be a distribution over deterministic testing strategies.
Unless otherwise indicated, we will use the term {\em testing strategy} (or just {\em strategy}) to refer to a deterministic testing strategy.

There are thus two sources of randomness for the random variables we consider: the probability distribution on the assignments $\pi \in \{0,1\}^n$ and the internal randomization of the strategies. When we compute probabilities and take expectations, this is done with respect to both these sources of randomness simultaneously. For a (possibly randomized) strategy $S$, let $\Cost{S}$ be such a random variable: the total cost incurred when running $S$ on a random assignment $\pi$.

A Boolean function $f = f\RB{x_1, \ldots, x_n}$ is called \emph{symmetric} if the function value only depends on $\sum_{i=1}^n x_i$. Consider some possibly randomized testing strategy $S$ for such a function. If its decisions only depend on how many 0s and 1s have been observed so far in the test results (not depending on {\em which} tests resulted in those 0s and 1s), and possibly the internal randomization, then we say that $S$ is {\em elementary}.
Such a strategy $S$ can be equivalently represented by a directed acyclic graph (DAG) by merging nodes in which the same number of 0s and the same number of 1s have been observed. In contrast to the representation as a binary decision tree, whose size can generally not be bounded by a polynomial in $n$, this DAG has size polynomial in $n$.
We call this the decision DAG (representation) of the elementary strategy $S$. For a strategy $S$ with a fixed representation (as a binary decision tree or, in the case of an elementary strategy, a DAG), we use $V(S)$ to denote the set of nodes in that representation.

{\em Pruning a node} of a  (tree or general DAG) representation of a possibly randomized testing strategy means terminating at that node.
That is, in the case of a tree representation, the whole subtree is removed and replaced by a leaf.
A {\em randomized pruning} of a node (with probability $\alpha$) means introducing a random choice at that node: With some probability $1-\alpha$, continue (as originally), and with probability $\alpha$, terminate immediately.
A (randomized) testing strategy $S'$ {\em is a (randomized) pruning} of a (possibly randomized) testing strategy $S$ if there is a sequence $S=S_1,S_2,\dots,S_k=S'$ for some $k\geq 1$ such that, for all $i\in[k-1]$, $S_{i+1}$ is produced from $S_i$ by 
a (possibly randomized) pruning of a node in a representation of $S_i$.
As with the definition of randomized testing strategies, a randomized pruning of a strategy $S$ could alternatively be defined as a distribution over (deterministic) prunings of $S$.

\subsection{Computing an optimal pruning}
\label{sec:sub:Prun}

The following definitions and statements are made for symmetric functions $f:\{0,1\}^n \rightarrow \{0,1\}$ (and only used when $f$ is a $k$-of-$n$ function). Given some $L\subseteq\{0,1\}$, let $U:\{0,1,*\}^n \rightarrow \{0,1\}$ be a utility function that assigns a utility value of $1$ to each partial assignment that is an $\ell$-certificate for some $\ell\in L$ and utility $0$ otherwise.
For $S$ a (possibly randomized) testing strategy for $f$, we define $U(S)$ to be $U(\sigma_S)$ where $\sigma_S$ is the random partial assignment produced by applying strategy $S$. Thus $U(S)$ is a random variable.
We call $U$ a certificate utility function (for $f$).

For a (possibly randomized) strategy $S$, let $\prunings(S)$ denote the set of all randomized prunings of $S$.
 Let $q_S:\mathbb{R}_{\geq 0}\to\mathbb{R}$ with $$q_S(B):=\max\, \CB{\Ex{U(S')} \,|\, S'\in \prunings(S) \wedge \mathbb{E}[\cost(S')]\leq B}.$$ That is, $q_S$ is the maximum expected utility of any randomized pruning of $S$ having expected cost at most $B$.

Toward computing this function, given a (possibly elementary) strategy $S$ and a node $v$ in the corresponding decision tree or DAG, define the random variable $\util(S,v)$ as follows.
If $S$ visits $v$, and an $\ell$-certificate has not yet been found before conducting the test at $v$, then it is $\util(S)$; otherwise, 
it is $0$.
Similarly, define the random variable $\text{cost}(S,v)$: If $S$ visits $v$, then it is the cost for conducting the test at $v$ and all tests that $S$ performs after that; if $S$ does not visit $v$, then it is $0$.
Clearly, for the root $r$, we have $\util(S,r)=\util(S)$ and $\text{cost}(S,r)=\text{cost}(S)$.

We first present a lemma which establishes that, for an elementary strategy, it does not matter whether we prune the corresponding decision tree or the corresponding decision DAG. This is a very intuitive statement since, to obtain the DAG from the tree, one only merges equivalent nodes, but we give a proof in \Cref{app:pruninglemmas} for completeness. The lemma is needed to obtain algorithms that run in time polynomial in the size of the input.

\begin{lemma}\label{lem:properprunings}
    Let $S$ be an elementary testing strategy for a symmetric function.
    Let $R$ be a (randomized) pruning of $S$.
    Then there exists a (randomized) pruning $R'$ of $S$ that is elementary, such that $\Ex{\Cost{R}} = \Ex{\Cost{R'}}$ and $\Ex{\Util{R}} = \Ex{\Util{R'}}$.
\end{lemma}

\begin{algorithm}[t]
    \caption{ComputePruningFunction($S$)}
    \label{alg:computepruningfunction}
    \SetAlgoVlined
    $S' \gets S$\;
    $\text{BP} \gets \{(\infty,\Ex{\Util{S}}),\RB{\Ex{\Cost{S}},\Ex{\Util{S}}}\}$ (set of breakpoints)\;
    \While{$\mathbb{E}[\cost(S')]>0$}{
        choose $v^\ast \in \arg\min_{v\in V(S')} \frac{\mathbb{E}[\util(S',v)\mid\text{$S'$ reaches $v$}]}{\mathbb{E}[\cost(S',v)\mid\text{$S'$ reaches $v$}]}$\;
        modify $S'$ by pruning at $v^\ast$\;
        $\text{BP} \gets \text{BP}\cup\{(\mathbb{E}[\cost(S')],\mathbb{E}[\util(S')])\}$\;
    }
    \Return{$\text{BP}$}
\end{algorithm}

\begin{algorithm}[t]
    \caption{ComputePrunedStrategy($S$, $B$)}
    \label{alg:computeprunedstrategy}    
    \SetAlgoVlined
    $S' \gets S$\;
    \While{$\mathbb{E}[\cost(S')]>B$}{
        choose $v^\ast \in \arg\min_{v\in V(S')} \frac{\mathbb{E}[\util(S',v)\mid\text{$S'$ reaches $v$}]}{\mathbb{E}[\cost(S',v)\mid\text{$S'$ reaches $v$}]}$\;
        $p \gets \min\left\{1,\frac{\mathbb{E}[\cost(S')]-B}{\mathbb{E}[\cost(S',v^\ast)]}\right\}$\;
        modify $S'$ by a randomized pruning (with probability $p$) at $v^\ast$\;
    }
    return $S'$
\end{algorithm}

Given strategy $S$ (in tree or DAG representation), the procedure ComputePruningFunction($S$) computes a representation of the function $q_S$.   As we prove below, $q_S$ is a piecewise linear function, and thus is represented by its breakpoints.  The procedure computes these breakpoints by maintaining a strategy $S'$ (initially $S$) and iteratively picking a node $v$ that minimizes the ratio of $\mathbb{E}[\util(S',v)\mid\text{$S'$ reaches $v$}]$ and $\mathbb{E}[\cost(S',v)\mid\text{$S'$ reaches $v$}]$. 
For each such $v$, it modifies $S'$ by making it stop at $v$, that is, by (deterministically) pruning node $v$.  
The strategies obtained in this way (until the empty strategy is reached) correspond to the breakpoints of $q_S$.  More generally, each value $B$ in the domain of $q_S$ corresponds to the strategy that would be obtained by allowing \emph{randomized} pruning of each node $v$ chosen in the procedure. 
The procedure ComputePrunedStrategy($S$, $B$) returns the strategy corresponding to $B$.
Pseudocode for the procedures is given in \Cref{alg:computepruningfunction} and \Cref{alg:computeprunedstrategy}, respectively.

We record the main properties of these procedures in the following lemma, proved in \Cref{app:pruninglemmas}.
We assume that the input strategy $S$ is given as input either as a binary decision tree or (when $S$ is an elementary strategy) as a DAG. The output representation has the same type as the input.

\begin{lemma}\label{lem:prune}
    Let $S$ be a testing strategy for a symmetric function $f$, and let $U$ be a certificate utility function for $f$. Then:
    \begin{itemize}
        \item[(i)] The function $q_S$ is piecewise linear, monotone, and concave.
        \item[(ii)] The breakpoints of $q_S$ are computed by ComputePruningFunction($S$). 
        Furthermore, for each $B\in\mathbb{R}_{\geq 0}$, a strategy $S'\in\prunings(S)$ with $\Ex{U(S')}=q_S(B)$ and $\mathbb{E}[\cost(S')]\leq B$ is computed by ComputePrunedStrategy($S$,$B$).
        \item[(iii)] ComputePruningFunction($S$) and ComputePrunedStrategy($S$,$B$) can
         be implemented to run in time polynomial in the size of the input representation of $S$.
    \end{itemize}
\end{lemma}

\section{An $O(1)$-Approximation Algorithm for MBT on Partition Matroids}
\label{sec:top-level}

In this section, we give the description and analysis of our $O(1)$-approximation algorithm for MBT on partition matroids. We describe our algorithm in a top--down manner and defer the description and analysis of subroutines to later sections. We call our algorithm ALG and prove the following theorem.

\begin{theorem}
\label{thm:maintheorem}
ALG is a polynomial-time $O(1)$-approximation algorithm for MBT on partition matroids, that is, the SBFE problem for functions $f=f^{\mathcal{M}}$ where $\mathcal{M}$ is a partition matroid.
\end{theorem}

ALG does this by combining elements of two strategies, the deterministic strategy \OPTone{} and the randomized strategy \KS, both of which are evaluation strategies for~$f^{\mathcal M}$ (when given sufficient budget), but with different goals: finding $1$-certificates and finding $0$-certificates, respectively. We describe and analyze these strategies in \Cref{sec:OPT1} and \Cref{sec:KS}, respectively. Apart from the underlying instance, both algorithms receive two parameters as input: A partial realization (of already observed values) and a budget that the algorithm is supposed to use (in expectation). For our present analysis of ALG, we state two key lemmas concerning these strategies here.

We have the following lemma about \OPTone{}.

\begin{restatable}{lemma}{optonelemma}
\label{OPT1BOptimal}
Let $\mathcal{M}$ be a matroid and let $f:=f^{\mathcal{M}}$.
Let $\sigma$ be a partial assignment to the input variables of $f$ and $B\in\mathbb{N}$. The per-realization cost of \OPTone{}($\sigma$,$B$) is at most $B$.
For every (possibly randomized) testing strategy $Z$ for the function $f_\sigma$ with per-realization cost at most $B$, the probability that \OPTone($\sigma$,$B$) has found a 1-certificate of $f_\sigma$ is at least the probability that $Z$ has found a 1-certificate of $f_\sigma$. Finally, \OPTone{} runs in polynomial time.
\end{restatable}

In contrast, the (more complicated) algorithm \KS{} is specifically designed for the case of partition matroids.
 We have the following lemma.

\begin{restatable}{lemma}{tablelemma}
\label{TableKeyLemma}
Let $\mathcal{M}$ be a partition matroid and let 
$f:=f^{\mathcal{M}}$.
Let $\sigma$ be a partial assignment to the input variables of $f$ and $B\in\mathbb{R}_{\geq 0}$. The expected cost of 
$\KS(\sigma,B)$ is at most $B$. The probability that $\KS(\sigma,B)$ finds a 0-certificate of $f_{\sigma}$ is at least $1-1/e$ times the probability that any (possibly randomized) testing strategy finds a 0-certificate of $f_{\sigma}$ within expected cost $B/6$. Finally, $\KS(\sigma,B)$ runs in polynomial time.
\end{restatable}

It can be shown that, due to the concavity of $q_S$ (\Cref{lem:prune}), this result implies an $O(1)$-approximation to MP0 on partition matroids. For technical reasons, we need this stronger statement to prove our main theorem.

\begin{algorithm}[t]
    \caption{ALG}
    \label{alg:main_alg}
    \SetAlgoVlined
    $\sigma \gets (*,\dots,*)$\;
    \For{$\ell=0,1,\dots$}{
        execute \OPTone{}$(\sigma,2^\ell)$\;    
        execute $\KS(\sigma,6\cdot 2^\ell)$\;
        update $\sigma$ to incorporate the outcomes of the tests just performed\;
        \If{$\sigma$ is a $v$-certificate for $v\in\{0,1\}$}{
        \Return{``$f^{\mathcal{M}}(x)=v$''}
        }
    }
\end{algorithm}

ALG operates in phases $\ell\in\{0,1,\ldots\}$.  In phase $\ell$, which starts with partial realization $\sigma$, ALG
executes \OPTone$(\sigma,2^\ell)$ followed by $\KS(\sigma,6\cdot 2^\ell)$.
However, during the execution of \KS{}, each time the next test has already been performed by \OPTone{}, 
ALG does not
actually perform the test again. (Indeed, our model does not allow re-performing tests.) Instead, it uses the result obtained when the test was previously performed by \OPTone{}.
We give the pseudocode in~\Cref{alg:main_alg}.

We analyze ALG. Toward this, let OPT be an optimal evaluation strategy for $f$, meaning it has minimum expected cost among all evaluation strategies for $f$.  Minimum expected cost can be achieved by a deterministic strategy, and we assume OPT is deterministic. We use the following definitions:

\begin{itemize}
    \item Let $\mu_{\ell}$ be the probability that ALG has not found a certificate of $f$ by the end of phase $\ell$.
    \item Let $\mu^*_{\ell}$ be the probability that the cost incurred by OPT is at least $2^{\ell}$, i.e., OPT does not find a certificate of $f$ before incurring cost $2^{\ell}$.
\end{itemize}

In related works~\cite{ImNZ16,EneNS17,ghuge2022non}, statements similar to the following have been shown. While our setting is different, the proof uses similar ideas.

\begin{lemma}
\label{lem:latency}
The following holds for all $\ell \geq 1$:
$$\mu_{\ell} \leq \frac1e\cdot \mu_{\ell-1} + \mu_{\ell}^*\,.$$
\end{lemma}

\begin{proof}

Let $\sigma$ be a random variable whose value is the partial realization representing the outcomes of the tests performed by ALG in phases 0 through $\ell-1$.
Let $W$ be the set of all such partial realizations that are not certificates for $f$.   
Thus, if $\sigma \in W$, ALG has not finished by the end of phase $\ell-1$. 
Note that, because ALG is randomized, $\sigma$ is determined by both the realizations of the $x_i$ and by the random choices made by ALG.
In what follows, we will compare what happens in ALG during phase $\ell$ to what happens in OPT within cost $2^{\ell}$, conditioned on $\sigma$.

We will assume $\sigma\in W$.
Then ALG's
task going forward is to find a certificate of the induced function $f_{\sigma}$. 
Clearly, since $f=f^\mathcal{M}$ for partition matroid $\mathcal{M}$, $f_{\sigma}=f^{\mathcal{M}'}$ for a partition matroid $\mathcal{M}'$.
Thus \Cref{OPT1BOptimal} and \Cref{TableKeyLemma} apply to $f_{\sigma}$.
Let OPT$_{\sigma}$ denote the strategy obtained from OPT by modifying it so that each time OPT performs a test on an $x_i$ such that $\sigma_i \neq *$, instead skip the test and proceed as if the test had been performed and the outcome was $\sigma_i$. Clearly, OPT$_{\sigma}$ is an evaluation strategy for $f_{\sigma}$.

Now consider any fixed $\pi \in \{0,1\}^n$ that is an extension of $\sigma$.  
If OPT executed on $x$ finds a 0-certificate (1-certificate) of $f$ within cost $2^{\ell}$, then OPT$_{\sigma}$ executed on $x$ clearly finds a 0-certificate (1-certificate) of $f_{\sigma}$ within cost $2^{\ell}$.  Thus conditioned on $\sigma$, the probability that OPT$_{\sigma}$ finds a 0-certificate (1-certificate) of $f_{\sigma}$ within cost $2^{\ell}$ is greater than or equal to the probability that OPT finds a 0-certificate (1-certificate) of $f$ within cost $2^{\ell}$.

Conditioned on $\sigma$,
ALG executes strategy
$\KS(\sigma,6\cdot 2^\ell)$ in phase $\ell$. 
Thus 
it follows from \Cref{TableKeyLemma} that
\begin{align}
\begin{split}
\label{eq:sum0}
&\Pr[\mbox{ALG finds } \text{a 0-certificate of $f$ in phase } \ell \mid \sigma] \\
= &\Pr[\mbox{ALG finds } \text{a 0-certificate of $f_{\sigma}$ in phase } \ell \mid \sigma] \\
 \geq &\left(1-\frac1e\right)\cdot \Pr[\mbox{OPT}_{\sigma} \mbox{ finds } \text{a 0-certificate of $f_{\sigma}$ within cost $2^{\ell}$} \mid \sigma] \\
 \geq &\left(1-\frac1e\right)\cdot \Pr[\text{OPT  finds a 0-certificate of $f$ within cost $2^{\ell}$} \mid \sigma]\,.
\end{split}
\end{align}
Similar reasoning using \Cref{OPT1BOptimal} gives us a bound for finding 1-certificates.
Conditioned on $\sigma$, in phase $\ell$, ALG executes the first $2^{\ell}$ tests that would be performed by \OPTone{} on $f_{\sigma}$.
A 1-certificate of $f_{\sigma}$ induces a 1-certificate of $f$.
Thus by \Cref{OPT1BOptimal}, it follows that
\begin{align}
\label{eq:sum1}
\begin{split}
& \Pr[\mbox{ALG finds } \mbox{a 1-certificate of $f$ in phase } \ell \mid \sigma]    \\  \geq & \Pr[\mbox{OPT finds a 1-certificate of $f$ within cost }2^{\ell} \mid\sigma]\,.
\end{split}
\end{align}

From \Cref{eq:sum0} and \Cref{eq:sum1}, and using the fact that finding a 1-certificate and finding a 0-certificate are disjoint events (for any strategy), we have
\begin{align*}
\begin{split}
&\Pr[\text{ALG finds a certificate of $f$ in phase } \ell \mid \sigma]   \\
= & \phantom{+}\;\Pr[\text{ALG finds a 0-certificate of $f$ in phase } \ell \mid \sigma]\\
& + \Pr[\text{ALG finds a 1-certificate of $f$ in phase } \ell \mid \sigma] \\
\geq &\left(1-\frac1e\right)\cdot\Pr[\text{OPT finds a certificate of $f$ within cost }2^{\ell} \mid\sigma]\,.
\end{split}
\end{align*}
From this, it follows that
\begin{align}
\label{eq:anycert}
\begin{split}
&\Pr[\mbox{ALG }\text{finds a certificate of $f$ in phase } \ell] \\
 = &\sum_{\sigma \in W} \Pr[\sigma] \cdot \Pr[\mbox{ALG } \mbox{finds a } \text{certificate of $f$ in phase } \ell\mid\sigma] \\  \geq &\left(1-\frac1e\right)\cdot \sum_{\sigma \in W}  \Pr[\sigma] \cdot  \Pr[\mbox{OPT finds a } \text{certificate of $f$ within cost } 2^\ell \mid\sigma]\\
 = &\left(1-\frac1e\right)\cdot \Pr[A \mbox{ and } B]\,,
\end{split}
\end{align}
where $A$ is the event that ALG does not find a certificate of $f$ by the end of phase $\ell-1$, and $B$ is the event that OPT finds a certificate of $f$ within cost $2^{\ell}$.  

Now consider the first and last lines of \Cref{eq:anycert}.
The quantity in the first line is equal to $\mu_{\ell-1} - \mu_{\ell}$.
To lower bound the last line, note that the union bound implies $\Pr[A \mbox{ and } B] \geq \Pr[A] - \Pr[\neg B]$.  
We also have that $\Pr[A]=\mu_{\ell-1}$.
By the definition of $B$, $\Pr[\neg B]$ is the probability that OPT incurs cost at least $2^{\ell}+1$, and thus 
$\Pr[\neg B] \leq \mu_{\ell}^*$.
Therefore, the quantity in the last line is greater than or equal to $(1-\nicefrac1e)\cdot (\mu_{\ell-1}-\mu^*_{\ell})$: $$\mu_{\ell-1}-\mu_{\ell} \geq  \left(1-\frac1e\right)\cdot (\mu_{\ell-1} - \mu^*_{\ell}).$$ By rearranging we get
$$
\mu_{\ell} \leq \frac1e \cdot\mu_{\ell-1} + \left(1-\frac1e\right)\cdot \mu^*_{\ell} \leq \frac1e\cdot \mu_{\ell-1} + \mu^*_{\ell}\,,
$$
as claimed.
\end{proof}

We are now ready to prove \Cref{thm:maintheorem}.
The proof follows from \Cref{lem:latency} (and the cost bounds of \Cref{OPT1BOptimal,TableKeyLemma}), and it is fairly standard (see, e.g.,~\cite{EneNS17,ghuge2022non}). We repeat the proof here for completeness.

\begin{proof}[Proof of \Cref{thm:maintheorem}]
    By \Cref{OPT1BOptimal} and \Cref{TableKeyLemma}, the expected cost of phases 0 through $\ell$ of ALG is at most $7\cdot \sum_{j=0}^{\ell} 2^{j} \leq 14 \cdot 2^{\ell}$. 
ALG ends in phase $0$ with probability $1-\mu_0$ and in phase $\ell \geq 1$ with probability $\mu_{\ell-1} - \mu_{\ell}$. As a consequence,

\begin{equation}
\label{ubALG}
\mathbb{E}[\cost(\text{ALG})] \leq 14\cdot (1 - \mu_0) + \sum_{\ell=1 }^\infty 14 \cdot 2^{\ell}(\mu_{\ell-1} - \mu_{\ell}) = 14 + 14\cdot \sum_{\ell=0}^{\infty} 2^{\ell}\mu_{\ell}.
\end{equation}

Similarly, by the definition of $\mu_{\ell}^*$, we can bound the cost of OPT as

\begin{equation}
\label{lbOPT}
\mathbb{E}[\cost(\text{OPT})] \geq \sum_{\ell=0}^\infty 2^{\ell}\cdot (\mu^*_{\ell} - \mu^*_{\ell+1}) \geq \mu^*_{0} + \frac{1}{2}\cdot \sum_{\ell =1}^\infty 2^{\ell} \mu^*_{\ell} = 1 + \frac{1}{2}\cdot \sum_{\ell=1}^\infty 2^{\ell} \mu^*_{\ell},
\end{equation}
where in the last equality we use the fact that $\mu_0^*=1$.

We define $\Gamma := \sum_{\ell=0}^\infty 2^{\ell} \mu_{\ell}$. Then we have 

\begin{align*}
        \Gamma = \sum_{\ell=0}^\infty 2^{\ell} \mu_{\ell} &\leq \mu_0 + \frac{1}{e}\cdot \sum_{\ell=1}^\infty 2^{\ell} \mu_{\ell-1} + \sum_{\ell=1}^\infty 2^{\ell} \mu_{\ell}^*\\
        &\leq \mu_0 + \frac{1}{e}\cdot \sum_{\ell=1}^\infty 2^{\ell} \mu_{\ell-1} + 2\cdot (\mathbb{E}[\cost(\text{OPT})]-1)\\
        &= \mu_0 + 2\cdot\frac{1}{e}\cdot \bigg(\sum_{\ell=0}^\infty 2^{\ell} \mu_{\ell}\bigg) + 2\cdot (\mathbb{E}[\cost(\text{OPT})]-1)\\
        &\leq \frac{2}{e} \cdot \Gamma + 2\cdot \mathbb{E}[\cost(\text{OPT})]-1,
\end{align*}
where the first inequality follows from \Cref{lem:latency}, the second one from \Cref{lbOPT}, and the last one from the fact that $\mu_0 \leq 1$.
It follows that  $\Gamma \leq 2\cdot (1-\frac{2}{e})^{-1} \cdot \mathbb{E}[\cost(\text{OPT})]$.  From \Cref{ubALG}, we conclude that $\mathbb{E}[\cost(\text{ALG})] \leq 14+14\cdot (1-\frac{2}{e})^{-1}\cdot \mathbb{E}[\cost(\text{OPT})]=O(1)\cdot \mathbb{E}[\cost(\text{OPT})]$. 

Finally, the bound on the running time follows from the bounds in \Cref{OPT1BOptimal} and \Cref{TableKeyLemma} as well as the definition of \Cref{alg:main_alg}.
\end{proof}

\section{Finding a $1$-Certificate Optimally on General Matroids}
\label{sec:OPT1}

The goal of this section is to define \OPTone{} and to prove \Cref{OPT1BOptimal}, which we restate here for convenience.

\optonelemma*

Although we will use it in evaluating the function $f=f^{\mathcal{M}}$ for $\mathcal{M}=(E,\mathcal{I})$ a partition matroid, the results in this section also hold for $\mathcal{M}$ an arbitrary matroid.

The algorithm \OPTone{}, given a partial realization $\sigma$, greedily attempts to construct a basis $S$ of $\mathcal{M}_{\sigma}$, where $\mathcal{M}_{\sigma} = (E, \mathcal{I}_{\sigma})$ is the matroid induced by $\sigma$.  It begins by initializing $S$ to be the empty set.  It considers the tests on $x_i$ in decreasing $p_i$ order, performing test $x_i$ if it is independent of the elements already in $S$, and then adding $i$ to $S$ if the test on $x_i$ has outcome 1.  It stops testing as soon as $S$ is a basis, meaning that $f^{\mathcal{M}_\sigma}(x)=1$ and thus $f^{\mathcal{M}}(x)=1$.  If $S$ still isn't a basis after all tests have been processed,  then $f^{\mathcal{M}}(x)=0$. (Note that a $0$-certificate may have been found earlier, but we ignore this for simplicity.)
We give the pseudocode in \Cref{alg:OPTOne}.

\begin{algorithm}[t]
    \caption{\OPTone{}($\sigma,B$)}
    \label{alg:OPTOne}
    \SetAlgoVlined
    $S \gets \emptyset$\;
    $c \gets 0$\;
    $L \gets \text{list of tests $x_i$ with $\sigma_i = \ast$ in decreasing order of $p_i$}$\;
    \For{$x_i$ in $L$}{
        \If{$S \cup \CB{x_i} \, \in \, \mathcal{I}_{\sigma}$}{
        perform test on $x_i$\;
        $c \gets c+1$\;
        \If{$x_i = 1$}{
        $S\gets S\cup\{x_i\}$\;
        }
        }
        \If{$S$ is a basis of $\mathcal{M}_{\sigma}$}{
        \Return{``$f_\sigma(x)=1$''}
        }
        \If{$c = B$}{
            \textbf{break}
        }
    }
    \Return{``No 1-certificate of $f_\sigma$ found within cost $B$''}
\end{algorithm}

\begin{proof}
[Proof of \Cref{OPT1BOptimal}]

We show the lemma by induction on $B$. 
First note that $f_\sigma$ is identical to $f^{\mathcal{M}_{\sigma}}$ (with $\mathcal{M}_{\sigma}$ also being a matroid).
We may thus assume w.l.o.g.\ that $\sigma=(*,\dots,*)$. In the following, we will therefore drop $\sigma$ from the notation \OPTone{}($\sigma,B$) for simplicity as well as from $\mathcal{M}_{\sigma}$ and $f_{\sigma}$. Now fix some $Z$. It is also w.l.o.g.\ to assume that $Z$ is a deterministic strategy that maximizes the probability of finding a 1-certificate.
Finally, again w.l.o.g., assume that the input variables of $f$, namely $x_1,\ldots,x_n$, are numbered so that $p_1 \geq p_2 \geq \ldots \geq p_n$.

Note that the case $\mathrm{rank}(\mathcal{M})=0$ is trivial.
Thus assume $\mathrm{rank}(\mathcal{M})\geq 1$. By this assumption, there exists some $x_\ell$ that is the first test performed by \OPTone{}($B$) (note that $\ell\neq 1$ is possible because $\mathcal{M}$ may contain loops, i.e., elements $x_i$ such that $\{x_i\}$ is not independent). Let $x_j$ be the first test performed by  $Z$. Assume w.l.o.g.\ that $x_j$ is not a loop either. 

For the base case, suppose $B=1$.  Then \OPTone($B$) performs the single test on $x_{\ell}$ whereas $Z$ performs the single test on $x_j$.
If $j=\ell$, \OPTone($B$) and $Z$ are identical, and the statement holds trivially.  Suppose $j \neq \ell$. The sets $\{x_j\}$ and $\{x_{\ell}\}$ are both independent.  Since all bases of $\mathcal{M}$ have the same size, either both $\{x_j\}$ and $\{x_{\ell}\}$ are bases of $\mathcal{M}$, or neither is.  If both are bases, then \OPTone($B$) 
finds a 1-certificate with probability $p_{\ell}$ while $Z$ finds a 1-certificate with probability $p_j$.
By the definition of \OPTone{} and the fact that $x_j$ is not a loop, $p_{\ell} \geq p_j$.  If neither $\{x_j\}$ nor $\{x_{\ell}\}$ is a basis, then both \OPTone($B$) and $Z$ have zero probability of finding a 1-certificate.
This completes the proof of the base case.

For the induction step, suppose $B > 1$.
Let $\sigma_0$ be the partial realization that sets $x_j=0$ and let $\sigma_1$ be the partial realization that sets $x_j=1$.
If the first test of $Z$, on $x_j$, doesn't yield a 1-certificate of $f$, $Z$ continues testing in order to find a 1-certificate of $f_{\sigma_0}$ (if $x_j=0$) or $f_{\sigma_1}$ (if $x_j=1$).
It does so in a way that maximizes its probability of obtaining a 1-certificate within remaining budget $B-1$.  
By the induction hypothesis, this probability is also maximized by executing \OPTone($B-1$) on either $f_{\sigma_0}$ or $f_{\sigma_1}$, as appropriate.  
Thus, w.l.o.g., we may assume that $Z$ executes \OPTone($B-1$) after performing its first test.

If $j = {\ell}$, then $Z$ and \OPTone($B$) are identical, and \OPTone($B$) also maximizes the probability of obtaining a 1-certificate within budget $B$. So suppose that $j \neq {\ell}$.  In that case, after testing $x_j$, if $Z$ doesn't yet have a 1-certificate, the next variable it considers testing is $x_{\ell}$ (regardless of whether $x_j=1$ or $x_j=0$).  We consider the modification of $Z$ that interchanges the roles of $x_j$ and $x_{\ell}$.  We call this $\tilde{Z}$.  We now show that the probability that $\tilde{Z}$ finds a 1-certificate of $f$ is at least the probability that $Z$ finds a 1-certificate of $f$.
This will complete the proof of the lemma: Since the first test of $\tilde{Z}$ is $x_{\ell}$, it will again follow by induction that \OPTone{} maximizes the probability of finding a 1-certificate within budget $B$.

We distinguish two cases:
\begin{itemize}
    \item[Case 1:] The variables $x_j$ and $x_{\ell}$ are independent in $\mathcal{M}$
(i.e., $\{x_j,x_{\ell}\}$ is an independent set).

In this case, neither $\{x_j\}$ nor  $\{x_{\ell}\}$ is a basis of $\mathcal{M}$.  Since $B > 1$, $Z$,
and $\tilde{Z}$ always perform their first two tests, those on $x_j$ and $x_{\ell}$ (in opposite order).  For each pair of possible outcomes of those two tests, $Z$ and $\tilde{Z}$ then follow the same testing strategy.  Therefore, $Z$ and $\tilde{Z}$ have the same probability of finding a 1-certificate.

    \item[Case 2:] $x_j$ and $x_{\ell}$ are dependent. 
    
As argued above, both $\{x_j\}$ and $\{x_{\ell}\}$ are bases of $\mathcal{M}$ or neither is. 
Suppose both are. 
Then since $B > 1$, $Z$ and $\tilde{Z}$ will both obtain a 1-certificate within their first two tests unless $x_{\ell}=x_j=0$.  In that case, after performing their tests on $x_j$ and $x_{\ell}$, $Z$ and $\tilde{Z}$ subsequently follow the same strategy.  
Therefore, $Z$ and $\tilde{Z}$ have the same probability of finding a 1-certificate.

So suppose neither $\{x_j\}$ nor $\{x_{\ell}\}$ is a basis of $\mathcal{M}$.
Then both $Z$ and $\tilde{Z}$ begin by \textit{considering} whether to test $x_j$ and $x_{\ell}$, in opposite order.  Specifically, $Z$ begins by testing $x_j$.  If $x_j=1$, it skips testing $x_{\ell}$ next, because of the dependence.  If $x_j=0$, it tests $x_{\ell}$ next.  Strategy $\tilde{Z}$ begins by testing $x_{\ell}$.  If $x_{\ell}=1$, it skips testing $x_j$ next, because of the dependence.  If $x_{\ell}=0$, it tests $x_j$ next.

For $s,t \in \{0,1\}$, let $P_{st}$ be the probability that $x_j=s$ and $x_{\ell} = t$. Also, for $u,v \in \{0,1,*\}$, let $f_{uv}=f_{\sigma'}$, where $\sigma'$ is the partial realization that assigns value $u$ to $x_j$ if $u \neq *$, and assigns value $v$ to $x_{\ell}$ if $v \neq *$. 

The probability that $Z$ finds a 1-certificate of $f$ is
\begin{align*}
&(P_{11}+P_{10})\cdot\Pr[\mbox{OPT}_1(B-1)
 \mbox{ finds 1-certificate of } f_{1*}]\\
 + &P_{01}
\cdot \Pr[\mbox{OPT}_1(B-2) \mbox{ finds 1-certificate of }f_{01}]\\  + &P_{00}\cdot \Pr[\mbox{OPT}_1(B-2) \mbox{ finds 1-certificate of } f_{00}].
\end{align*}

The probability that $\tilde{Z}$ finds a 1-certificate of $f$ is
\begin{align*}
&(P_{11}+P_{01}) \cdot \Pr[\mbox{OPT}_1(B-1)
 \mbox{ finds 1-certificate of }f_{*1}]\\
 + &P_{10}
\cdot \Pr[\mbox{OPT}_1(B-2) \mbox{ finds 1-certificate of }f_{10}]\\ + &P_{00}\cdot \Pr[\mbox{OPT}_1(B-2) \mbox{ finds 1-certificate of } f_{00}].
\end{align*}

We now show that the first quantity, for $Z$, is bounded above by the second quantity, for $\tilde{Z}$.  
In matroid theory, if two elements $e,e'$ of a matroid are such that $e$ and $e'$ are not loops, and $\{e,e'\}$ is a dependent set, then $e$ and $e'$ are called parallel elements of the matroid.  If $e$ and $e'$ are parallel elements of a matroid, then for every basis $I$ of the matroid that contains $e$, the set $(I\backslash \{e\} )\cup \{e'\}$ is also a basis of the matroid (cf.\ \cite{Oxley}, Proposition 1.7.2).  Therefore, for every independent set $I$ of the matroid that contains $e$, the set $(I\setminus \{e\} )\cup \{e'\}$ is also an independent set, and the symmetric statement is true for independent sets $I$ containing $e'$.  

Thus 
the functions $f_{1*}$ and $f_{*1}$ correspond to minors $\mathcal{M}_{1*}$ and $\mathcal{M}_{*1}$ of $\mathcal{M}$, both having the same collection of independent sets, none of which contains $x_j$ or $x_{\ell}$.  It follows that  for any budget $B'$, the probability that \OPTone($B'$) finds a 1-certificate is the same for $f_{*1}$, $f_{1*}$, $f_{01}$, and $f_{10}$.  
Also,
since the probability that \OPTone($B'$) finds a 1-certificate of a function is non-decreasing in $B'$, and because $P_{10} \leq P_{01}$,
it follows that the first quantity above, for $Z$, is bounded above by the second, for $\tilde{Z}$.  Thus the probability that $\tilde{Z}$ finds a 1-certificate within budget $B$ is at least the probability that $Z$ does, as claimed.

\end{itemize}

Finally, the bound on the running time is obvious.
\end{proof}

\section{Finding a $0$-Certificate Approximately Optimally on Partition Matroids}
\label{sec:KS}

In this section we present $\KS{}(\sigma,B)$, the approximation algorithm we use to maximize the probability of finding a 0-certificate of a partition matroid, under an expected budget constraint.  We prove \Cref{TableKeyLemma}, which we restate here for convenience.  

\tablelemma*

The pseudocode for \KS{} is presented in \Cref{alg:KSSub}.
\KS{} works by computing appropriate prunings of the strategy we call IO (or InsideOut), applied to each of $g_{1,\sigma},\dots,g_{d,\sigma}$. (Recall that $g_j$ is the $k_j$-of-$n_j$ function corresponding to the $j$-th partition class of the partition matroid, and $g_{j,\sigma}$ is the function that $\sigma$ induces on it.) We denote these strategies by $\text{IO}_{1,\sigma},\dots,\text{IO}_{d,\sigma}$.  IO is an elementary strategy, and in our algorithm, it is represented by a DAG.  

We now present a lemma about the quality of prunings of IO. 
We defer further discussion of IO, and the proof of the lemma, to \Cref{sec:IO}.

\begin{restatable}{lemma}{expectedbudgetlemma}
\label{lem:IOapprox}
    Let $\mathcal{M}$ be a uniform matroid and let $f:=f^{\mathcal M}$.
    Equivalently, let $f$ be a $k$-of-$n$ function.
    Let $S$ be a (possibly randomized) testing strategy for $f$ such that $S$ has expected cost $B$, and it finds a 0-certificate of $f$ with probability $P$.
    There exists a randomized pruning $S'$ of the elementary strategy IO with expected cost at most $6 \cdot B$, which finds a 0-certificate of $f$ with probability at least $P$.
\end{restatable}

Recall that, given a budget $B$ and a testing strategy $S$ for $f$, the function \text{ComputePrunedStrategy} (defined in \Cref{sec:sub:Prun}) can be used to generate 
the randomized pruning of $S$ that maximizes the probability of finding a 0-certificate of $f$ within expected cost at most $B$, among all such prunings.
Using \Cref{lem:prune}, we hence obtain a polynomial-time approximation algorithm for MP0 on uniform matroids which works by computing a randomized pruning of elementary strategy IO.

To determine the budgets $B_1,\dots,B_d$ to use for pruning the strategies $\text{IO}_{1,\sigma},\dots,\text{IO}_{d,\sigma}$, algorithm \KS{} solves a continuous knapsack problem. To this end, \KS{} first calls the procedure \text{ComputePruningFunctions} (defined in \Cref{sec:sub:Prun}) on $\text{IO}_{1,\sigma},\dots,\text{IO}_{d,\sigma}$ to obtain functions $q_1,\dots,q_d$, respectively. Starting at $B_1=\dots=B_d=0$ and stopping when all budget has been invested, it continuously distributes the budget among the $d$ partition classes using a continuous-greedy approach to maximize the sum of utilities $\sum_{j\in[d]}q_j(B_j)$. We give pseudocode in \Cref{alg:KSSub}, using $$\partial_+ q\RB{a} := \lim_{x \rightarrow a, x > a} \frac{q\RB{x}-q\RB{a}}{x-a}$$ to denote the right derivative of a function $q$ on $\mathbb{R}$ at position $a$. Note that
\KS($\sigma$,$B$) runs in polynomial time
because the number of breakpoints of each $q_j$ is polynomial (cf.\ \Cref{lem:prune}).

We now prove \Cref{TableKeyLemma}.

\begin{algorithm}[t]
    \caption{\KS$\RB{\sigma,B}$}
    \label{alg:KSSub}
    \SetAlgoVlined
    $q_1, \ldots, q_d \gets \text{ComputePruningFunction}\RB{\text{IO}_{1,\sigma}},\dots,\text{ComputePruningFunction}\RB{\text{IO}_{d,\sigma}}$\;
    $B_1,  \ldots, B_d \gets 0$\;
    \While{$\sum_{j\in[d]} B_j < B$}{
        choose $j^\ast \in \arg\max_{j \in \SB{d}} \partial_+ q_j\RB{B_j}$\;
        $\Delta \gets \Delta > 0$ minimal such that $q_{j^\ast}$ has breakpoint at $B_{j^\star}+\Delta$ or $\sum_{j\in[d]}B_j+\Delta= B$\;
        $B_{j^\star}\gets B_{j^\star}+\Delta$\;
    }
    \For{$j \in \SB{d}$}{
        execute $\text{ComputePrunedStrategy}(\text{IO}_{j,\sigma},B_j)$\;
    }
\end{algorithm}

\begin{proof}[Proof of~\Cref{TableKeyLemma}]
The cost bound follows immediately from the definition of \KS($\sigma$,$B$).
We thus focus on the probability bound. Let OPT be a (possibly randomized) strategy with expected cost at most $B/6$.

Consider the values of $B_1,\dots,B_d$ at the end of the iteration of the main loop corresponding to $\KS(\sigma,B)$. For any $j\in[d]$, denote by $S_j$ the event that $\text{ComputePrunedStrategy}(\text{IO}_{j,\sigma},B_j)$ finds a 0-certificate for $g_{\sigma,j}$.
Similarly, denote by $B_j'$ the expected cost that OPT spends on $g_{\sigma,j}$, and denote by $T_j$ the event that OPT finds a 0-certificate for $g_{\sigma,j}$. We first notice that 
\begin{equation}\label{eq:lem3-dom}
    \sum_{j\in[d]}\Pr[S_j\mid\sigma]=\sum_{j\in[d]} q_j(B_j)\geq \sum_{j\in[d]} q_j(6\cdot B'_j)\geq \sum_{j\in[d]}\Pr[T_j\mid\sigma]\,.
\end{equation}
The equality simply follows by \Cref{lem:prune} (ii). For the first inequality, first notice that the randomized strategy $\KS(\sigma,B)$ maximizes $\sum_{j\in[d]} q_j(B_j)$ under the constraint $\sum_{j\in[d]}B_j=B$ (using concavity of the $q_j$; \Cref{lem:prune} (i)). Then the inequality follows because $B\geq 6\cdot \sum_{j\in[d]}\cdot B_j'$ and the $q_j$ are monotone (again \Cref{lem:prune} (i)). The second inequality simply follows by \Cref{lem:IOapprox}.

In the following, we are going to interpret events as random variables taking values $0$ or $1$. We let $S_1',\dots,S_d'$ be events that occur with the same marginal probabilities as $S_1,\dots,S_d$, respectively, but are correlated to maximize $\mathbb{E}[\max(S_1',\dots,S_d')]$. Note that $\max(S_1',\dots,S_d')$ is the event that at least one of $S_1',\dots,S_d'$ occurs, and that $\mathbb{E}[\max(S_1',\dots,S_d')]$ is maximized when $S_1',\dots,S_d'$ never occur simultaneously, unless their sum of probabilities is $1$ (in which case $\max(S_1',\dots,S_d')=1$ with probability $1$). Then we obtain
\begin{align*}
    &\phantom{=}\;\Pr[\text{\KS$(\sigma,B)$ finds $0$-certificate of $f_{\sigma}$}\mid\sigma] \\
    &= \mathbb{E}[\max(S_1,\dots,S_d)\mid\sigma] & \text{(Def.\ of $S_1,\dots,S_d$)}\\
    & \geq \left(1-\frac1e\right)\cdot\mathbb{E}[\max(S_1',\dots,S_d')\mid\sigma] & \text{(*)}\\
    & = \left(1-\frac1e\right)\cdot \min\left(1,\sum_{j\in[d]}\Pr[S_j\mid\sigma]\right) & \text{(Def.\ of $S_1',\dots,S_d'$)}\\
    & \geq \left(1-\frac1e\right)\cdot \min\left(1,\sum_{j\in[d]}\Pr[T_j\mid\sigma]\right) & \text{(\Cref{eq:lem3-dom})}\\
    & \geq \left(1-\frac1e\right)\cdot \Pr[\text{OPT finds $0$-cert.\ within exp.\ cost $B/6$}\mid\sigma]  & \text{(Def.\ of $T_1,\dots,T_d$, union bound)},
\end{align*}
where the step marked with (*) is a standard argument, but it also follows from invoking the correlation gap~\cite{AgrawalDSY10}.

Finally, the polynomial bound on the running time follows from combining the fact that IO is elementary, \Cref{lem:prune}, and the definition of \Cref{alg:KSSub}.
\end{proof}

\section{An $O(1)$-Approximation Algorithm for MP0 on Uniform Matroids}
\label{sec:IO}

In this section we formally define IO (InsideOut) and prove that
\Cref{lem:IOapprox}, which we restate here for convenience, holds.

\expectedbudgetlemma*

Let $f$ be a $k$-of-$n$ function.
Let $\mathcal{A}$ and $\mathcal{B}$ be disjoint ordered lists of random variables from the set $\CB{x_1, \ldots, x_n}$, such that each variable appears in one of the two lists. We first explain the round robin testing strategy 
RoundRobin$\RB{\mathcal{A},\mathcal{B},f}$
whose pseudocode is presented in \Cref{alg:RRob}.

\begin{algorithm}[t]
    \caption{RoundRobin$\RB{\mathcal{A},\mathcal{B},f}$}
    \label{alg:RRob}
    \SetAlgoVlined
    $\sigma \gets (*,\dots,*)$\;
    $i \gets 0$\;
    \While{$\sigma$ not a certificate of $f$}{
        perform $i$-th test from $\mathcal{A}$ (if it exists) and $i$-th test from $\mathcal{B}$ (if it exists)\;
        update $\sigma$ to include the results of the tests just performed\;
        $i \gets i+1$\;
    }
\end{algorithm}

We refer to the $i$-th iteration of the while loop in the pseudocode as the $i$-th \textit{round} of the round robin.  
If in some iteration $i$, only one of the two lists has an $i$-th element (which can happen if one list is longer than the other), then the round consists of just that one element. 
Also, RoundRobin$\RB{\mathcal{A},\mathcal{B},f}$ is written so that it satisfies what we call the {\em Round-Completion} (RC) property, meaning that the strategy cannot terminate during a round.
Thus, even if the value of $f$ can already be determined after testing the first random variable in a round, the test on the remaining random variable in the round, if it exists, must be performed before the strategy can terminate.
The Round-Completion property is not needed from an algorithmic perspective, but it simplifies our presentation and our analysis.
RoundRobin$\RB{\mathcal{A},\mathcal{B},f}$ evaluates $f$ since it tests, if necessary, all random variables and does not terminate without a certificate.

Our base strategy, InsideOut($f$) (cf.\ \Cref{alg:IO}), executes RoundRobin($\mathcal{L}_0$,$\mathcal{L}_1,f$), where $\mathcal{L}_0$ is the sequence of elements with $p_i \leq \frac12$ sorted by decreasing $p_i$, and 
$\mathcal{L}_1$ is the sequence of all elements in $\CB{x_1, \ldots, x_n}$ with $p_i > \frac12$ sorted by increasing $p_i$.
For random variables with the same distribution, we assume an arbitrary but fixed order.
In what follows, for succinctness we will usually drop the argument $f$ and just call the above procedures RoundRobin($\mathcal{A}$,$\mathcal{B}$) and InsideOut.

\begin{algorithm}[t]
    \caption{InsideOut$\RB{f}$}
    \label{alg:IO}
    \SetAlgoVlined
    $\mathcal{L}_0 \gets$ sequence of all elements in $\CB{x_1, \ldots, x_n}$ with $p_i \leq \frac12$ sorted by decreasing $p_i$\;
    $\mathcal{L}_1 \gets$ sequence of all elements in $\CB{x_1, \ldots, x_n}$ with $p_i > \frac12$ sorted by increasing $p_i$\;
    execute RoundRobin($\mathcal{L}_0$,$\mathcal{L}_1$,$f$)\;
\end{algorithm}

We call a strategy \textit{$j$-pseudo-IO} if it is the same as InsideOut, except that the first element $x_j$ of either $\mathcal{L}_0$ or $\mathcal{L}_1$ may not be in correct sorted order.
That is, $\mathcal{L}_0$ still contains the random variables with $p_i \leq \frac12$, and $\mathcal{L}_1$ still contains those with $p_i > \frac12$, but one of these two sequences starts with $x_j$ and the sorting criterion (decreasing $p_i$ for $\mathcal{L}_0$, and increasing $p_i$ for $\mathcal{L}_1$) only needs to be satisfied beginning with the second element in the sequence.

Using the fact that $f$ is a $k$-of-$n$ function, it is easy to show that IO is an elementary strategy (recall definition from~\Cref{sec:prelim}).
IO tests its elements in a fixed order, until the value of $f$ can be determined at the end of a round.
At the end of the $\ell$-th round, some fixed number $n_{\ell}$ of elements have been tested.
IO will terminate at the end of this round iff among those $n_{\ell}$ realizations there are exactly $k$ or $k+1$ 1's, with the remainder being 0's, or exactly $n-k+1$ or $n-k+2$ 0's, with the remainder being 1's.  Thus termination only depends on the number of 1's and number of 0's observed.  

\subsection{The analysis of IO}
\label{subsec:analysisofIO}

We start with some definitions.
Let $S$ be a (possibly randomized) testing strategy for a $k$-of-$n$ function $f$.
To keep notation concise, in this section we also allow $k \leq 0$ and $k > n$. In the first case, the value of $f$ is 1 on all realizations, and in the second, the value of $f$ is 0 on all realizations.
In the first case, $(*,*, \ldots, *)$ is a 1-certificate of $f$ and in the second,  it is a 0-certificate of $f$.  In both cases, the certificate is found without performing any tests.  Here the \textit{utility} of $S$, written $\Util{S}$, is a random variable with value $1$ if $S$ finds a 0-certificate of $f$ and with value $0$ otherwise.
For an integer $m$, we define $\Utilk{m}{S}$ to be the indicator random variable of the event that $S$ observes at least $m$ 0's (among the realizations of tested variables) before terminating\footnote{For convenience this includes the usage of $\Utilk{m}{S} = 1$ for $m \leq 0$.}. 
We use the short-hand $\kk := n-k+1$.
Finding a 0-certificate of a $k$-of-$n$ function means observing at least $\kk$ tests that evaluate to 0, so $\Util{S} \equiv \Utilk{\kk}{S}$.
We use $\qq[i] := 1 - p_i$ for the probability that the test on $x_i$ has outcome 0.
Finally, we use the term {\em (randomized) RC-pruning of IO} to mean a (randomized) pruning of IO obeying the Round-Completion property.

Before presenting the proof of \Cref{lem:IOapprox}, we give the following proof sketch.  At a high level, we prove the result by induction on $n$, strengthening the lemma statement by requiring $S'$ to be a randomized RC-pruning (not just a randomized pruning) of the elementary strategy IO.  
The randomized testing strategy $S$ corresponds to a distribution over deterministic strategies. 
Fix some $\hat{S}$ from the support of that distribution.
In our inductive argument, we specify a series of transformations on $\hat{S}$ that, in expectation, maintain its utility and increase its cost by only a bounded amount.    
We start by keeping the first element $x_i$ of $\hat{S}$ and using the induction hypothesis to replace the left subtree of the root of $\hat{S}$ by a randomized RC-pruning of IO (on the induced subinstance), and similarly for the right subtree.  
Thus both subtrees execute (randomized, pruned) versions of a round robin between the same two lists $\mathcal{L}_0$ and $\mathcal{L}_1$.
The resulting tree is not yet a randomized pruning of IO on the full instance, primarily because
\begin{enumerate}
    \item[(i)] the test on $x_i$ is performed before beginning the round robin and
    \item[(ii)] testing $x_i$ first may violate the testing order used by IO.
\end{enumerate}
We address (i) by moving $x_i$ into the first round of the subsequent round robin.  More particularly, if $p_i \leq \frac{1}{2}$ we prepend it to $\mathcal{L}_0$ otherwise we prepend it to $\mathcal{L}_1$.
This shifts the elements in the affected list to adjacent rounds, at constant additional cost.
The resulting strategy is a randomized RC-pruning of $i$-pseudo-IO.
To address (ii), we need to convert this $i$-pseudo-IO-based strategy into a randomized RC-pruning of standard IO, which requires that $x_i$ be moved to the correct position in its list.
The next, and most involved part of the proof, shows that this can be done in such a way that expected utility is preserved, and expected cost is not increased too much (cf. \Cref{lem:IOtreesurgery}).
Like the proof of \Cref{OPT1BOptimal} in~\Cref{sec:OPT1}, which shows the optimality of the Greedy algorithm for finding 1-certificates on arbitrary matroids, this proof proceeds by induction and includes case analysis.  However, 
the case analysis here is significantly more complex.
The proof of \Cref{OPT1BOptimal} relies on a standard type of  interchange argument which 
involves exchanging the roles of just two tests in a strategy, while here more involved restructuring of the strategy is required. 
This added complexity can be attributed to properties of IO (including its round structure), properties of the 0-certificates, and the fact that the argument here is in terms of \emph{expected} cost.

To analyze the increase in expected cost from moving $x_i$, we introduce an \textit{amortized cost} defined as follows.
For testing $x_i$ with $p_i \leq \frac12$, the amortized cost $\Acost{i}$ is $1$ if $x_i = 0$ and $0$ otherwise.
For $p_i > \frac12$, the amortized cost is $1$ if $x_i = 1$ and $0$ otherwise.
This makes $\Acost{i}$ a random variable depending on $x_i$.
We use the random variable $\Acost{S}$ for the amortized cost of a strategy $S$.
Amortized cost can be related to the (actual) cost within a constant factor in expectation:
For both $p_i \leq \frac12$ and $p_i > \frac12$, the expected amortized cost of the test on $x_i$ is at least $\frac12$ (and at most $1$).
Therefore, by linearity of expectation,
\begin{equation}
\label{eq:amortizedcost}
    \frac12 \cdot \Ex{\Cost{S}} \leq \Ex{\Acost{S}} \leq \Ex{\Cost{S}}
\end{equation}
for any strategy $S$.

We now present the technical lemma that we use in addressing (ii):

\begin{lemma}
    \label{lem:IOtreesurgery}
    Let $i \in [n]$ and let $\mathcal{I}_i$ denote the $i$-pseudo-IO strategy for $k$-of-$n$ function $f(x_1,\ldots,x_n)$\footnote{This makes $x_i$ the element that is potentially not in the correct position for IO.}.  
    Let $S$ be a randomized RC-pruning of $\mathcal{I}_i$, with expected amortized cost $\Ex{\Acost{S}} = B$. Then there exists a randomized RC-pruning $S''$ of the IO strategy for $f$, such that $S''$ has the following properties:
    \begin{itemize}
        \item $\Ex{\Acost{S''}} \leq B + 1$
        \item $\Ex{\Utilk{\kk}{S''}} \geq \Ex{\Utilk{\kk}{S}}$
    \end{itemize}
\end{lemma}

We defer the proof of \Cref{lem:IOtreesurgery} until 
\Cref{sec:IOtechnicallemma} and
proceed with proving \Cref{lem:IOapprox}. 
\begin{proof} [Proof of \Cref{lem:IOapprox}]

Using induction on $n$, the number of tests, we prove the following 
\textit{Induction Hypothesis}.

\bigskip
{\noindent \textit{Induction Hypothesis}}: Let $S$ be a (possibly randomized) testing strategy for $f(x_1,\ldots,x_n)$ with expected cost $B$ and expected utility $\PP$. For $C_1 \geq 6$, there exists a randomized RC-pruning $S'$ of IO, with $\Ex{\Acost{S'}} \leq \frac{C_1}{2} \cdot B$ and expected utility $\Ex{\Util{S'}} \geq \PP$.
\bigskip

Note that the induction hypothesis differs from the statement of \Cref{lem:IOapprox} in 
bounding the \emph{amortized} expected cost of the pruning.  It also requires the pruning to have the round completion property. By \Cref{eq:amortizedcost}, the induction hypothesis is at least as strong as the lemma statement.  

For $n = 1$, any randomized strategy performs the test on the single element with some probability, otherwise it stops immediately.  
This is a randomized  RC-pruning of IO (without any transformation necessary). By \Cref{eq:amortizedcost}, the expected amortized cost is at most the expected cost of $S$ and the statement holds if $C_1 \geq 2$.

Assume the induction hypothesis holds for $n-1 \geq 1$.  We will show that it then holds for $n$. 

To this end, consider the randomized strategy $S$ for $f$, which corresponds to a distribution over deterministic strategies.
Let $J$ be a set of indices such that
the $S_j$ with $j \in J$ are the deterministic strategies in the support of this distribution, where $S_j$ is selected with probability $P_j$. 
For each $j \in J$, we will construct an associated randomized RC-pruning of IO $R''_j$ such that
\begin{align}\Ex{\Acost{R''_j}} & \leq \frac{C_1}{2} \cdot \Ex{\Cost{S_j}}\qquad\qquad\text{and}\label{eq:inductionacost}\\
\Ex{\Utilk{\kk}{R''_j}} & \geq \Ex{\Utilk{\kk}{S_j}}\,.\label{eq:inductionutil}
\end{align}

Fix $j \in J$.
If $S_j$ does not test any random variables, it is trivially an RC-pruning of IO.
Set~$R''_j = S_j$.  In this case, $\Utilk{\kk}{S_j}=1$ if $\kk \leq 0$, otherwise $\Utilk{\kk}{S_j}=0$.
Then $0 = \Ex{\Acost{R''_j}} \leq \frac{C_1}{2} \cdot \Ex{\Cost{S_j}} = 0$ holds for $C_1 \geq 0$, and $\Ex{\Utilk{\kk}{R''_j}} \geq \Ex{\Utilk{\kk}{S_j}}$.

In the case where $S_j$ tests at least one element, let $x_i$ be the first element tested. 
If $\kk \leq 0$, then we can set $R''_j$ to be the strategy that does not test any random variables, yielding $\Utilk{\kk}{R''_j}=1$ and $\Ex{\Acost{R''_j}} = 0$. The case where $\kk \geq 1$ is shown in the following.
Let $S_j^0$ and $S_j^1$ denote the (possibly empty) substrategies that $S_j$ uses depending on whether $x_i$ has value $0$ or $1$.
Thus we have
\begin{equation}\label{eq:IOstart}
\Ex{\Cost{S_j}} = 1 + \qq[i] \cdot \Ex{\Cost{S_j^0}} + \RB{1-\qq[i]} \cdot \Ex{\Cost{S_j^1}}
\end{equation}
and
\begin{equation*}
    \Ex{\Utilk{\kk}{S_j}} = \qq[i] \cdot \Ex{\Utilk{\kk-1}{S_j^0}} + \RB{1-\qq[i]}\cdot \Ex{\Utilk{\kk}{S_j^1}}\,.
\end{equation*}

For either outcome of $x_i$, 
we have an induced instance of the Expected Budget problem on $n-1$ variables: If $x_i=0$, the induced instance is for a $k$-of-$(n-1)$ function, and if $x_i=1$, the induced instance is for a $(k-1)$-of-$(n-1)$ function.  By the induction hypothesis, 
we can thus replace $S_j^0$ by a randomized RC-pruning $R_j^0$ of IO (on the induced subinstance) that has expected
utility at least $\Ex{\Utilk{\kk-1}{S_j^0}}$ and $\Ex{\Acost{R_j^0}} \leq \frac{C_1}{2} \cdot \Ex{\Cost{S_j^0}}$, and $S_j^1$ by a randomized RC-pruning $R_j^1$ of IO (again on the induced subinstance) with expected utility
at least $\Ex{\Utilk{\kk}{S_j^1}}$ and $\Ex{\Acost{R_j^1}} \leq \frac{C_1}{2} \cdot \Ex{\Cost{S_j^1}}$.
We may assume that the distributions over prunings in the two branches are independent.
Let $R_j$ denote the resulting randomized strategy, and now consider $R_j$ as a replacement for $S_j$.

Since the amortized cost of testing $x_i$ is at most $1$, we have
\begin{align}
    \Ex{\Acost{R_j}} &\leq 1 + \qq[i] \cdot \Ex{\Acost{R_j^0}} + \RB{1-\qq[i]} \cdot \Ex{\Acost{R_j^1}} \label{eq:ind_outer1} \\
    &\leq 1 + \frac{C_1}{2} \SB{\qq[i] \cdot \Ex{\Cost{S_j^0}} + \RB{1-\qq[i]} \cdot \Ex{\Cost{S_j^1}}}\label{eq:ind_outer2}.
\end{align}
As long as $C_1 \geq 2$, this implies $\Ex{\Acost{R_j}} \leq \frac{C_1}{2} \cdot \Ex{\Cost{S_j}}$; however, $R_j$ is in general not (yet) a distribution over RC-prunings of IO, as mentioned in the proof exposition.

We first restore the Round-Completion property. Observe that all components of $R_j^0$ and $R_j^1$, even though they have different stopping criteria, are based on the same (possibly empty) lists $\mathcal{A}$ and $\mathcal{B}$, derived from the probabilities of the elements; the lists depend neither on the stopping criteria nor on the number of 0s required.
One of these two lists contains the random variables (other than $x_i$) with probability at most $\frac12$, and the other contains those (other than $x_i$) with probability greater than $\frac12$.
Let $\mathcal{A}$ be the former list if $p_i$ is at most $\frac12$, else let $\mathcal{A}$ be the latter list---i.e., in either case, $\mathcal{A}$ is the list containing the elements with probabilities lying in the same range as the probability of $x_i$.
Accordingly $\mathcal{B}$ denotes the other list.
Let $\mathcal{A}'$ denote the sequence of variables obtained by prepending $x_i$ to $\mathcal{A}$.
In the case that $\mathcal{B}$ is empty, any of the deterministic components of $R_j$ is already based on the \enquote{round structure} $\CB{x_i}\CB{x_{a_1}}\CB{x_{a_2}}\ldots$, where braces denote the rounds of testing with the possibility of stopping afterwards, and therefore is already an $i$-pseudo-IO strategy adhering to the Round-Completion property.
If $\mathcal{B}$ is nonempty, then any deterministic component of $R_j$ has a round structure of the form $\CB{x_i}\CB{x_{b_1}}\CB{x_{b_2}}\ldots$ or $\CB{x_i}\CB{x_{b_1},x_{a_1}}\CB{x_{b_2},x_{a_2}}\ldots$ (depending on whether $\mathcal{A}$ is empty) where we index the rounds with $\ell$; in round $\ell$ with $\ell \geq 2$, variables $x_{b_{\ell-1}}$ and $x_{a_{\ell-1}}$ are evaluated if they exist and $R_j$ might stop, depending on the outcomes and internal randomization.

The required round structure for a pseudo-IO strategy with Round Completion, on the other hand, is given by $\CB{x_i,x_{b_1}}\CB{x_{b_2}}\ldots$ or $\CB{x_{i},x_{b_1}}\CB{x_{a_1},x_{b_2}}\ldots$. 
Fix a deterministic component of $R_j$, and a realization $\sigma$. 
Consider the deterministic strategy obtained by modifying the component of $R_j$ to instead adhere to this required round structure for a pseudo-IO, by testing the same elements in the same order as before, and any additional elements necessary to fulfill the RC property.
For any realization $\sigma$, the following properties hold after this modification of the round structure:
\begin{itemize}
    \item All random variables evaluated by the original strategy are also evaluated.
    \item At most one additional random variable is evaluated.
    \item The Round-Completion property is fulfilled with respect to sequences $\mathcal{A}'$ and $\mathcal{B}$.
\end{itemize}
Lastly, any testing round conducted after a certificate has been found is removed.
The modified strategy is therefore an RC-pruning of $i$-pseudo-IO $\mathcal{I}_i$ with additional expected amortized cost of at most $1$ (from at most one additional test) and at least the same expected utility (any certificate found will still be found after the modification).
Apply this transformation to each deterministic component of $R_j$ and denote the resulting strategy by $R'_j$, which is now a distribution over RC-prunings of $i$-pseudo-IO $\mathcal{I}_i$ and thus a randomized RC-pruning of $\mathcal{I}_i$ with
\begin{align}\label{eq:ind_preIO}
    \Ex{\Acost{R'_j}} &\leq 1 + \Ex{\Acost{R_j}}\\
    \Ex{\Utilk{\kk}{R'_j}} & \geq \Ex{\Utilk{\kk}{R_j}} .
\end{align}

By \Cref{lem:IOtreesurgery}, there exists a randomized RC-pruning $R''_j$ of IO with the following properties:
\begin{align}
    \Ex{\Acost{R''_j}} &\leq 1 + \Ex{\Acost{R'_j}}\label{eq:ind_inner}\\
    \Ex{\Utilk{\kk}{R''_j}} & \geq \Ex{\Utilk{\kk}{R'_j}} .
\end{align}

By combining \nameCrefs{eq:ind_inner}
 \labelcref{eq:ind_inner}, \labelcref{eq:ind_preIO}, \labelcref{eq:ind_outer2} and \labelcref{eq:IOstart}, we obtain
\begin{align}
    \label{eq:IOoverallacost}
    \Ex{\Acost{R''_j}} &\leq 2 + 1+ \frac{C_1}{2} \SB{\qq[i] \cdot \Ex{\Cost{S_j^0}} + \RB{1-\qq[i]} \cdot \Ex{\Cost{S_j^1}}} \\
    &\leq \frac{C_1}{2} \cdot \SB{1 + \qq[i] \cdot \Ex{\Cost{S_j^0}} + \RB{1-\qq[i]} \cdot \Ex{\Cost{S_j^1}}}\\
    &\leq \frac{C_1}{2} \cdot \Ex{\Cost{S_j}}\label{eq:IOcore}
\end{align}
where the second inequality holds as long as $\frac{C_1}{2} \geq 3$.

To finish the proof, recall the initial randomized strategy $S$, which selects deterministic strategies $S_j$ with probability $P_j$.
Consider strategy $S''$ defined by selecting strategy $R''_j$ with probability $P_j$.
$S''$ is a distribution over randomized RC-prunings of IO, therefore it is a randomized RC-pruning of IO itself.
We have shown that for each $j \in J$, \Cref{eq:inductionacost,eq:inductionutil} hold.
By linearity of expectation applied over $J$, this implies the induction hypothesis and, together with \Cref{eq:amortizedcost}, \Cref{lem:IOapprox} holds for any $C_1 \geq 6$.
\end{proof}

\subsubsection{Proof of \Cref{lem:IOtreesurgery}}
\label{sec:IOtechnicallemma}

It remains to present the proof of \Cref{lem:IOtreesurgery}.
We include two illustrative cases here while the remaining cases are covered in the appendix.

\begin{proof} [Proof of \Cref{lem:IOtreesurgery}]
Let $\mathcal{A} = \RB{x_{i}, x_{a_1}, \ldots, x_{a_m}}$ and $\mathcal{B} = \RB{x_{b_1}, \ldots, x_{b_\ell}}$ be the two lists on which $\mathcal{I}_i$ performs its round robin, so $x_i$ is the element that is not necessarily in the correct position in list $\mathcal{A}$, from the definition of $\mathcal{I}_i$.
Note that this element is identical for any of the RC-prunings of $\mathcal{I}_i$ and that $m$ and $\ell$ can also be $0$.
The proof is by induction on $m$.

If $\kk \leq 0$, we can set $S''$ to be the strategy that does not test any random variables. This is an RC-pruning of IO and the lemma statement follows immediately from $\Utilk{\kk}{S''}=1$ and $\Acost{S''} = 0$.
If $m = 0$, list $\mathcal{A}$ is already in the correct sorted order and the lemma clearly holds.
If $m > 0$ and $x_i$ is already in its correct position from the definition of InsideOut, the lemma also holds.

For the remaining cases, $x_{a_1}$ in $\mathcal{A}$ and $|\frac12 - p_i| \geq |\frac12-p_{a_1}|$.
For the induction step, 
let $M \geq 1$ and
assume the lemma statement holds for all instances with $m < M$.  
 We will show the lemma holds for $m=M$.  Consider an instance with $m = M$.  List $\mathcal{A}$ contains both $x_i$ and $x_{a_1}$ and consists of $m+1$ elements altogether.
In what follows, for $k \in \{1,2\}$ we will refer to variable $x_{b_k}$ as \emph{live} if $k \leq \ell$ (the length of $\mathcal{B}$), i.e., if $x_{b_k}$ exists as an element of $\mathcal{B}$.  We also refer to $x_i$ and $x_{a_1}$, both in $\mathcal{A}$, as live.  

Let $J$ be the set of indices such that
the $S_j$ with $j \in J$ are the deterministic RC-prunings of $\mathcal{I}_i$ in the support of the distribution over deterministic RC-prunings of $\mathcal{I}_i$ that corresponds to the randomized strategy $S$.  By linearity of expectation, it suffices to show that for every $S_j$, there exists a randomized RC-pruning $R_j$ of IO such that
\begin{align}
\Ex{\Acost{R_j}} &\leq \Ex{\Acost{S_j}} + 1\label{eq:TScost}\\
\Ex{\Utilk{\kk}{R_j}} &\geq \Ex{\Utilk{\kk}{S_j}}\label{eq:TSutil}.
\end{align}

The result will then follow by replacing each $S_j$ in the support of $S$ by $R_j$ (not modifying the probability associated with $S_j$); the resulting strategy is a distribution over randomized RC-prunings of IO, and is thus itself a randomized RC-pruning of IO. 

We now fix $j$. If $S_j$ does not test any elements, then it is already a (trivial) RC-pruning of IO.
So suppose that $S_j$ performs at least one round of testing.

Since $S_j$ is an RC-pruning of $\mathcal{I}_i$, its first-round elements are the live elements of $\{x_i,x_{b_1}\}$ and (if carried out) its second-round elements are the live elements of $\{x_{a_1},x_{b_2}\}$.  

We do not construct $R_j$ from $S_j$ directly. Instead, 
we construct replacements for $S_j$ after conditioning on fixed but arbitrary values for the variables $x_{b_1}$ and $x_{b_2}$ (if they are live), and we then
combine the replacements of $S_j$ to form $R_j$.
Intuitively, the conditioning ensures that a test on either $x_{b_1}$ or $x_{b_2}$ has only one possible outcome.  

Let $\sigma$ denote the partial realization representing the chosen realizations of these two elements.
Conditioned on $\sigma$, in the first 2 rounds, $S_j$ can depend only on the value of $x_i$ and the value of $x_{a_1}$ (as illustrated e.g., on the left side of
\Cref{tab:tree11}, in the first two levels of the tree). 
We construct a replacement in the form of a randomized RC-pruning of IO, also conditioned on $\sigma$. By definition of the IO strategy, 
this randomized RC-pruning tests the live elements in $\{x_{a_1},x_{b_1}\}$ in the first round, and the live elements of $\{x_{a_2},x_{b_2}\}$ in the second round (assuming that round is carried out).

We show below that the desired \Cref{eq:TScost,eq:TSutil} above hold for this randomized RC-pruning, conditioned on $\sigma$.
This is done by case analysis, where each of the (as we see later) seven cases corresponds to the stopping behavior---with or without a certificate---of $S_j$ (conditioned on $\sigma$), after completion of the first testing round\footnote{Specifically, the distinction between cases, and the associated replacements, do not depend on the realizations of the tested elements in the second round.}.  For the case analysis, we treat $x_{b_1}$ and $x_{b_2}$ as both being live. The arguments easily extend if this does not hold because the case analysis conditions on fixed values for these variables. 

By the law of total expectation over the four combinations of values for $x_{b_1}$ and $x_{b_2}$, this implies that these inequalities hold overall, as long as the different replacements can be merged into a strategy $R_j$ in a well-defined way.
As we will discuss later, this can in fact be done, because in our replacements, decisions (on the next test and on whether to stop) do not depend on the values of  $x_{b_1}$ or $x_{b_2}$ until and unless they have been tested and thus their values are known.  

Thus we fix the values of $x_{b_1}$ and $x_{b_2}$, and break the analysis in two levels into several cases---on the lower level as described above according to the properties of $S_j$ in its first two rounds of testing, and on the top level on whether $\qq[i] \geq \frac12$.
We take into account the values of $x_{b_1}$ and $x_{b_2}$ by defining $\kk_1 := \kk -\RB{1-x_{b_1}}$ and $\kk_2 := \kk - \RB{2-x_{b_1}-x_{b_2}}$.

For each case, we perform a transformation on the strategy $S_j$, which either directly gives us a randomized RC-pruning of IO,  or swaps elements $x_i$ and $x_{a_1}$ so that the element in the wrong position is now at a lower level, allowing us to apply the induction hypothesis for the subinstance.
We will show that the inequalities in the induction hypothesis hold after the transformation.  

Consider $S_j$.
We will analyze cost and utility changes explicitly for its first 2 rounds.  
Let $T^0$ denote the substrategy of $S_j$ if $x_i = 0$ and $T^1$ if $x_i = 1$; let $T^{uv}$ for $u,v \in \CB{0,1}$ be the strategies on the induced subinstances after performing the first two testing rounds for the corresponding combination of outcomes of $x_i$ and $x_{a_1}$ (which includes the strategy of stopping with or without a certificate without doing any testing); we also use $T'^{\;0}$ and $T'^{\;1}$ for the substrategies of a strategy $S'_j$ representing an intermediate step in the transformation, for realizations $0$ and $1$ of $x_{a_1}$ in the first round.

\subsubsection*{Case 1: $\qq[i] \geq \frac12$}
In what follows, summarized as Case 1, we assume $\qq[i] \geq \frac12$,  and thus $\qq[a_1] \leq \qq[i]$.
Note that by definition of $\acost$, this implies $\Ex{\Acost{a_1}} = \qq[a_1]$ and $\Ex{\Acost{i}} = \qq[i]$.

Recall that the strategy that does not test any random variables has been handled upfront.
We can thus distinguish a number of subcases by properties of the substrategies $T^0$ and $T^1$ after the first round of tests.  More particularly, for each substrategy, we classify it according to  whether it (a)
terminates immediately without any further tests because a 0-certificate was found as a result of the first round of $S_j$, (b) terminates without further tests and without having found a 0-certificate, or (c) performs tests on $x_{a_1}$ and $x_{b_2}$, and possibly more after that. 
We denote these three alternatives as \textit{Certificate}, \textit{Abort} and \textit{Continue}, and use their combinations as identifiers for the resulting cases---for example, \textit{Certificate-Continue} denotes the case where $T^0$ does not test any elements because a 0-certificate has already been found and $T^1$ continues with at least one testing round.

There are thus 3 possible properties for each of $T_0$ and $T_1$, and 9
combinations of properties of $T_0$ and $T_1$ combined.
However, for an RC-pruning $S_j$ of pseudo-IO, it is impossible for $T_1$ to have the \textit{Certificate} property if $T_0$ did not also have that property (since we are searching for 0-certificates), which reduces the number of subcases to 7.

We give two subcases of Case 1 here -- Case 1.1 (\textit{Continue-Continue}), which illustrates the general approach, and Case 1.2 (\textit{Continue-Abort}), which requires a stopping decision with probability between $0$ and $1$.
The other subcases of Case 1, with varying complexity, are shown in the appendix. Of those, one case (\textit{Abort-Continue}) is special: It is suboptimal to stop without a certificate if $x_i = 0$ but continue if $x_i = 1$, so it can be reduced to two other cases.
The 7 subcases of Case 2 ($\qq[i] < \frac12$) are also covered in the appendix. Unfortunately, those subcases do not immediately follow by symmetry with Case 1, because utility is only obtained by finding $\kk$ 0's. 

\underline{Case 1.1} (\textit{Continue-Continue}):
\begin{figure}
\begin{minipage}{.48\textwidth}
    \centering
    \scalebox{0.75}{
        \begin{forest}
            for tree ={edge={very thick,-Stealth},s sep = .5cm}
            [{$\CB{x_i,x_{b_1}}$},innernode,
                [{$\CB{x_{a_1},x_{b_2}}$},innernode,edge label={node[midway,above left] {$x_i = 0\,\,$}}
                    [{$\,\,T^{00}\,\,$},leaftree, edge label={node[midway,above left] {$x_{a_1} = 0$}} 
                    ] 
                    [{$\,\,T^{01}\,\,$},edge label={node[midway,above right] {$x_{a_1} = 1$}},leaftree
                    ]
                ]
                [{$\CB{x_{a_1},x_{b_2}}$},innernode,edge label={node[midway,above right] {$\,\,x_i=1$}} 
                    [{$\,\,T^{10}\,\,$},leaftree, edge label={node[midway,above left] {$x_{a_1}=0$}} 
                    ] 
                    [{$\,\,T^{11}\,\,$},edge label={node[midway,above right] {$x_{a_1}=1$}},leaftree
                    ]
                ] 
            ]
        \end{forest}
    }
\end{minipage}%
\hfill\vline\hfill
\begin{minipage}{.48\textwidth}
    \centering
    \scalebox{0.75}{
        \begin{forest}
            for tree ={edge={very thick,-Stealth},s sep = .5cm}
            [{$\CB{x_{a_1},x_{b_1}}$},innernode,
                [{$\CB{x_i,x_{b_2}}$},innernode,edge label={node[midway,above left] {$x_{a_1} = 0\,\,$}}
                    [{$\,\,T^{00}\,\,$},leaftree, edge label={node[midway,above left] {$x_i = 0$}} 
                    ] 
                    [{$\,\,T^{10}\,\,$},edge label={node[midway,above right] {$x_i = 1$}},leaftree
                    ]
                ]
                [{$\CB{x_i,x_{b_2}}$},innernode,edge label={node[midway,above right] {$\,\,x_{a_1}=1$}} 
                    [{$\,\,T^{01}\,\,$},leaftree, edge label={node[midway,above left] {$x_i=0$}} 
                    ] 
                    [{$\,\,T^{11}\,\,$},edge label={node[midway,above right] {$x_i=1$}},leaftree
                    ]
                ] 
            ]
        \end{forest}
    }
\end{minipage}

\caption{Case 1.1 (\textit{Continue-Continue}). $S_j$ on the left, $S'_j$ on the right (note swapped roles of $T^{01}$ and $T^{10}$ so that these subtrees are reached with the same probability as before).}
\label{tab:tree11}
\end{figure}
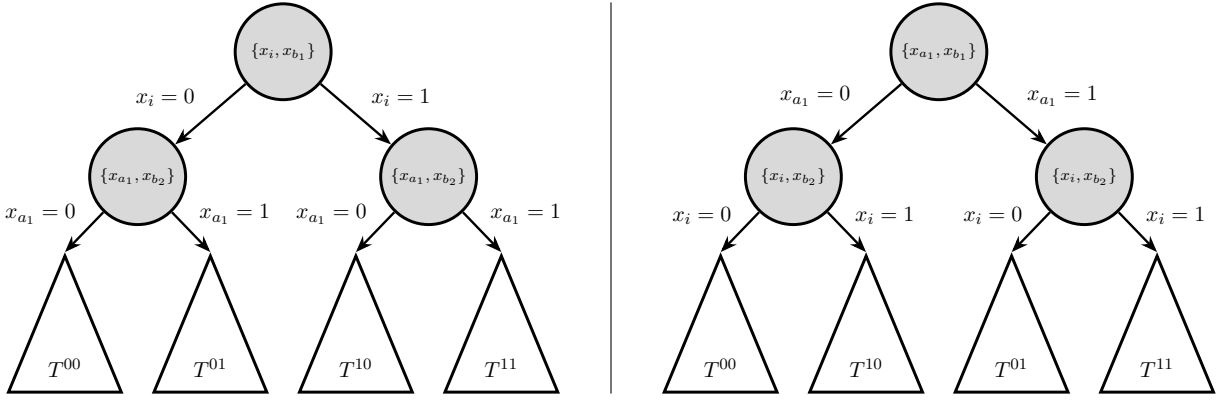

This is the case where two rounds of testing are executed, regardless of the value of $x_i$.
We give an illustration of the top of the decision tree for this case in \Cref{tab:tree11}, showing the tested variables in the first two rounds. Depending only on the outcomes of tests on $x_i$ and $x_{a_1}$ (since we've conditioned on fixed outcomes of tests on $x_{b_1}$ and $x_{b_2}$), the strategy then continues execution with subtree (substrategy) $T^{11}$, $T^{10}$, $T^{01}$, or $T^{00}$. 
Consider $S'_j$, obtained from $S_j$ by swapping the position of elements $x_i$ and $x_{a_1}$, as well as the positions of subtrees $T^{01}$ and $T^{10}$.
In $S'_j$, each of the 4 subtrees after the second round is reached with the same probability as in $S_j$.  Also, each is reached after having obtained the same number of 0 outcomes from the observations in rounds 1 and 2, and thus is executed on the same subinstance of the original problem. 
It follows that expected amortized cost and expected utility remain unchanged.
Observe that for $S'_j$, the elements in the first round are now in the position they should be to fulfill the properties of a pruning of IO.

The strategies $T'^{\,1}$ and $T'^{\,0}$ on the subinstances after the first round of testing in $S'_j$ are now prunings of a pseudo-IO strategy for those subinstances, which has $x_i$ as the out-of-order element at the start of a list containing $M-1$ other elements.
By the induction hypothesis, there exists a randomized pruning $R^0$ of IO (defined on the subinstance) with, in expectation, the same or higher $\util_{\kk_1-1}$ as $T'^{\,0}$ and additional expected amortized cost of at most $1$, and a randomized pruning $R^1$ of IO with, in expectation, same or higher $\util_{\kk_1}$ as $T'^{\,1}$ and, again, an additional expected amortized cost of at most $1$.

Let $R_j$ be the strategy that results from $S'_j$ by replacing $T'^{\,0}$ by $R^0$ and $T'^{\,1}$ by $R^1$ (using the cross product of the distributions in $R^0$ and  $R^1$).
$R_j$ is therefore a distribution over prunings of IO because all elements are now in the proper order.

For the utility, we have:
\begin{align*}
    &\Ex{\Utilk{\kk}{R_j}} = \qq[a_1] \cdot \Ex{\Utilk{\kk_1-1}{R^0}} + \RB{1-\qq[a_1]} \cdot \Ex{\Utilk{\kk_1}{R^1}} \\
        &\quad\geq \qq[a_1] \cdot \Ex{\Utilk{\kk_1-1}{T'^{\,0}}} + \RB{1-\qq[a_1]} \cdot \Ex{\Utilk{\kk_1}{T'^{\,1}}}\\
        &\quad= \qq[a_1] \cdot \SB{\qq[i] \cdot \Ex{\Utilk{\kk_2-2}{T^{00}}} + \RB{1-\qq[i]} \cdot \Ex{\Utilk{\kk_2-1}{T^{10}}}}\\
        &\qquad+ \RB{1-\qq[a_1]} \cdot \SB{\qq[i] \cdot \Ex{\Utilk{\kk_2-1}{T^{01}}} + \RB{1-\qq[i]} \cdot \Ex{\Utilk{\kk_2}{T^{11}}}}\\
        &\quad= \qq[i] \cdot \SB{\qq[a_1] \cdot \Ex{\Utilk{\kk_2-2}{T^{00}}} + \RB{1-\qq[a_1]} \cdot \Ex{\Utilk{\kk_2-1}{T^{01}}}}\\
        &\qquad+ \RB{1-\qq[i]} \cdot \SB{\qq[a_1] \cdot \Ex{\Utilk{\kk_2-1}{T^{10}}} + \RB{1-\qq[a_1]} \cdot \Ex{\Utilk{\kk_2}{T^{11}}}}\\
        &\quad= \Ex{\Utilk{\kk}{S_j}}
\end{align*}
where the induction hypothesis is used in the inequality.
The equalities follow from the definition of the expressions and comparing coefficients.
For the expected amortized cost:
\begin{alignat*}{1}
    &\Ex{\Acost{R_j}} = \Acost{b_1} + \Ex{\Acost{a_1}} + \qq[a_1] \cdot \Ex{\Acost{R^0}} + \RB{1-\qq[a_1]} \cdot \Ex{\Acost{R^1}} \\
        &\quad\leq \Acost{b_1} + \Ex{\Acost{a_1}} + \qq[a_1] \cdot \SB{\Ex{\Acost{T'^{\,0}}}+1} + \RB{1-\qq[a_1]} \cdot \SB{\Ex{\Acost{{T'^{\,1}}}}+1}\\
        &\quad=\Acost{b_1} +\Acost{b_2} + \Ex{\Acost{a_1}+\Acost{i}}+1\\
        &\qquad+ \qq[a_1] \cdot \SB{\qq[i] \cdot \Ex{\Acost{{T^{00}}}} + \RB{1-\qq[i]} \cdot \Ex{\Acost{T^{10}}}} \\
        &\qquad + \RB{1-\qq[a_1]} \cdot \SB{\qq[i] \cdot \Ex{\Acost{{T^{01}}}} + \RB{1-\qq[i]} \cdot \Ex{\Acost{T^{11}}}}\\
        &\quad=\Acost{b_1} +\Acost{b_2} + \Ex{\Acost{i}+\Acost{a_1}}+1\\
        &\qquad+\qq[i] \cdot \SB{\qq[a_1] \cdot \Ex{\Acost{{T^{00}}}} + \RB{1-\qq[a_1]} \cdot \Ex{\Acost{{T^{01}}}}} \\
        &\qquad + \RB{1-\qq[i]} \cdot \SB{\qq[a_1] \cdot \Ex{\Acost{{T^{10}}}} + \RB{1-\qq[a_1]} \cdot \Ex{\Acost{{T^{11}}}}}\\
        &\quad= \Ex{\Acost{S_j}} + 1
\end{alignat*}
where the inequality holds by induction hypothesis, the first, second and last equality by definition of the expressions and the third equality by comparing the coefficients.
For $\ell \in \CB{0,1}$, the arguments and estimates work in the same way.

\underline{Case 1.2} (\textit{Continue-Abort}):
\begin{figure}
\begin{minipage}[b]{.48\textwidth}
    \centering
    \scalebox{0.75}{
        \begin{forest}
            for tree ={edge={very thick,-Stealth},s sep=0.5cm}
            [{$\CB{x_i,x_{b_1}}$},innernode,s sep = 3cm
                [{$\CB{x_{a_1},x_{b_2}}$},innernode,edge label={node[midway,above left] {$x_i = 0\,\,$}}
                    [{$\,\,T^{00}\,\,$},edge label={node[midway,above left] {$x_{a_1} = 0$}},leaftree
                    ]
                    [{$\,\,T^{01}\,\,$},edge label={node[midway,above right] {$x_{a_1}=1$}},leaftree
                    ]
                ]
                [{},endnodef, child anchor=,edge label={node[midway,above right] {$\,\,x_i = 1\,\,$}}
                ]
            ]
        \end{forest}
    }
\end{minipage}%
\hfill\vline\hfill
\begin{minipage}[b]{.48\textwidth}
    \centering
    \scalebox{0.75}{
        \begin{forest}
            for tree ={edge={very thick,-Stealth},s sep=.5cm}
            [{$\CB{x_{a_1},x_{b_1}}$},innernode,
                [,nonode,tier=stopa,edge={-},edge label={node[midway,above left] {$x_{a_1} = 0\,\,$}}
                    [{$\CB{x_i,x_{b_2}}$},innernode,tier=contb,
                        [{$T^{00}$},leaftree, edge label={node[midway,above left] {$x_i =0$}}
                        ] 
                        [{$T^{01}$},edge label={node[midway,above right] {$x_i = 1$}},leaftree
                        ]
                    ]
                ]
                [{$1-\alpha$},stopnode,edge label={node[midway,above right] {$\,\,x_{a_1} = 1\,\,$}}
                    [{$\CB{x_i,x_{b_2}}$},tier=contb,innernode,edge={dotted}
                        [{$\,\,T^{01}\,\,$},leaftree, edge label={node[midway,above left] {$x_i = 0$}} 
                        ]
                        [{},endnodef,edge label={node[midway,above right] {$\,\,x_i = 1\,\,$}}
                        ]
                    ]
                ]
            ]
        \end{forest}
    }

\end{minipage}

\caption{Case 1.2 (\textit{Continue-Abort}). $S_j$ on the left; $S'_j$ on the right, defined in the text as random selection over two deterministic strategies ($S''_j$ and $\tilde{S}''_j$), and can be depicted in this way. The octagonal nodes correspond to stopping without a 0-certificate (with certainty or with probability $1-\alpha$ and continuing otherwise). In this case, $\alpha := \frac{\qq[i] - \qq[a_1]}{\qq[i] \cdot \RB{1-\qq[a_1]}}$.}
\label{tab:tree12}
\end{figure}
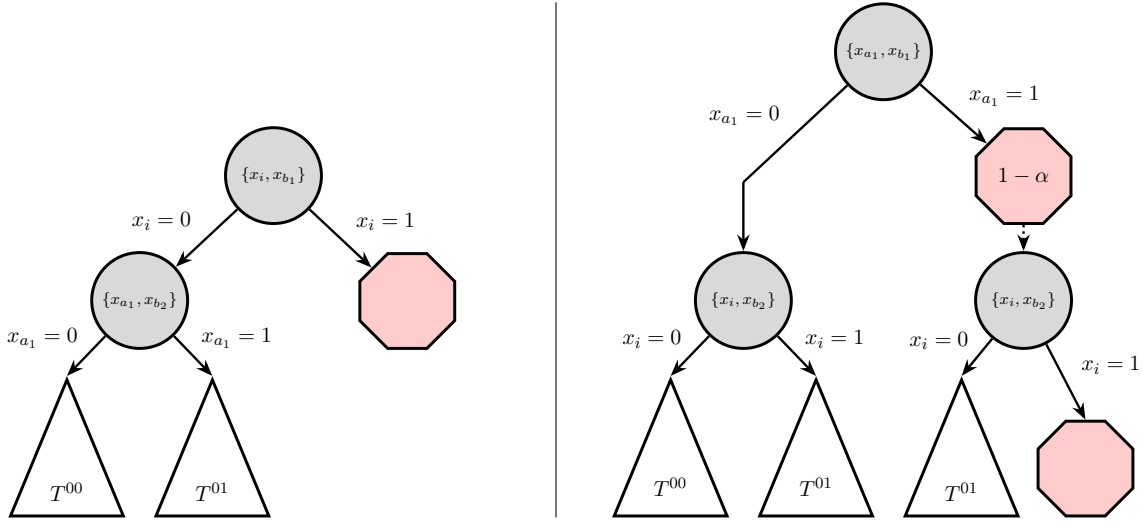

This is the case where evaluation continues if $x_i = 0$ (including that $T^{00}$ and $T^{01}$ may possibly not test any elements) and stops without a 0-certificate if $x_i = 1$; implicitly, this means $\kk \geq 2$.
The first transformation step is to swap elements $x_i$ and $x_{a_1}$.
However, this decreases the probability of reaching subtree $T^{01}$ and thus the expected utility since $\qq[i] \cdot \RB{1-\qq[a_1]} \geq \qq[a_1] \cdot \RB{1-\qq[i]}$ because $\qq[i] \geq \qq[a_1]$.
To compensate, we define two deterministic strategies, both with $x_{a_i}$ swapped to the first round, and choose randomly between them. 
The first, denoted by $\tilde{S}''_j$, is identical to $S_j$ except for swapping elements $x_i$ and $x_{a_i}$.
It is chosen with probability $1-\alpha$ (with $\alpha$ as given below).
The second strategy, denoted by $S''_j$, instead of stopping after discovering that $x_{a_1}=1$, continues to the second round. Following the second round, if $x_i=0$, it executes $T^{01}$ (on the same subinstance with the same number of 0s found so far), else if $x_i=1$, it stops without a certificate.
Strategy $S''_j$ is chosen with probability $\alpha$.  The resulting randomized strategy $S'_j$, composed of $\tilde{S}''_j$ and $S''_j$, is depicted in a single tree on the right side of \Cref{tab:tree12}.
Again, $x_{a_1}$ is now in the position it has in IO.
We set $\alpha$ to be $\frac{\qq[i] - \qq[a_1]}{\qq[i] \cdot \RB{1-\qq[a_1]}} \in \RB{0,1}$.  As shown below, this ensures that the overall expected utility is preserved.

We denote the substrategies after the first round for $x_{a_1} = 0$ in $S'_j$ with $T'^{\,0}$ (the same for $S''_j$ and $\tilde{S}''_j$), and for $x_{a_1} = 1$ by $T'^{\,1}$ for $S''_j$ and by $\tilde{T}'^{\,1}$ for $\tilde{S}''_j$.
Note that $T'^{\,0}$, $T'^{\,1}$ and $\tilde{T}'^{\,1}$ are all prunings of pseudo-IO on the respective subinstances; $\tilde{T}'^{\;1}$ (which tests no elements) is trivially already a pruning of IO.
They are also RC-prunings, since $S_j$ adheres to the round-completion property.
By the induction hypothesis, we can replace $T'^{\,0}$ and $T'^{\,1}$ by distributions $R^0$ and $R^1$ over prunings of IO on the subinstances while preserving expected utility and bounding any additional expected amortized cost.
The resulting strategy $R_j$ now follows the order required for IO and is a distribution over prunings.

We calculate the expected utility and expected amortized cost:
\begin{align*}
    &\Ex{\Utilk{\kk}{R_j}} = \qq[a_1] \cdot \Ex{\Utilk{\kk_1-1}{R^0}} + \RB{1-\qq[a_1]} \cdot \SB{\alpha \cdot \Ex{\Utilk{\kk_1}{R^1}} + \RB{1-\alpha} \cdot 0}\\
    &\quad \geq \qq[a_1] \cdot \Ex{\Utilk{\kk_1-1}{T'^{\,0}}} + \frac{\qq[i] - \qq[a_1]}{\qq[i]} \cdot \Ex{\Utilk{\kk_1}{T'^{\,1}}} \\ 
    &\quad = \qq[a_1] \cdot \SB{\qq[i] \cdot \Ex{\Utilk{\kk_2-2}{T^{00}}} + \RB{1-\qq[i]} \cdot \Ex{\Utilk{\kk_2-1}{T^{01}}}} + \frac{\qq[i] - \qq[a_1]}{\qq[i]} \cdot \qq[i] \cdot \Ex{\Utilk{\kk_2-1}{T^{01}}}\\
    &\quad = \qq[i] \cdot \qq[a_1] \cdot \Ex{\Utilk{\kk_2-2}{T^{00}}} + \qq[i] \cdot \RB{1-\qq[a_1]} \cdot \Ex{\Utilk{\kk_2-1}{T^{01}}} \\
    &\quad = \Ex{\Utilk{\kk}{S_j}} \, .
\end{align*}
In the inequality, we use that $\tilde{T}'^{\,1}$ does not contribute to utility, the guarantee on expected utility from the induction hypothesis for the other cases and we plug in the definition of $\alpha$.
The second equality follows from the definition of the substrategies.
For the third equality, we rearrange the terms algebraically to match the structure of $S_j$.

Estimating amortized cost is more intricate; depending on the realization of $x_{b_1}$ and $x_{b_2}$, additional amortized cost may occur from swapping $x_i$ and $x_{a_1}$:
\begin{align*}
    &\Ex{\Acost{R_j}} = \Ex{\Acost{a_1}} + \Acost{b_1} \\
    &\qquad+ \qq[a_1] \cdot \Ex{\Acost{R^0}} + \RB{1-\qq[a_1]} \cdot \SB{\alpha \cdot \Ex{\Acost{R^1}} + \RB{1-\alpha} \cdot 0}\\
    &\quad\leq \Ex{\Acost{a_1}} + \Acost{b_1} + \qq[a_1] \cdot \RB{1 + \Ex{\Acost{T'^{\,0}}}} + \RB{1-\qq[a_1]} \cdot \alpha \cdot \RB{1 + \Ex{\Acost{T'^{\,1}}}}\\
    &\quad= \Ex{\Acost{a_1}} + \Acost{b_1} \\
    &\qquad+ \qq[a_1] \cdot \SB{\Ex{\Acost{i}} + \Acost{b_2} + \qq[i] \cdot \Ex{\Acost{T^{00}}} + \RB{1-\qq[i]} \cdot \Ex{\Acost{T^{01}}}}\\
    &\qquad+ \RB{1-\qq[a_1]} \cdot \alpha \cdot \SB{\Ex{\Acost{i}} + \Acost{b_2} + \qq[i] \cdot \Ex{\Acost{T^{01}}}}\\
    &\qquad+1 \cdot \SB{\qq[a_1] + \alpha \cdot \RB{1-\qq[a_1]}}\\
    &\quad= \Ex{\Acost{a_1}} + \Acost{b_1} \\
    &\qquad+ \qq[i] \cdot \qq[a_1] \cdot \Ex{\Acost{T^{00}}} + \SB{\qq[a_1] \cdot \RB{1-\qq[i]} + \RB{\qq[i] - \qq[a_1]}} \cdot \Ex{\Acost{T^{01}}}\\
    &\qquad+ \SB{\qq[a_1] + \alpha \cdot \RB{1-\qq[a_1]}} \cdot \SB{\Ex{\Acost{i}}+\Acost{b_2}} + \SB{\qq[a_1] + \alpha \cdot \RB{1 - \qq[a_1]}}\\
    &\quad= \Ex{\Acost{i}} + \Acost{b_1}\\
    &\qquad+ \qq[i] \cdot \SB{\Ex{\Acost{a_1}} + \Acost{b_2} + \qq[a_1] \cdot \Ex{\Acost{T^{00}}} + \RB{1-\qq[a_1]} \cdot \Ex{\Acost{T^{01}}}}\\
    &\qquad+ \RB{1-\qq[i]} \cdot \Ex{\Acost{a_1}} - \SB{1 - \qq[a_1] - \alpha \cdot \RB{1-\qq[a_1]}} \cdot \Ex{\Acost{i}}\\
    &\qquad+ \SB{\qq[a_1] + \alpha \cdot \RB{1-\qq[a_1]} - \qq[i]} \cdot \Acost{b_2} + \SB{\qq[a_1] + \alpha \cdot \RB{1 - \qq[a_1]}}\\
    &\quad\leq \Ex{\Acost{S_j}} + 1 \, .
\end{align*}
In the first equality, we use the strategy definition and that $\tilde{T}'^{\,1}$ does not cause any amortized cost.
The first inequality uses the cost bound given by the induction hypothesis for replacing the pseudo-IO prunings.
The second equality plugs in the definition of $T'^{\,0}$ and $T'^{\,1}$ and rearranges.
The third and fourth inequality perform further rearrangements and use the definition of $\alpha$.
In the last step, we retrieve the expected amortized cost of $S_j$ and upper-bound the quantities in the remaining two lines of the expression by $0$ and $1$ respectively, the calculations for which are deferred to \Cref{app:surgerynurse}.
As before, the results carry over for the case $\ell \in \CB{0,1}$.

This concludes the presentation and analysis of the two example subcases of Case 1.  Analysis of the remaining subcases of Case 1, 
and of Case 2 ($\qq[i] < \frac12$), is deferred to \Cref{app:continuesurgery}.

Having conditioned on $\sigma$ in the case analysis, we now argue that the replacements 
can in fact be merged into a strategy $R_j$ in a well-defined way.  
Since all proposed replacements are (randomized) RC-prunings of IO, the underlying choice of elements to test is the same.
Thus, issues with a well-defined $R_j$ can only arise if a replacement stops (with non-zero probability) before completion of the second round---after that, the 
outcomes of $x_{b_1}$ and $x_{b_2}$ are known.
None of the proposed replacements in any of the cases stop before the first round.
Some replacements do stop with non-zero probability after the first round, possibly dependent on the value of $x_{a_1}$; however, we only use such replacements for cases in which the original strategy also stops, for at least one of the two possible values of $x_i$, after the first round.
Recall that the cases are distinguished based on their stopping behavior in the first two rounds:
For a fixed  value of $x_{b_1}$, if the original strategy $S_j$ stops after the first round when $x_{b_2}=0$, it will also stop in the same way after the first round if $x_{b_2}=1$. Therefore, for fixed value of $x_{b_1}$, the relevant case in our case analysis is not affected by the value of $x_{b_2}$.
This in turn implies that the original strategies will be replaced with (randomized) strategies that have the same stopping behaviors for $x_{b_2}=1$ and $x_{b_2}=0$ after the first round. Consequently, the proposed replacement strategies do not depend on the outcomes of testing $x_{b_1}$ and $x_{b_2}$ before those elements have been queried, and we obtain a well-defined strategy $R_j$.

Overall, this implies that the induction hypothesis is also correct for $m=M$, concluding the induction step and therefore proving \Cref{lem:IOtreesurgery}.
\end{proof}

\section{Conclusion}
A natural question suggested by our work is whether a constant-factor approximation algorithm, for MBT on partition matroids, can also be achieved in the setting where testing $x_i$ incurs an associated cost $c_i$.
Indeed, for the problem of finding 1-certificates for general matroids, it is 
straightforward to generalize \OPTone{} for this setting by ordering the tests according to the ratio $p_i/c_i$.
However, it is not at all clear how to generalize the InsideOut strategy in a way that would be a constant-factor approximation for MP0.
Also, the current analysis in \Cref{sec:IO} is already highly technical, and relies heavily on the unit-cost assumption.  
It therefore appears that handling arbitrary costs will require new ideas both for the design of the algorithm and for its analysis.

On a conceptual level, this paper introduces MBT as a new class of SBFE problems, as well as introducing the budget-in-expectation problem.
We hope that 
our results and techniques---in particular, considering budget-in-expectation---help pave the way towards (approximately) solving MBT on general matroids and to obtaining algorithms for other SBFE problems.

\section*{Acknowledgments.}
We thank Stefan Walzer for early discussions on difficult instances for related SBFE problems, and the anonymous reviewers for providing helpful suggestions and feedback.  Lisa Hellerstein was partially supported by NSF Award 1909335. Kevin Schewior was in part supported by the Independent Research Fund Denmark, Natural Sciences, grants DFF-0135-00018B and DFF-4283-00079B.

\newpage

\bibliographystyle{acm}
\bibliography{bib/bibliography}

\newpage
\appendix

\section{Missing Material from \Cref{sec:Intro}}\label{app:counterexamples}

\subsection{Counterexample for pruning optimal evaluation strategy for MP0 on uniform matroids}

\begin{lemma}
    Let $\Util{A}$ denote the random variable indicating the event that strategy $A$ finds a $0$-certificate.
    For any positive integer $M$, there exists an instance of a $k$-of-$n$ function and a strategy $R$ with the following properties:
    \begin{itemize}
        \item $\Ex{\Util{S'}} \leq \frac{1}{M} \cdot \Ex{\Util{R}}$ for any randomized pruning $S'$ of $S$ with $\Ex{\Cost{S'}} \leq \Ex{\Cost{R}}$ 
        \item $\Ex{\Cost{S''}} \geq M \cdot \Ex{\Cost{R}}$ for any randomized pruning $S''$ of $S$ with $\Ex{\Util{S''}} \geq \Ex{\Util{R}}$ 
    \end{itemize}
    where $S$ is the optimal evaluation strategy.

\end{lemma}
\begin{proof}
    Consider the following instance with $n = 1 + \frac{1}{\varepsilon}$ and $k=n-1$; element 1 has probability $p_1 = 1 - \eps$, and the remaining $n-1$ elements have probability $p_2 = \ldots = p_n = 1 - \eps^2$ of being active.
    We will later set $\eps$ appropriately as a function of $M$.

    Let $R$ be the following (pruned) strategy:
    Start with the test on $x_1$. If $x_1 = 0$, continue with full evaluation (i.e., either until a second inactive variable is found or until all remaining elements have been queried).
    If $x_1 = 1$, stop. The expected cost of running $R$ consists of a contribution of 1 from $x_1$ and at most $\eps$ from each of the remaining elements, since they are inspected with probability at most $\eps$.
    Therefore, $\Ex{\Cost{R}} \leq 1 + \RB{n-1}\cdot\eps = 2$.
    The strategy tree of $R$ has $n-1 = \frac{1}{\eps}$ leaves ending with a $0$-certificate.
    The probability of reaching each of these leafs is at least $\eps \cdot \RB{1-\eps^2}^{n-2} \cdot \eps^2 \geq \eps \cdot \RB{1-\RB{n-1}\eps^2} \cdot \eps^2 = \eps^3 \cdot \RB{1-\eps}$, so $\Ex{\Util{R}} \geq \eps^2 \cdot \RB{1-\eps}$.

    $S$ as the optimal evaluation strategy does the following: Start with the test on $x_2$.
    If $x_2 = 1$, continue with testing $x_3$ and so on until reaching a variable that is inactive or until elements $x_2$ to $x_n$ have been queried (and $S$ stops without a $0$-certificate);
    after the first inactive element, continue with the test on $x_1$ and, if $x_1 = 1$, query all remaining elements until a second inactive variable is found or all elements have been tested.
    
    Let $\tilde{S}$ be a pruning of $S$.
    Let $m \in \CB{1, \ldots, n-1}$ be the number of active elements that $\tilde{S}$ would query if it has not yet found a single inactive variable before stopping without a certificate.
    We can generously ignore any cost incurred after finding the first inactive element and obtain a lower bound on $\Ex{\Cost{\tilde{S}}}$ of $m \cdot \RB{1-\eps^2}^{m-1} \geq m \cdot \RB{1-m\eps^2}$ because each of those elements incurs a cost 1 from testing if it is reached in the tree.
    Conditioned on the first element selected ($x_2$) being inactive, there is one leaf ending with a $0$-certificate reached with probability $\eps$ and $n-2$ leafs ending with a $0$-certificate, each of which reached with probability at most $\eps^2$.
    By the union bound, the highest probability of finding a $0$-certificate we can hope for is at most $\eps + \RB{n-2} \cdot \eps^2 \leq 2\eps$;
    conditioned on the $i$th selected element being the first to be inactive, that value only decreases.
    By the law of total expectation and upper-bounding the probability that the $i$th element being the first inactive one by $1 \cdot \eps^2$, we obtain $\Ex{\Util{\tilde{S}}} \leq m \cdot \eps^2 \cdot 2\eps = 2 \cdot m \cdot \eps^3$.
    Overall we have:
    \begin{equation}\label{eq:evalratiobound}
        \frac{\Ex{\Cost{\tilde{S}}}}{\Ex{\Util{\tilde{S}}}} \geq \frac{m \cdot \RB{1-m\eps^2}}{2\cdot m\cdot\eps^3} = \frac{1}{\eps^3}\cdot\frac{1-m\eps^2}{2} \geq \frac{1}{\eps^3}\cdot\frac{1-\eps}{2}
    \end{equation}
    By linearity of expectation, this bound on the ratio also holds for any distribution over prunings of $S$.

    From \Cref{eq:evalratiobound}, randomized pruning $S'$ of $S$ with $\Ex{\Cost{S'}} \leq \Ex{\Cost{R}} \leq 2$ finds a $0$-certificate with probability at most $\Ex{\Util{S'}} \leq \eps^3 \frac{4}{1-\eps}$; at the same time, $\Ex{\Util{R}} \geq \eps^2 \cdot \RB{1-\eps}$, so $\frac{\Ex{\Util{R}}}{\Ex{\Util{S'}}} \geq \frac{1}{\eps} \cdot \frac{\RB{1-\eps}^2}{4}$.
    Choosing $\eps \leq \frac{1}{8M}$, we obtain $\Ex{\Util{S'}} \leq \frac1M \cdot \Ex{\Util{R}}$, proving the first lemma statement.
    \Cref{eq:evalratiobound} also implies that any distribution $S''$ over prunings of $S$ with $\Ex{\Util{S''}} \geq \Ex{\Util{R}} \geq \eps^2 \cdot \RB{1-\eps}$ incurs cost $\Ex{\Cost{S''}} \geq \frac{1}{\eps} \cdot \frac{\RB{1-\eps}^2}{2} \geq 2M \geq M \cdot \Ex{\Cost{R}}$ for $\eps \leq \frac{1}{8M}$, yielding the second lemma statement.
\end{proof}

\subsection{Approximation factor achievable via the submodular goal value approach of Deshpande et al.}
\label{app:submodulargoal}

Consider an SBFE problem for a certain type of Boolean function $f$.  The approach used by Deshpande et al.\ to obtain a good approximation algorithm for this problem is based on constructing a {\em goal} function $g$ associated with the function $f$ being evaluated~\cite{deshpande2016approximation}. Goal function $g$ is (state-dependent) monotone and submodular.  It associates integer utility values to the information obtained from tests, such that it attains a fixed utility value $Q$ (called the goal value of $g$) precisely when the obtained information is sufficient to determine the value of $f$.  In this way, the SBFE problem for $f$ can be reduced to the Stochastic Submodular Cover problem for utility function $g$.  There are potentially many different goal functions $g$ that can be constructed for $f$, with different properties, including different values for $Q$.

Deshpande et al.\ solve the resulting Stochastic Submodular Cover problem on $g$ using one of two greedy algorithms.   
The first, Adaptive Greedy, is an $O(\log Q)$-approximation algorithm for the Stochastic Submodular Cover problem~\cite{deshpande2016approximation,Hellerstein2021tight,cuiNagarajan-SOSA23}, and yields 
an $O(\log Q)$-approximation for
the associated SBFE problem. 
An alternative is to use Dual Adaptive Greedy, for which Deshpande et al.\ proved a different approximation bound depending on other properties of $g$.  The definition of this bound is somewhat technical, so we will just refer to it as $\alpha$.  

The approach of Deshpande et al.\ to  achieving good approximation bounds for SBFE problems is therefore as follows: try to design $g$ such that either $O(\log Q)$ or $\alpha$ is small, then use Adaptive Greedy or Dual Adaptive Greedy, with respect to $g$, as the approximation algorithm for evaluating $f$.  Using this approach, they achieve an $O(\log n)$ approximation algorithm when $f$ is a Boolean function represented by a decision tree of size polynomial in $n$ (by constructing $g$ with goal value $Q$ that is polynomial in $n$), and a constant-factor approximation algorithm when $f$ is a Boolean linear threshold functions (by constructing $g$ with constant-size $\alpha$). 

If $f$ is a Boolean function whose value depends on all $n$ of its input variables, the goal value $Q$ of any goal function $g$ for $f$ is at least $n$~\cite{bach18submodular}.  It follows that the approach above using Adaptive Greedy and its $O(\log Q)$ bound is not useful in achieving constant-factor approximations for SBFE problems.  In fact, Deshpande et al.\ show the following both for a linear function $f$, and for a function $f$ that is a disjunction of
disjoint $k_i$-of-$n_i$ functions, for $k_i=n_i$ (i.e., a read-once DNF formula): any goal function $g$ for $f$ must have goal value $Q$ that is exponential in $n$, so $\log Q$ is linear in $n$.  Since the SBFE problem for the latter type of function is effectively equivalent to MBT on partition matroids, this rules out achieving {\em any} sublinear approximation based on the $O(\log Q)$ bound for evaluating these functions $f$.

This still leaves open the possibility that for $f$ a disjunction (or conjunction) of $k_i$-of-$n_i$ functions, 
$g$ might be constructed so that $\alpha$ is constant, as Deshpande et al.\ did for linear threshold functions. 
However, it also follows immediately from results of Deshpande et al.\ that 
for {\em any} Boolean function $f$, 
when the probabilities $p_j$ associated with the tests are all equal to $1/2$, $\alpha$ upper bounds the ratio DGREEDY/(2$\cdot$CERT). Here DGREEDY is the expected cost of using Dual Adaptive Greedy (with respect to any goal function $g$ for $f$) as the strategy for evaluating $f$, and CERT is the minimum expected cost of certifying the value of $f(x)$ (i.e., the expected minimum offline cost for evaluating $f$ on random $x$), when the $p_j$ are all equal to $1/2$.\footnote{Similarly, $\log Q$ upper bounds AGREEDY/(2 $\cdot$ CERT) when the $p_j$ are all equal to $\frac{1}{2}$, where AGREEDY is the expected cost of Adaptive Greedy.}
In~\cite{AllenHKU13}, it was shown (Theorem 6) that for any constant $\beta$, where $0 < \beta < 1$, there is a function $f$ that is a disjunction of $k_i$-of-$n_i$ functions, with $k_i=n_i$, such that the ratio OPT/CERT
 is $\Omega(n^{\beta}/\log n)$.  Here OPT is the minimum expected cost of any adaptive strategy for evaluating $f$.  Since Dual Adaptive Greedy is such a strategy, $\alpha = \Omega(n^{\beta})$ for every $0 < \beta < 1$.  This rules out achieving any $o(n^{\beta})$ approximation bound for the MBT problem on partition matroids via Dual Adaptive Greedy and its $\alpha$ bound.

\section{Missing Material from~\Cref{sec:sub:Prun}}
\label{app:pruninglemmas}

\begin{proof}[Proof of \Cref{lem:properprunings}.]
    Consider the elementary testing strategy $S$ and two nodes $u$ and $v$ in the binary decision tree corresponding to $S$ that are reached with the same count of active and inactive elements.
    While the choice and/or order of elements leading up to $u$ and $v$ as well as the subinstances induced at $u$ and $v$ may be different, the substrategies $S(u)$ and $S(v)$ rooted at $u$ and $v$, respectively, are equivalent as $S$ is elementary.
    In particular, the sets of tests in the subtrees are identical at $u$ and $v$.
    $S(v)$ is a well-defined substrategy for the subinstance at $u$, as is $S(u)$ for the subinstance at $v$, and, as $f$ is a symmetric function, either substrategy returns a certificate if run at $v$ if and only if it returns a certificate if  run at $u$.
    Accordingly, $\Exc{\Util{S,u}}{S \text{ reaches }u} = \Exc{\Util{S,v}}{S \text{ reaches }v}$.
    Further, for any pruning $\tilde{S}$ of $S$ that reaches both $u$ and $v$, the substrategies $\tilde{S}(u)$ and $\tilde{S}(v)$ rooted at $u$ and $v$ can both be seen as prunings of the same substrategy of $S$.
    They differ only in terms of stopping, not in the choice or sequence of elements since they were obtained by pruning an elementary strategy.

    Consider now the (the finite set of) prunings $R_j$ of $S$ from the support of the distribution corresponding to randomized strategy $R$, with $j \in J$ for the appropriate set $J$, and consider the following procedure:
    Fix $j \in J$, and fix tuple $\tau = \RB{s,t}$ with $s, t \in \CB{0, \ldots, n}, s+t \leq n$.
    We say that $R_j$ is elementary for tuple $\tau$ if it continues with the same testing strategy at any node with exactly $s$ active and $t$ inactive elements.

    If $R_j$ is not already elementary for $\tau$, do the following. 
    Let $V =  V\RB{R_j, \tau}$ be the (possibly empty) set of nodes in the binary decision tree corresponding to $R_j$ with exactly $s$ active and $t$ inactive elements observed.
    For $v \in V$, let $R_j(v)$ be the (deterministic) testing strategy used by $R_j$ after having tested the variable at $v$.
    Let $\Pty{R_j,v} > 0$ the probability of arriving at $v$, and set $\Pty{R_j,V} := \sum_{v \in V} \Pty{R_j,v}$.
    For each $v$, let $R'^v_j$ be the testing strategy that is obtained from $R_j$ by continuing with $R_j(v)$ at each node in $V$.
    Note that this is well-defined, only changes the decision tree in subtrees rooted at nodes in $V$ and, in particular, for any $u \in V$, $\Pty{R_j,u} = \Pty{R'^v_j,u}$.
    $R'^v_j$ is a pruning of $S$, and by construction it is elementary for $\tau$.

    Let $R'_j$ be the distribution over prunings $\CB{R'^v_j \,\vert\, v \in V}$ that selects $R'^v_j$ with probability $\frac{\Pty{R_j,v}}{\Pty{R_j,V}}$.
    This transformation preserves expected utility and cost:
    \begin{align*}
        &\Ex{\Util{R'_j}} = \sum_{v \in V} \frac{\Pty{R_j,v}}{\Pty{R_j,V}} \cdot \SB{\tilde{U} + \Pty{R_j,V} \cdot \Exc{\Util{R_j,v}}{R_j\text{ reaches }v}}\\
        &\quad= \tilde{U} + \sum_{v \in V} \Pty{R_j, v} \cdot \Exc{\Util{R_j,v}}{R_j\text{ reaches }v} = \Ex{\Util{R_j}}\,\,,
    \end{align*} for the appropriate value $\tilde{U}$.
    Analogously, we obtain $\Ex{\Cost{R'_j}} = \Ex{\Cost{R_j}}$.

    $R'_j$ is a distribution over prunings, each of which is elementary for $\tau$.
    Replacing $R_j$ by $R'_j$ in $R$ this for each $j \in J$, we obtain $R'$ as a distribution over distributions over prunings, and thus itself a distribution over prunings also elementary for $\tau$.

    This operation (for tuple $\tau$) does not destroy being elementary for other tuples: 
    Let $\preceq$ denote the partial order induced by integer comparison on tuples.
    Let $R_j$ be a distribution over prunings of $S$ elementary for some other tuple $\tau'$. If $\tau$ and $\tau'$ are not comparable w.r.t.\ $\preceq$, then none of the subtrees at $V\RB{R_j, \tau'}$ are changed as they are disjoint from the subtrees at $V\RB{R_j,\tau}$.
    
    If $\tau' \prec \tau$ and the prunings in $R_j$ are elementary for $\tau'$, the subtrees at $V\RB{R_j, \tau'}$ are identical; the construction of $R'^v_j$ makes the same replacement in each of those subtrees.
    Therefore, the subtrees at  $V(R'^v_j,\tau')$ are still identical and thus $R'^v_j$ is still elementary for $\tau'$ (since $\tau' \prec \tau$, nodes corresponding to $\tau'$ that were pruned in $R_j$ remain pruned in $R'^v_j$).
    
    Else, if $\tau \prec \tau'$, every node in $V\RB{R_j,\tau'}$ is part of a subtree rooted in  $V\RB{R_j,\tau}$.
    In $R'^v_j$, each such subtree in $V\RB{R_j,\tau}$ is replaced by $R_j(v)$. 
    Since $R_j$ was elementary for $\tau'$, all substrategies within $R_j(v)$ at nodes corresponding to $\tau'$ are identical; therefore, after the replacement, any substrategy at $V(R'^v_j,\tau')$ is (still) identical to that substrategy, including any substrategy at a node in $V(R'^v_j,\tau')$ but not in $V(R_j,\tau')$ that had previously been pruned somewhere in the path (but that is now reached with non-zero probability).
    
    Therefore, after iteratively applying the above operation for each possible tuple, having preserved expected cost and utility, the resulting randomized strategy $R'$ is a distribution over prunings of $S$ where each pruning is elementary for any tuple; thus, it is itself a randomized pruning of $S$ that is elementary .
\end{proof}

In the proof of \Cref{lem:prune}, we will use the following obvious fact.

\begin{fact}\label{fact:algorithms-same}
    For all strategies $S$ and budgets $B$, the value of the function output by ComputePruning\-Function($S$) at $B$ is precisely $\Ex{U(\text{ComputePrunedStrategy}(S,B))}$. 
\end{fact}

We now give the proof of the lemma.

\begin{proof}[Proof of \Cref{lem:prune}.]

    Regarding (i), by definition, $q_S$ is clearly monotone.
    Concavity and piecewise linearity follow from the observation that each randomized pruning $\tilde{S}$ of $S$ can be seen as a distribution over strategies from the finite set of deterministic prunings of $S$. With linearity of expectation on cost and utility, this makes $\ExIl{\cost(\tilde{S})}$ and $\ExIl{\util(\tilde{S})}$ convex combinations of the respective values for the deterministic prunings of $S$.
    
    For concavity, more formally, assume for contradiction that $q_S$ is not concave, i.e., there exist $B_{\ell}, B_{r} \in \mathbb{R}_{\geq 0}$ and $\alpha \in \RB{0,1}$ such that $q_S((1-\alpha)B_\ell+\alpha B_r) < (1-\alpha)\cdot q_S(B_\ell) + \alpha \cdot q_S(B_r)$. Let $S_\ell, S_r \in \prunings(S)$ with $\ExIl{\cost(S_\ell)} \leq B_\ell$, $\ExIl{\cost(S_r)} \leq B_r$, $\ExIl{\util(S_\ell)} = q_S(B_\ell)$ and $\ExIl{\util(S_r)} = q_S(B_r)$, and let $S^\ast \in \prunings(S)$ be the strategy obtained by selecting $S_\ell$ with probability $1-\alpha$ and $S_r$ with probability $\alpha$. By linearity of expectation, $\ExIl{\util(S^\ast)} = (1-\alpha)\cdot q_S(B_\ell) + \alpha \cdot q_S(B_r)$ and $\ExIl{\cost(S^\ast)} \leq (1-\alpha) B_\ell + \alpha B_r$, and thus $q_S((1-\alpha)B_\ell+\alpha B_r) \geq (1-\alpha)\cdot q_S(B_\ell) + \alpha \cdot q_S(B_r)$, in contradiction to the assumption that $q_S$ is not concave.

    We show part (ii) for the binary decision tree representation of $S$; together with \Cref{lem:properprunings}, the proof extends to DAG representations by viewing the pruning of the DAG as a pruning of the corresponding strategy tree.
    
    The proof works by induction on the height of $S$. The base case, in which $S$ is empty, is trivial.
    For the induction step, let $k$ be the height of $S$ and fix some budget $B$. By \Cref{fact:algorithms-same}, it suffices to show that ComputePrunedStrategy($S$,$B$) computes a strategy $S'\in\prunings(S)$ maximizing $\ExIl{U(S')}$ subject to $\mathbb{E}[\cost(S')]\leq B$.
    
    To see that the latter claim is true, consider a strategy $S^*\in\prunings(S)$ that does maximize $\ExIl{U(S^*})$ under the constraint $\mathbb{E}[\cost(S^*)]\leq B$.
    We will transform $S^*$ into the strategy that is computed by ComputePruned\-Strategy($S$, $B$) without decreasing the utility and without increasing the cost.
    To this end, let $p$ be the probability that $S^*$wo du  stops immediately, $x_i$ be the first variable tested by $S^*$, and $S_0^*$ and $S_1^*$ be the left and right subtree of $S^*$, respectively. Analogously, let $S_0$ and $S_1$ be the left and right subtree of $S$, respectively.

    First, replace $S_j^*$ in $S^*$ by ComputePrunedStrategy($S_j$, $\mathbb{E}[\cost(S_j^*)]$), for $j\in\{0,1\}$. By induction, this keeps both the expected cost and utility of $S^*$ identical.
    Again by induction and optimality of $S^*$, the quantities $p$, $y_0:=\mathbb{E}[\cost(S_0^*)]$, and $y_1:=\mathbb{E}[\cost(S_1^*)]$ are chosen such that its utility $$(1-p)\cdot ((1-p_i)\cdot q_{S_0}(y_0)+ p_i\cdot q_{S_1}(y_1))$$ is maximized
    under the constraint that its expected cost $$(1-p)\cdot(1+(1-p_i)\cdot y_0 + p_i\cdot y_1)=B.$$
    Since $q_{S_0}$ and $q_{S_1}$ are monotone concave functions, it is not difficult to see that the optimal values can be computed by the following procedure.
    Start with $a:=\mathbb{E}[\cost(S_0)]$, $b:=\mathbb{E}[\cost(S_1)]$, and $p:=1$, and continuously decrease the variables in two phases until the budget constraint is fulfilled.
    Specifically, in the Phase 1, decrease a variable $y_z$ with $z\in\{0,1\}$ such that the right derivative $\dot{q}_{S_z}^-(y_z)$ is minimal, breaking ties arbitrarily but only switching at breakpoints. This is done until the aforementioned quantity reaches the ratio between the objective and the LHS of the budget constraint, i.e.,
    \begin{equation}
        \frac{(1-p_i)\cdot q_{S_0}(y_0)+ p_i\cdot q_{S_1}(y_1)}{1+(1-p_i)\cdot y_0 + p_i\cdot y_1}\leq\min\{\dot{q}_{S_0}^-(y_0),\dot{q}_{S_1}^-(y_1)\}.
    \end{equation}
    (Note that this inequality may also be fulfilled in the beginning.) As long as this inequality remains true with equality, Phase 1 can be continued or ended, again choosing arbitrarily (but only changing one's mind at breakpoints). If the inequality would cease to be true with equality by continuing Phase 1, Phase 1 has to be ended. In Phase 2, simply decrease $p$ until the budget constraint is fulfilled. It is easy to see that there is a one-to-one correspondence between this procedure and the procedure ComputePrunedStrategy($S$, $B$), where the first phase corresponds to pruning below the root and the second phase corresponds to pruning at the root. Hence, ComputePrunedStrategy($S$, $B$) is also an optimal solution to the above optimization problem.
    
    Lastly, note that it is obvious that both algorithms run in polynomial time since the number of iterations of their main loops is at most $|V(S)|$, implying (iii).
\end{proof}

\section{Missing Material from \Cref{sec:IO}}

\subsection{Continuation of proof of \Cref{lem:IOtreesurgery}}\label{app:continuesurgery}

\subsubsection*{Continuation of Case 1: $\qq[i] \geq \frac12$}
Recall that in Case 1, $\qq[i] \geq \frac12$ and thus $\qq[a_1] \leq \qq[i]$, $\Ex{\Acost{a_1}} = \qq[a_1]$ and $\Ex{\Acost{i}} = \qq[i]$.

\underline{Case 1.3} (\textit{Certificate-Continue}):
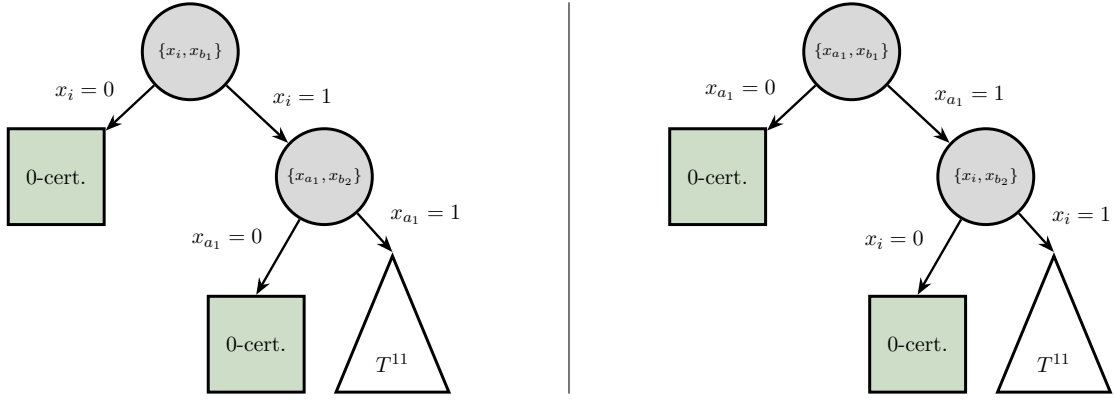
\begin{figure}
\begin{minipage}[b]{.48\textwidth}
    \centering
    \scalebox{0.75}{
        \begin{forest}
            for tree ={edge={very thick,-Stealth},s sep=0.5cm}
            [{$\CB{x_i,x_{b_1}}$},innernode,s sep = 3cm
                [{$0$-cert.},endnodes,child anchor=,edge label={node[midway,above left] {$x_i = 0\,\,$}}
                ]
                [{$\CB{x_{a_1},x_{b_2}}$},innernode, child anchor=,edge label={node[midway,above right] {$\,\,x_i = 1\,\,$}}
                    [{$0$-cert.},endnodes,edge label={node[midway,above left] {$\,\,x_{a_1}= 0\,\,$}}
                    ]
                    [{$T^{11}$},leaftree,edge label={node[midway,above right] {$\,\,x_{a_1}= 1\,\,$}}
                    ]
                ]
            ]
        \end{forest}
    }
\end{minipage}%
\hfill\vline\hfill
\begin{minipage}[b]{.48\textwidth}
    \centering
    \scalebox{0.75}{
        \begin{forest}
            for tree ={edge={very thick,-Stealth},s sep=0.5cm}
            [{$\CB{x_{a_1},x_{b_1}}$},innernode,s sep = 3cm
                [{$0$-cert.},endnodes,child anchor=,edge label={node[midway,above left] {$x_{a_1} = 0\,\,$}}
                ]
                [{$\CB{x_i,x_{b_2}}$},innernode, child anchor=,edge label={node[midway,above right] {$\,\,x_{a_1} = 1\,\,$}}
                    [{$0$-cert.},endnodes,edge label={node[midway,above left] {$\,\,x_{i}= 0\,\,$}}
                    ]
                    [{$T^{11}$},leaftree,edge label={node[midway,above right] {$\,\,x_{i}= 1\,\,$}}
                    ]
                ]
            ]
        \end{forest}
    }

\end{minipage}

\caption{Case 1.3 (\textit{Certificate-Continue}). $S_j$ on the left, $S'_j$ on the right.}
\label{tab:tree13}
\end{figure}

This is the case where $S_j$, after taking into account the outcome of $x_{b_1}$ tested in round 1, requires one more 0 for a 0-certificate.
Let $S'_j$ be the strategy obtained by swapping the position of elements $x_i$ and $x_{a_1}$, as shown in \Cref{tab:tree13}.
Since only one 0 from either $x_i$ or $x_{a_1}$ suffices to find a certificate, we terminate successfully except for the case where both results are $1$, where an arbitrary (possibly empty) substrategy $T^{11}$ is carried out.
The probability for these events is the same in $S_j$ and $S'_j$, however, some additional expected amortized cost is incurred in $S'_j$ because the test on $x_{b_2}$ is carried out with higher probability than before.
Let $T'^{\,1}$ be the substrategy of $S'_j$ for the case $x_{a_1} = 1$.
$T'^{\,1}$ is a pruning of pseudo-IO with shorter lists, so by the induction hypothesis, we can replace it with a distribution $R^1$ over prunings of IO on the subinstance without, in expectation, decreasing utility and with bounded additional (amortized) cost.
The strategy $R_j$ obtained by replacing $T'^{\,1}$ in $S'_j$ with $R^1$ is a distribution over prunings of IO because all elements are now in the appropriate order.
It easily follows from the induction hypothesis that $\Ex{\Utilk{\kk}{R_j}} \geq \Ex{\Utilk{\kk}{S_j}}$.
Expected amortized cost changes as follows:
\begin{align*}
    &\Ex{\Acost{R_j}} = \Ex{\Acost{a_1}} + \Acost{b_1} + \RB{1 - \qq[a_1]} \cdot \Ex{\Acost{R^1}}\\
    &\quad\leq \Ex{\Acost{a_1}} + \Acost{b_1} + \RB{1 - \qq[a_1]} \cdot \SB{1 + \Ex{\Acost{T'^{\;1}}}}\\
    &\quad=\Ex{\Acost{a_1}} + \Acost{b_1} + \RB{1-\qq[a_1]} \cdot \SB{\Ex{\Acost{i}} + \Acost{b_2} + \RB{1-\qq[i]} \cdot \Ex{\Acost{T^{11}}}}\\
    &\qquad+ \RB{1-\qq[a_1]}\cdot 1\\
    &\quad=\Ex{\Acost{i}} + \Acost{b_1} + \RB{1-\qq[i]} \cdot \SB{\Ex{\Acost{a_1}} + \Acost{b_2} + \RB{1-\qq[a_1]}\cdot \Ex{\Acost{T^{11}}}}\\
    &\qquad- \qq[a_1] \cdot \Ex{\Acost{i}} + \qq[i] \cdot \Ex{\Acost{a_1}}\\
    &\qquad+ \Acost{b_2} \cdot \SB{\RB{1-\qq[a_1]}-\RB{1-\qq[i]}} + \RB{1-\qq[a_1]}\cdot 1\\
    &\quad\leq \Ex{\Acost{S_j}} + \SB{1 + \qq[i] - 2\qq[a_1]}\\
    &\quad\leq \Ex{\Acost{S_j}} + 1
\end{align*}
In the first inequality, we apply the bound on expected amortized cost from the induction hypothesis.
In the third equality we rearrange to retrieve the expected amortized cost of $S_j$; in the second inequality, we use that $- \qq[a_1] \cdot \Ex{\Acost{i}} + \qq[i] \cdot \Ex{\Acost{a_1}} = 0$ and that $\Acost{b_2} \leq 1$.
For the last inequality, we use $\qq[i] \leq 1$ and $2\qq[a_1] \geq 1$.
For $\ell \in \CB{0,1}$, the arguments and estimates work in the same way. This concludes Case 1.3.

\underline{Case 1.4} (\textit{Abort-Continue}):
\begin{figure}
\begin{minipage}[b]{.48\textwidth}
    \centering
    \scalebox{0.75}{
        \begin{forest}
            for tree ={edge={very thick,-Stealth},s sep=0.5cm}
            [{$\CB{x_i,x_{b_1}}$},innernode,s sep = 3cm
                [{},endnodef, child anchor=,edge label={node[midway,above left] {$\,\,x_i = 0\,\,$}}
                ]
                [{$\CB{x_{a_1},x_{b_2}}$},innernode,edge label={node[midway,above right] {$x_i = 1\,\,$}}
                    [{$\,\,T^{10}\,\,$},edge label={node[midway,above left] {$x_{a_1} = 0$}},leaftree
                    ]
                    [{$\,\,T^{11}\,\,$},edge label={node[midway,above right] {$x_{a_1}=1$}},leaftree
                    ]
                ]
            ]
        \end{forest}
    }
\end{minipage}%
\hfill\vline\hfill
\begin{minipage}[b]{.48\textwidth}
    \centering
    \scalebox{0.75}{
        \begin{forest}
            for tree ={edge={very thick,-Stealth},s sep=.5cm}
            [{$\CB{x_i,x_{b_1}}$},innernode,s sep=3cm
                [{$1-\alpha$},stopnode,edge label={node[midway,above left] {$\,\,x_{i} = 0\,\,$}}
                    [{$\CB{x_{a_1},x_{b_2}}$},innernode,edge=dotted,tier=contb
                        [{$\,\,T'^{\,01}\,\,$},edge label={node[midway,above left] {$x_{a_1} = 0$}},leaftree
                        ]
                        [{$\,\,T'^{\,11}\,\,$},edge label={node[midway,above right] {$x_{a_1}=1$}},leaftree
                        ]
                    ]
                ]
                [,nonode,edge={-},edge label={node[midway,above right] {$x_{i} = 1\,\,$}}
                    [{},endnodef, tier=contb
                    ]
                ]
            ]
        \end{forest}
    }

\end{minipage}

\caption{Case 1.4 (\textit{Abort-Continue}). $S_j$ on the left; $S'_j$ on the right, defined in the text as distribution over two deterministic strategies ($S''_j$ and $\tilde{S}''_j$) and can be represented as shown above: If $x_i = 0$ in the first round, $S'_j$ stops with probability $1-\alpha$ for $\alpha := \frac{1-\qq[i]}{\qq[i]}$ or, with the remaining probability, continues the way that $S_j$ continues after $x_i = 1$ (possibly finding a certificate earlier).}
\label{tab:tree14}
\end{figure}
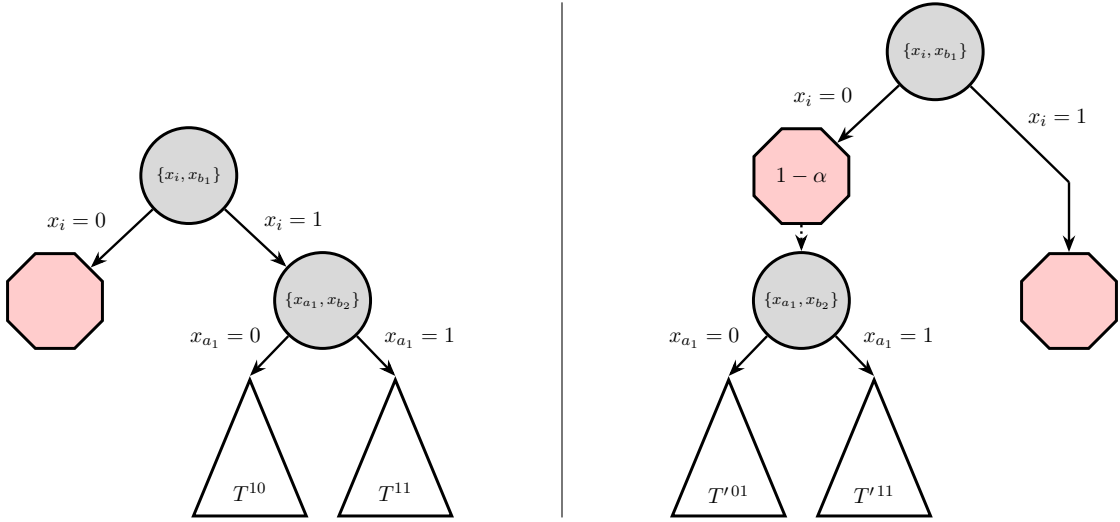

In this case, $S_j$ stops without a certificate after the first 0 and continues otherwise (as shown in \Cref{tab:tree14}).
Stopping without a certificate after observing a 0 from the first element is intuitively not the best we can do. Indeed, $S_j$ is (weakly) dominated by the following strategy, where $T^1$ is the substrategy of $S_j$ for $x_i = 1$:

With probability $\alpha=\frac{1-\qq[i]}{\qq[i]} \leq 1$, do the following strategy called $S''_j$:
Perform the first testing round. If $x_i = 1$, stop (without a certificate). If $x_i =0$, carry on with round $2$ (tests on $x_{a_1}$ and $x_{b_2}$).
Depending on the outcome of $x_{a_1}$, continue with $T'^{\,10}$ if $x_{a_1} = 0$ and $T'^{\,11}$ otherwise.
Substrategies $T'^{\,11}$ and $T'^{\,10}$ are derived from $T^{11}$ and $T^{01}$, modified to stop after the round in which the certificate has been found.
This modification is necessary because by our definition, the Round-Completion property requires the strategy to stop after the round in which a certificate was found and $T'^{\,11}$ and $T'^{\,10}$ start with an additional 0 from testing $x_i$ compared to $T^{11}$ and $T^{10}$.
The transformation weakly increases utility and weakly decreases expected amortized cost as possibly more certificates are reached at possibly less amortized cost. The resulting substrategy for $x_i = 0$ is denoted by $T'^{\,0}$, and is a pruning of IO on the subinstance. This makes $S''_j$ a pruning of $i$-pseudo-IO.

With the remaining probability $1-\alpha$, we perform tests on $x_i$ and $x_{b_1}$ and stop without a certificate.
We denote this strategy by $\tilde{S}''_j$.
It is also a pruning of $i$-pseudo-IO.

The overall resulting randomized strategy is denoted by $S'_j$, itself a randomized pruning of $\mathcal{I}_i$.
Observe that $T'^{\,0}$ is carried out with probability $\alpha \cdot \qq[i] = 1-\qq[i]$ in $S'_j$, which is the same probability that $T^1$ is carried out in $S_j$.
Expected amortized cost for $x_i$ and $x_{b_1}$ do not change in $S'_j$ compared to $S_j$, yielding $\Ex{\Acost{S'_j}} \leq \Ex{\Acost{S_j}}$.

Both prunings of $\mathcal{I}_i$, $S''_j$ matches Case 1.2 and $\tilde{S}''_j$ matches Case 1.5 (subsequently defined and analyzed). Therefore, this case reduces to those cases, using the replacements proposed therein. 
For $\ell \in \CB{0,1}$, the arguments and estimates work in the same way.

Note that the expected amortized cost of not testing any elements instead of $\tilde{S}''_j$ would have been lower at the same utility.
However, by using $\tilde{S}''_j$ we avoid a stopping decision possibly dependent on the outcome of $x_{b_1}$ without having selected that test yet.
The replacements from Cases 1.2 and 1.5 are randomized RC-prunings of IO that do not terminate before the end of the first round, so they can be used for a well-defined overall strategy (cf. argument on obtaining an overall strategy for all possible outcomes of $x_{b_1}$ and $x_{b_2}$ after the case analysis in the main part of the paper).

\underline{Case 1.5} (\textit{Abort-Abort}):
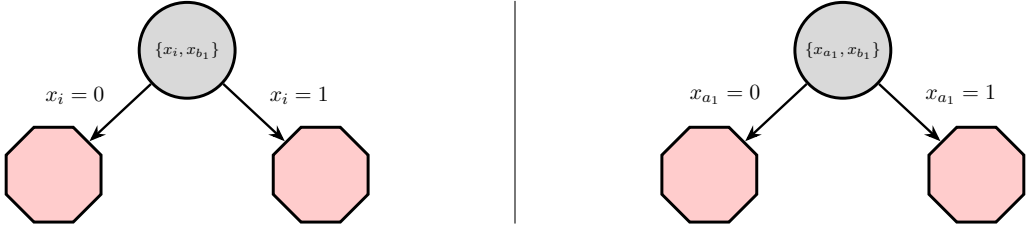
\begin{figure}
\begin{minipage}[b]{.48\textwidth}
    \centering
    \scalebox{0.75}{
        \begin{forest}
            for tree ={edge={very thick,-Stealth},s sep=0.5cm}
            [{$\CB{x_i,x_{b_1}}$},innernode,s sep = 3cm
                [{},endnodef, child anchor=,edge label={node[midway,above left] {$\,\,x_i = 0\,\,$}}
                ]
                [{},endnodef, child anchor=,edge label={node[midway,above right] {$\,\,x_i = 1\,\,$}}
                ]
            ]
        \end{forest}
    }
\end{minipage}%
\hfill\vline\hfill
\begin{minipage}[b]{.48\textwidth}
    \centering
    \scalebox{0.75}{
        \begin{forest}
            for tree ={edge={very thick,-Stealth},s sep=0.5cm}
            [{$\CB{x_{a_1},x_{b_1}}$},innernode,s sep = 3cm
                [{},endnodef, child anchor=,edge label={node[midway,above left] {$\,\,x_{a_1} = 0\,\,$}}
                ]
                [{},endnodef, child anchor=,edge label={node[midway,above right] {$\,\,x_{a_1} = 1\,\,$}}
                ]
            ]
        \end{forest}
    }

\end{minipage}

\caption{Case 1.5 (\textit{Abort-Abort}). $S_j$ on the left, $S'_j$ on the right.}
\label{tab:tree15}
\end{figure}

This is the case where $S_j$ stops without a 0-certificate for both realizations of $x_i$, as depicted in \Cref{tab:tree15}.
We obtain $S'_j$ by replacing the test on $x_i$ with that on $x_{a_1}$.
The resulting strategy is a pruning of IO.
The utility is $0$ in both cases.
The expected amortized cost only decreases: $\Ex{\Acost{S'_j}} = \Ex{\Acost{a_1}} + \Acost{b_1} = \qq[a_1] + \Acost{b_1} \leq \qq[i] + \Acost{b_1} = \Ex{\Acost{i}} + \Acost{b_1} = \Ex{\Acost{S_j}}$.
For $\ell \in \CB{0,1}$, the arguments and estimates work in the same way.
Therefore, selecting $R_j = S'_j$, the induction hypothesis also holds in this case.

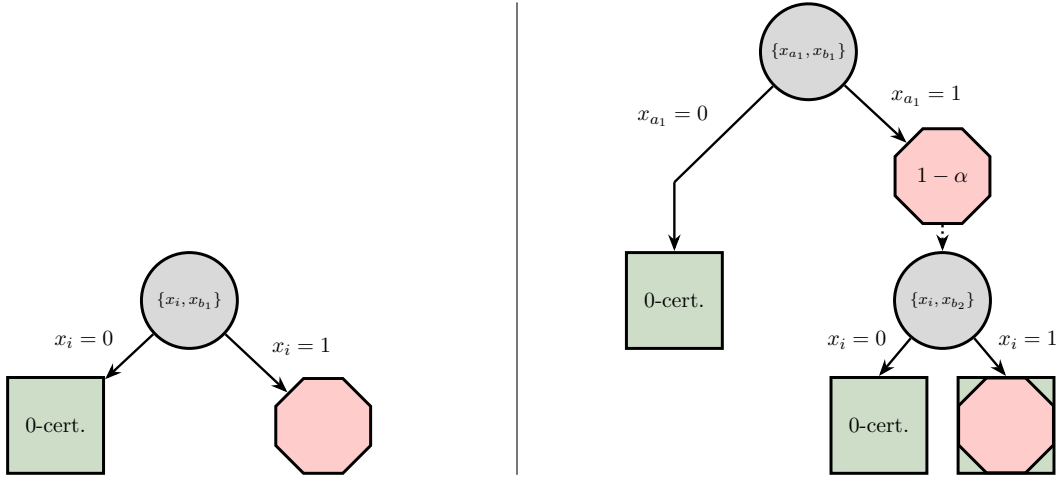
\begin{figure}[h]
\begin{minipage}[b]{.48\textwidth}
    \centering
    \scalebox{0.75}{
        \begin{forest}
            for tree ={edge={very thick,-Stealth},s sep=0.5cm}
            [{$\CB{x_i,x_{b_1}}$},innernode,s sep = 3cm
                [{$0$-cert.},endnodes,child anchor=,edge label={node[midway,above left] {$\,\,x_i = 0\,\,$}}
                ]
                [{},endnodef, child anchor=,edge label={node[midway,above right] {$\,\,x_i = 1\,\,$}}
                ]
            ]
        \end{forest}
    }
\end{minipage}%
\hfill\vline\hfill
\begin{minipage}[b]{.48\textwidth}
    \centering
    \scalebox{0.75}{
        \begin{forest}
            for tree ={edge={very thick,-Stealth},s sep=.5cm}
            [{$\CB{x_{a_1},x_{b_1}}$},innernode,s sep=3cm
                [,nonode,edge={-},edge label={node[midway,above left] {$x_{a_1} = 0\,\,$}}
                    [{$0$-cert.},endnodes, tier=contb
                    ]
                ]            
                [{$1-\alpha$},stopnode,edge label={node[midway,above right] {$\,\,x_{a_1} = 1\,\,$}}
                    [{$\CB{x_i,x_{b_2}}$},innernode,edge=dotted,tier=contb
                        [{$0$-cert.},endnodes,edge label={node[midway,above left] {$x_{i}=0$}}
                        ]
                        [{a},name=special,edge label={node[midway,above right] {$x_{i}=1$}},endnodes,
                        ]
                    ]
                ]
            ]
            \node at (special) [stopnodebase] {};
        \end{forest}
    }
\end{minipage}

\caption{Case 1.6 (\textit{Certificate-Abort}). $S_j$ on the left; $S'_j$ on the right, defined in the text as a distribution over two deterministic strategies. It can be represented as in this diagram: If $x_{a_1} = 1$ in the first round, with probability $1-\alpha$ for $\alpha := \frac{\qq[i] - \qq[a_1]}{\qq[i] \cdot \RB{1-\qq[a_1]}}$, execution stops without a certificate; with the remaining probability a second testing round ($x_i$ and $x_{b_2}$) is conducted.
The bottom right node denotes that the outcome depends on the (fixed) value $x_{b_2}$: if $x_{b_2} = 0$, we have a certificate (square), otherwise not (octagon).}
\label{tab:tree16}
\end{figure}

\underline{Case 1.6} (\textit{Certificate-Abort}):
This is the case where $S_j$, after taking into account the outcome of $x_{b_1}$, requires one more 0 for a 0-certificate; if $x_i = 1$, it stops without the certificate(as depicted in \Cref{tab:tree16}).

The first transformation step is again to replace the test on $x_i$ in the first round by $x_{a_1}$; denote the resulting strategy with $\tilde{S}''_j$.
However, as in Case 1.2, this reduces the probability of finding a 0-certificate in the first round.
To compensate, we only run $\tilde{S}''_j$ with probability $1-\alpha$ for $\alpha := \frac{\qq[i] - \qq[a_1]}{\qq[i] \cdot \RB{1-\qq[a_1]}} \leq 1$.
With the remaining probability, $\alpha$, we run the following strategy $S''_j$:
Also start with $x_{a_1}$ and $x_{b_1}$ in the first round.
However, as opposed to $\tilde{S}''_j$, in the case $x_{a_1} = 1$, instead of stopping, continue with a second testing round ($x_i$ and $x_{b_2}$).
As only a single 0 is missing, if $x_i = 0$, $S''_j$ has found a certificate. If $x_i = 1$, it stops (with the result depending on the test on $x_{b_2}$: if $x_{b_2} = 0$, it stops with a certificate; otherwise, without).

Let $S'_j$ be the resulting randomized strategy, based on $S''_j$ and $\tilde{S}''_j$.
Observe that $S'_j$ is already a randomized pruning of IO: Since at most two testing rounds are conducted, the correct order of elements is already guaranteed, without the need for further replacements based on invoking the induction hypothesis.

It remains to bound expected utility and expected amortized cost.
The bounds are shown for the case $x_{b_2} = 1$; since the realization of $x_{b_2}$ has no effect on $S_j$, the results imply the same bounds for $x_{b_2} = 0$ (higher utility with same bound on the cost).
\begin{align*}
    &\Ex{\Utilk{\kk}{S'_j}} \geq \qq[a_1] + \alpha \cdot \RB{1-\qq[a_1]}\cdot \qq[i]\\
    &\quad = \qq[a_1] + \qq[i] - \qq[a_1] \\
    &\quad= \Ex{\Utilk{\kk}{S_j}}
\end{align*}
The first inequality uses the construction of $S'_j$ and assumes $x_{b_2} = 1$ (lowest possible utility).
\begin{align*}
    &\Ex{\Acost{S'_j}} = \Ex{\Acost{a_1}} + \Acost{b_1} + \alpha \cdot \RB{1-\qq[a_1]} \cdot \RB{\Ex{\Acost{i}} + \Acost{b_2}}\\
    &\quad= \Ex{\Acost{i}} + \Acost{b_1} + \Ex{\Acost{a_1}} - \RB{1-\alpha\cdot \RB{1-\qq[a_1]}}\cdot\Ex{\Acost{i}} \\
    &\qquad+ \alpha \cdot \RB{1-\qq[a_1]} \cdot \Acost{b_2}\\
    &\quad\leq \Ex{\Acost{S_j}} + \SB{\qq[a_1] - \qq[i] + \alpha \cdot \RB{1-\qq[a_1]} \cdot \qq[i]} + \SB{\alpha \cdot \RB{1-\qq[a_1]}}\\
    &\quad\leq \Ex{\Acost{S_j}} + 0 + \SB{1-\frac{\qq[a_1]}{\qq[i]}}\\
    &\quad\leq \Ex{\Acost{S_j}} + 1
\end{align*}

In the first inequality, we use that $\Ex{\Acost{i}} = \qq[i]$, $\Ex{\Acost{a_1}} = \qq[a_1]$ and $\Acost{b_2} \leq 1$.
In the second inequality, we apply the definition of $\alpha$.
As argued, this holds for any realization of $x_{b_1}$ and $x_{b_2}$, and also in case those elements do not exist (i.e., if $\ell \in \CB{0,1}$).
By setting $R_j = S'_j$ we thus conclude that the induction hypothesis holds for Case 1.6, too.

\underline{Case 1.7} (\textit{Certificate-Certificate}):
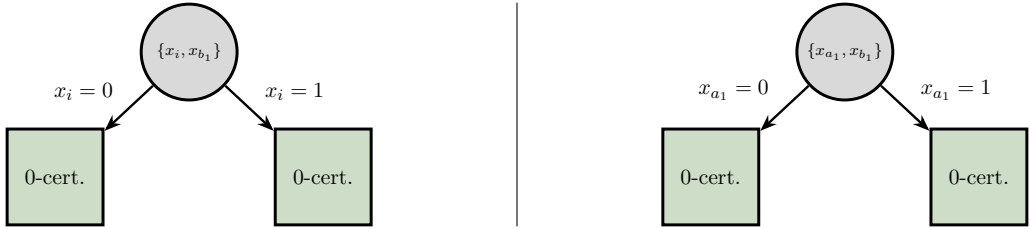
\begin{figure}
\begin{minipage}[b]{.48\textwidth}
    \centering
    \scalebox{0.75}{
        \begin{forest}
            for tree ={edge={very thick,-Stealth},s sep=0.5cm}
            [{$\CB{x_i,x_{b_1}}$},innernode,s sep = 3cm
                [{$0$-cert.},endnodes, child anchor=,edge label={node[midway,above left] {$\,\,x_i = 0\,\,$}}
                ]
                [{$0$-cert.},endnodes, child anchor=,edge label={node[midway,above right] {$\,\,x_i = 1\,\,$}}
                ]
            ]
        \end{forest}
    }
\end{minipage}%
\hfill\vline\hfill
\begin{minipage}[b]{.48\textwidth}
    \centering
    \scalebox{0.75}{
        \begin{forest}
            for tree ={edge={very thick,-Stealth},s sep=0.5cm}
            [{$\CB{x_{a_1},x_{b_1}}$},innernode,s sep = 3cm
                [{$0$-cert.},endnodes, child anchor=,edge label={node[midway,above left] {$\,\,x_{a_1} = 0\,\,$}}
                ]
                [{$0$-cert.},endnodes, child anchor=,edge label={node[midway,above right] {$\,\,x_{a_1} = 1\,\,$}}
                ]
            ]
        \end{forest}
    }

\end{minipage}

\caption{Case 1.7 (\textit{Certificate-Certificate}). $S_j$ on the left, $S'_j$ on the right.}
\label{tab:tree17}
\end{figure}

As in the case analysis we are only considering instances with $\kk \geq 1$, this case can only occur if $x_{b_1} = 0$ (and thus $\ell \geq 1$) and $\kk=1$. 
$S_j$ is not a pruning of IO, so we replace it with the strategy $S'_j$ that starts with tests on $x_{a_1}$ and $x_{b_1}$ in the first round, and will also find a certificate for both outcomes of $x_{a_1}$ (see also \Cref{tab:tree17}). Note that $S'_j$ is a pruning of IO.
The utility is $1$ in both cases.
Analogously to Case 1.5, it is easy to see that 
$\Ex{\Acost{S'_j}} \leq \Ex{\Acost{S_j}}$.
The induction hypothesis holds when setting $R_j = S'_j$. Again, the bounds also work for $\ell = 1$ (as argued above, Case 1.7 cannot occur if $\ell =0$).

This completes the analysis of Case 1.

\subsubsection*{Case 2: $\qq[i] < \frac12$}
In this case, we assume $\qq[i] < \frac12$.
This implies $\qq[a_1] \geq \qq[i]$, and further, $\Ex{\Acost{a_1}} = \RB{1-\qq[a_1]}$ and $\Ex{\Acost{i}} = \RB{1-\qq[i]}$ by the definition of $\acost$. 
Arguing in analogy to Case 1, overall 7 subcases need to be considered. These are analyzed in the following.

\underline{Case 2.1} (\textit{Continue-Continue}):
This is again the case in which the first two rounds are always completed.
We apply the same transformations and the same bounds as in Case 1.1, which do not rely on using properties of the value ranges of the probabilities.
Therefore, the calculations carry over directly.

\underline{Case 2.2} (\textit{Continue-Abort}):
\begin{figure}
\begin{minipage}[b]{.48\textwidth}
    \centering
    \scalebox{0.75}{
        \begin{forest}
            for tree ={edge={very thick,-Stealth},s sep=0.5cm}
            [{$\CB{x_i,x_{b_1}}$},innernode,s sep = 3cm
                [{$\CB{x_{a_1},x_{b_2}}$},innernode,edge label={node[midway,above left] {$x_i = 0\,\,$}}
                    [{$\,\,T^{00}\,\,$},edge label={node[midway,above left] {$x_{a_1} = 0$}},leaftree
                    ]
                    [{$\,\,T^{01}\,\,$},edge label={node[midway,above right] {$x_{a_1}=1$}},leaftree
                    ]
                ]
                [{},endnodef, child anchor=,edge label={node[midway,above right] {$\,\,x_i = 1\,\,$}}
                ]
            ]
        \end{forest}
    }
\end{minipage}%
\hfill\vline\hfill
\begin{minipage}[b]{.48\textwidth}
    \centering
    \scalebox{0.75}{
        \begin{forest}
            for tree ={edge={very thick,-Stealth},s sep=.5cm}
            [{$\CB{x_{a_1},x_{b_1}}$},innernode,s sep = 3cm
                [{$\CB{x_i,x_{b_2}}$},innernode,tier=contb,edge label={node[midway,above left] {$\,\,x_{a_1} = 0\,\,$}}
                    [,nonode,tier=stopa,edge={-},edge label={node[midway,above left] {$x_{i} = 0\,\,$}}
                        [{$T^{00}$},leaftree, tier=bot]
                    ] 
                    [{$1-\alpha$},stopnode,edge label={node[midway,above right] {$\,\,x_{i} = 1\,\,$}}
                        [{$T^{01}$},leaftree,tier=bot,draw,edge=dotted
                        ]
                    ]
                ]
                [{},endnodef, child anchor=,edge label={node[midway,above right] {$\,\,x_{a_1} = 1\,\,$}}
                ]
            ]
        \end{forest}
    }

\end{minipage}

\caption{Case 2.2 (\textit{Continue-Abort}). $S_j$ on the left; $S'_j$ on the right, can defined as a randomized strategy that stops early with probability $1-\alpha$ for $\alpha := \frac{\qq[i] \cdot \RB{1-\qq[a_1]}}{\qq[a_1] \cdot \RB{1-\qq[i]}}$ if $x_{a_1}=0$ and $x_i = 1$.}
\label{tab:tree22}
\end{figure}
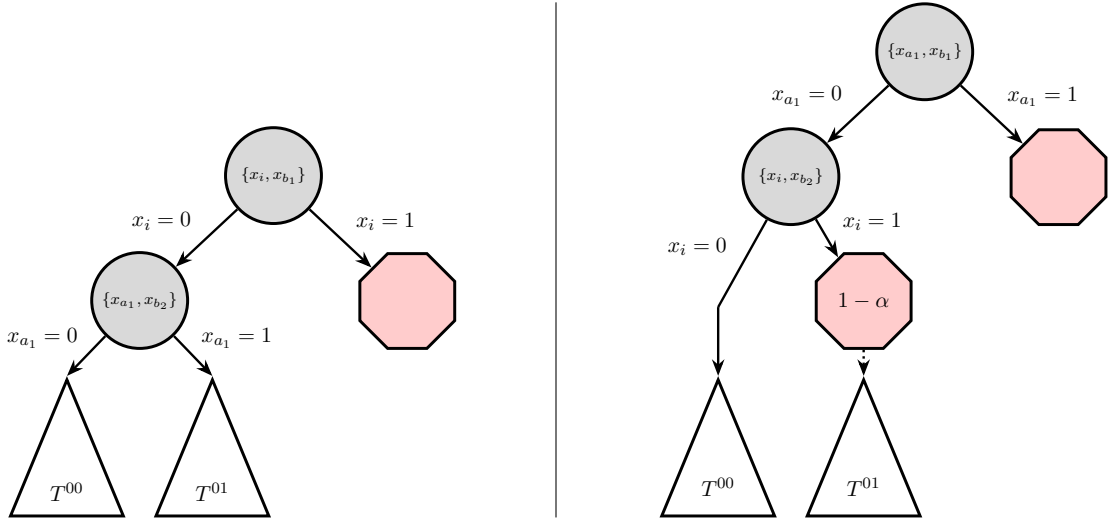

This is the case where $S_j$ stops without a certificate after the first round if $x_i = 1$ and continues otherwise, as depicted in \Cref{tab:tree22}.
We transform $S_j$ into $S'_j$, which will be a randomization over two strategies, $S''_j$ and $\tilde{S}''_j$; both start from $S_j$ by swapping $x_i$ and $x_{a_1}$, and if $x_{a_1} = 1$, directly stop without a certificate after the first round, as in the original strategy. 

If $x_{a_1} =0$, then $S''_j$, chosen with probability $\alpha := \frac{\qq[i] \cdot \RB{1-\qq[a_1]}}{\qq[a_1] \cdot \RB{1-\qq[i]}} \leq 1$, continues with the following substrategy, denoted by $T'^{\,0}$: Continue with the tests on $x_i$ and $x_{b_2}$; if $x_i = 0$, continue with $T^{00}$, otherwise with $T^{01}$.
With the remaining probability $1-\alpha$, strategy $\tilde{S}''_j$ continues with $\tilde{T}'^{\,0}$, defined as follows: Also test elements $x_i$ and $x_{b_2}$; however, after that, only continue if $x_i = 0$ (with $T^{00}$) and stop evaluation without a certificate otherwise.

Let $S'_j$ be the resulting randomized strategy based on $S''_j$ and $\tilde{S}''_j$.
Note that it is ensured that $T^{00}$ and $T^{01}$ are executed on the same respective subinstances in $S_j$ and $S'_j$; by choice of $\alpha$, $T^{00}$ and $T^{01}$ are reached with the same overall probability before and after the transformation to preserve utility.

By construction of $S''_j$ and $\tilde{S}''_j$, $T'^{\,0}$ and $\tilde{T}'^{\,0}$ are randomized RC-prunings of $i$-pseudo-IO on the subinstance obtained after the first round for $x_{a_1} = 0$.
By applying the induction hypothesis, we can replace both with randomized prunings $R^0$ and $\tilde{R}^0$ of IO (on the subinstance), respectively, to obtain the resulting randomized overall strategy $R_j$.
In $R_j$, all elements are in the desired order and it is a randomized pruning of IO.
In the following, we bound utility and amortized cost.

\begin{align*}
        &\Ex{\Utilk{\kk}{R_j}} = \alpha \cdot \qq[a_1] \cdot \Ex{\Utilk{\kk_1-1}{R^0}} + \RB{1-\alpha} \cdot \qq[a_1] \cdot \Ex{\Utilk{\kk_1-1}{\tilde{R}^0}}\\
        &\quad\geq \qq[a_1] \cdot \SB{\alpha \cdot \Ex{\Utilk{\kk_1-1}{T'^{\,0}}} + \RB{1-\alpha}\cdot \Ex{\Utilk{\kk_1-1}{\tilde{T}'^{\,0}}}}\\
        &\quad= \qq[a_1] \cdot \SB{\qq[i] \cdot \Ex{\Utilk{\kk_2-2}{T^{00}}} + \alpha \cdot \RB{1-\qq[i]} \cdot \Ex{\Utilk{\kk_2-1}{T^{01}}}}\\
        &\quad= \qq[i] \cdot \qq[a_1] \cdot \Ex{\Utilk{\kk_2-2}{T^{00}}} + \qq[i] \cdot \RB{1-\qq[a_1]} \cdot \Ex{\Utilk{\kk_2-1}{T^{01}}}\\
        &\quad= \Ex{\Utilk{\kk}{S_j}}\, .
\end{align*}

The first equality uses the definition of $R_j$, $R^0$ and $\tilde{R}^0$; in the first inequality, we rearrange and apply the induction hypothesis (and the corresponding replacements).
The second equality uses the construction of $T'^{\,0}$ and $\tilde{T}'^{\,0}$ (conditioned on $x_i = 0$, $T^{00}$ is carried out independently of the internal randomization).
The third equality applies the definition of $\alpha$.

\begin{align*}
    &\Ex{\Acost{R_j}} = \Ex{\Acost{a_1}} + \Acost{b_1} + \alpha \cdot \qq[a_1] \cdot \Ex{\Acost{R^0}} + \RB{1-\alpha} \cdot \qq[a_1] \cdot \Ex{\Acost{\tilde{R}^0}}\\
        &\quad\leq \Ex{\Acost{a_1}} + \Acost{b_1} + \alpha \cdot \qq[a_1] \cdot \SB{\Ex{\Acost{T'^{\,0}}} + 1} + \RB{1-\alpha} \cdot \qq[a_1] \cdot \SB{\Ex{\Acost{\tilde{T}'^{\,0}}} + 1}\\
        &\quad= \Ex{\Acost{a_1}} + \Acost{b_1} + \qq[a_1] \cdot 1\\
        &\qquad+ \qq[a_1] \cdot \SB{\Ex{\Acost{i}} + \Acost{b_2} + \qq[i] \cdot \Ex{\Acost{T^{00}}} + \alpha \cdot \RB{1-\qq[i]} \cdot \Ex{\Acost{T^{01}}}} \\
        &\quad= \Ex{\Acost{i}} + \Acost{b_1}\\
        &\qquad+ \qq[i] \cdot \SB{\Ex{\Acost{a_1}} + \Acost{b_2} + \qq[a_1] \cdot \Ex{\Acost{T^{00}}}} + \qq[a_1] \cdot \RB{1-\qq[i]} \cdot \alpha \cdot \Ex{\Acost{T^{01}}} \\
        &\qquad+ \SB{\RB{1-\qq[i]} \cdot \Ex{\Acost{a_1}} - \RB{1-\qq[a_1]} \cdot \Ex{\Acost{i}}}+ \RB{\qq[a_1] - \qq[i]} \cdot \Acost{b_2} + \qq[a_1]\\
        &\quad\leq \Ex{\Acost{i}} + \Acost{b_1}\\
        &\qquad+ \qq[i] \cdot \SB{\Ex{\Acost{a_1}} + \Acost{b_2} + \qq[a_1] \cdot \Ex{\Acost{T^{00}}} + \RB{1-\qq[a_1]} \cdot \Ex{\Acost{T^{01}}}}\\
        &\qquad+ 0 + 2 \qq[a_1] - \qq[i]\\
        &\quad\leq \Ex{\Acost{S_j}} + 1\, .
\end{align*}
The first equality applies the construction of $R_j$; note that for $x_{a_1}=1$, no additional cost is incurred after the first round.
The first inequality uses the induction hypothesis.
In the second equality, we use the definition of $T'^{\,0}$ and $\tilde{T}'^{\,0}$ ($T^{00}$ is run by both), and re-arrange the terms in the third.
The second inequality uses the definition of $\alpha$, the fact that $\Ex{\Acost{a_1}} = \RB{1-\qq[a_1]}$ and $\Ex{\Acost{i}} = \RB{1-\qq[i]}$ within the scope of Case 2, and the bound $\Acost{b_2} \leq 1$.
In the last step, we retrieve the expected amortized cost of $S_j$ and the induction hypothesis follows.
As before, the results carry over for the case $\ell \in \CB{0,1}$.

\underline{Case 2.3} (\textit{Certificate-Continue}):
This is again the case where a single 0 is needed (after taking into account the outcome of $x_{b_1}$ tested in round 1)---compare Case 1.3. Let $S'_j$ be the strategy obtained by swapping the position of elements $x_i$ and $x_{a_1}$, see \Cref{tab:tree13}.
Since only one 0 from either $x_i$ or $x_{a_1}$ suffices to find a certificate, we terminate successfully except for the case where both results are $1$, where some substrategy $T^{11}$ is carried out.
The probability for these events is the same in $S_j$ and $S'_j$, and, as opposed to Case 1.3, the expected amortized cost of the test on $b_2$ decreases.
Let $T'^{\,1}$ be the substrategy of $S'_j$ for the case $a_1 = 1$.
$T'^{\,1}$ is an RC-pruning of $\mathcal{I}_i$, so we can replace it with a randomized pruning $R^1$ of IO on the subinstance without decreasing expected utility and with bounded additional expected (amortized) cost.
The strategy $R_j$ obtained by replacing $T'^{\,1}$ with $R^1$ is a randomized pruning of IO because all elements are now in the appropriate order.
Overall, we have $\Ex{\Utilk{\kk}{R_j}} \geq \Ex{\Utilk{\kk}{S_j}}$ with the same argument as in Case 1.3 (the value ranges of probabilities and amortized cost are not used).
Regarding expected amortized cost, we observe:
\begin{align*}
    &\Ex{\Acost{R_j}} = \Ex{\Acost{a_1}} + \Acost{b_1} + \RB{1 - \qq[a_1]} \cdot \Ex{\Acost{R^1}}\\
    &\quad\leq \Ex{\Acost{a_1}} + \Acost{b_1} + \RB{1 - \qq[a_1]} \cdot \SB{\Ex{\Acost{T'^{\,1}}} + 1}\\
    &\quad=\Ex{\Acost{a_1}} + \Acost{b_1} + \RB{1-\qq[a_1]} \cdot \SB{\Ex{\Acost{i}} + \Acost{b_2} + \RB{1-\qq[i]} \cdot \Ex{\Acost{T^{11}}}}\\
    &\qquad+ \RB{1-\qq[a_1]}\cdot 1\\
    &\quad=\Ex{\Acost{i}} + \Acost{b_1} + \RB{1-\qq[i]} \cdot \SB{\Ex{\Acost{a_1}} + \Acost{b_2} + \RB{1-\qq[a_1]}\cdot \Ex{\Acost{T^{11}}}}\\
    &\qquad- \qq[a_1] \cdot \Ex{\Acost{i}} + \qq[i] \cdot \Ex{\Acost{a_1}}\\
    &\qquad+ \Acost{b_2} \cdot \SB{\RB{1-\qq[a_1]}-\RB{1-\qq[i]}} + \RB{1-\qq[a_1]}\cdot 1\\
    &\quad \leq \Ex{\Acost{S_j}} +  \RB{1+\Acost{b_2}} \cdot \RB{\qq[i] - \qq[a_1]} + \RB{1-\qq[a_1]} \cdot 1\\
    &\quad\leq \Ex{\Acost{S_j}} + 1\, .
\end{align*}
In the first inequality, we apply the bound on expected amortized cost from the induction hypothesis.
In the third equality we rearrange to isolate the expected amortized cost of $S_j$; in the second inequality, we use that $\Ex{\Acost{i}}  = 1-\qq[i]$ and $\Ex{\Acost{a_1}} = 1-\qq[a_1]$.
The last inequality holds because $\qq[i] - \qq[a_1] \leq 0$.
For $\ell \in \CB{0,1}$, the arguments and estimates work in the same way.

\underline{Case 2.4} (\textit{Abort-Continue}):
\begin{figure}
\begin{minipage}[b]{.48\textwidth}
    \centering
    \scalebox{0.75}{
        \begin{forest}
            for tree ={edge={very thick,-Stealth},s sep=0.5cm}
            [{$\CB{i,b_1}$},innernode,s sep = 3cm
                [{},endnodef, child anchor=,edge label={node[midway,above left] {$\,\,x_i = 0\,\,$}}
                ]
                [{$\CB{a_1,b_2}$},innernode,edge label={node[midway,above right] {$x_i = 1\,\,$}}
                    [{$\,\,T^{10}\,\,$},edge label={node[midway,above left] {$x_{a_1} = 0$}},leaftree
                    ]
                    [{$\,\,T^{11}\,\,$},edge label={node[midway,above right] {$x_{a_1}=1$}},leaftree
                    ]
                ]
            ]
        \end{forest}
    }
\end{minipage}%
\hfill\vline\hfill
\begin{minipage}[b]{.48\textwidth}
    \centering
    \scalebox{0.75}{
        \begin{forest}
            for tree ={edge={very thick,-Stealth},s sep=.5cm}
            [{$\CB{i,b_1}$},innernode,
                [,nonode,edge={-},edge label={node[midway,above left] {$x_{i} = 1\,\,$}}
                    [{$\CB{a_1,b_2}$},innernode,tier=contb
                        [{$\,\,T'^{\,10}\,\,$},edge label={node[midway,above left] {$x_{a_1} = 0$}},leaftree
                        ]
                        [{$\,\,T'^{\,11}\,\,$},edge label={node[midway,above right] {$x_{a_1}=1$}},leaftree
                        ]
                    ]
                ]            
                [{$1-\alpha$},stopnode,edge label={node[midway,above right] {$\,\,x_{i} = 0\,\,$}}
                    [{$\CB{a_1,b_2}$},innernode,edge=dotted,tier=contb
                        [{$\,\,T^{10}\,\,$},edge label={node[midway,above left] {$x_{a_1} = 0$}},leaftree
                        ]
                        [{$\,\,T^{11}\,\,$},edge label={node[midway,above right] {$x_{a_1}=1$}},leaftree
                        ]
                    ]
                ]
            ]
        \end{forest}
    }

\end{minipage}

\caption{Case 2.4 (\textit{Abort-Continue}). $S_j$ on the left; $S'_j$ on the right, defined in the text as random selection over two deterministic strategies ($S''_j$ and $\tilde{S}''_j$) and can be represented as shown above: If $x_i = 0$, $S'_j$ continues the way that $S_j$ continues after $x_i = 1$ (possibly finding a certificate earlier). If $x_i=1$, it continues the same way as $S_j$ with probability $\alpha := \frac{1-2\qq[i]}{1-\qq[i]}$ or stops without a certificate.}
\label{tab:tree24}
\end{figure}
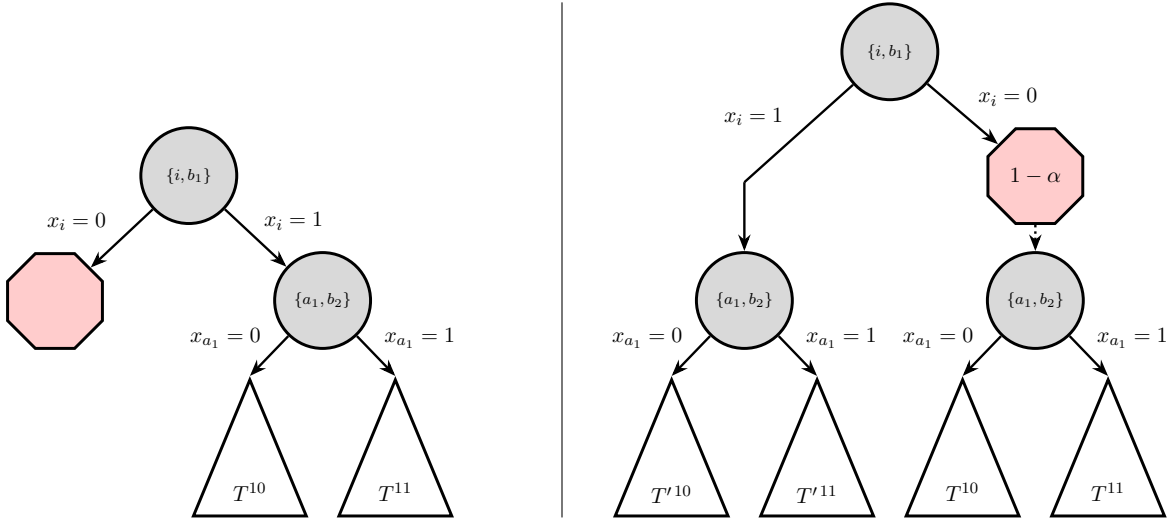

As in Case 1.4, we argue that $S_j$ can be replaced by a distribution over prunings of $\mathcal{I}_i$ that have already been considered and thus reduces to those cases.
Again, the overall idea is to continue evaluating after $x_i = 0$ instead of $x_i = 1$ at, in expectation, lower amortized cost and higher utility.
However, since $\RB{1-\qq[i]} \geq \qq[i]$, we need to continue evaluating after $x_i = 1$ at least with some probability in order to sustain utility.
Let $T^1$ be the substrategy of $S_j$ for $x_i = 1$ with substrategies $T^{11}$ and $T^{10}$.

$S'_j$ (see also \Cref{tab:tree24}) is constructed as follows:
With probability $\alpha := \frac{1-2\qq[i]}{1-\qq[i]} \leq 1$, do the following, denoted by $S''_j$:
Perform round 1 with tests on $x_i$ and $x_{b_1}$, as before.
If $x_i = 0$, continue with $T'^{\,0}$ obtained by modifying $T^1$ in the following way:
Terminate as soon as, at the end of a round, $\kk_1-1$ variables with value $0$ have been found, instead of $\kk_1$ (recall that now we have the additional 0 from $x_i$ available and need one 0 fewer for a certificate).
Let $T'^{\,10}$ and $T'^{\,11}$ be the substrategies that $T'^{\,0}$ uses after the second round for $x_{a_1} = 0$ and $x_{a_1} = 1$, respectively.
This way, $T'^{\,0}$ fulfills the Round-Completion property.
Also, a certificate is found on every realization for which $T^1$ terminates with a certificate, yielding higher or equal $\util_{\kk_1-1}$.
At the same time, the expected amortized cost does not increase.
If $x_i = 1$, continue with $T^1$ as before.

With the remaining probability $1-\alpha$, also continue with $T'^{\,0}$ if $x_i = 0$ and stop without a certificate otherwise. Denote this strategy with $\tilde{S}''_j$.

The probability $\alpha$ is chosen in such a way that the tests on $x_{a_1}$ and $x_{b_2}$ are carried out with the same probability $1-\qq[i] = \qq[i] + \alpha \cdot \RB{1-\qq[i]}$ in $S_j$ and $S'_j$ and therefore make the same contribution to expected amortized cost.
Also, in $S'_j$, $T^{11}$ or $T'^{\,11}$ as well as $T^{10}$ or $T'^{\,10}$ are reached with the same respective probability as $T^{11}$ and $T^{10}$ in $S_j$, with weakly lower expected amortized cost and weakly higher expected utility.
Overall, $\Ex{\Acost{S_j}} \geq \Ex{\Acost{S'_j}}$ and $\Utilk{\kk}{S_j} \leq \Utilk{\kk}{S'_j}$.
Further, $S'_j$ is a randomization over $S''_j$ and $\tilde{S}''_j$, both prunings of $\mathcal{I}_i$, matching Cases 2.1 and 2.2. Therefore, this case can be reduced to those two cases. 

Similarly to Case 1.4, the replacing strategies obtained from the substitutions used in from Cases~2.1 and 2.2 are randomized RC-prunings of IO that do not terminate before the end of the first round, so they can be used for obtaining a well-defined overall strategy (cf. argument on obtaining an overall strategy for all possible outcomes of $x_{b_1}$ and $x_{b_2}$ after the case analysis in the main part of the paper).

\underline{Case 2.5} (\textit{Abort-Abort}):
This is the case where $S_j$ stops without a 0-certificate for both outcomes of the test on $x_i$ (cf.\ \Cref{tab:tree15}).
We obtain $S'_j$ by replacing element $x_i$ with $x_{a_1}$.
The resulting strategy is a pruning of IO.
The utility is always $0$ in both cases.
The expected amortized cost does not increase: $\Ex{\Acost{S'_j}} = \Ex{\Acost{a_1}} + \Acost{b_1} = 1-\qq[a_1] + \Acost{b_1} \leq 1-\qq[i] + \Acost{b_1} = \Ex{\Acost{i}} + \Acost{b_1} = \Ex{\Acost{S_j}}$.
For $\ell \in \CB{0,1}$, the arguments and estimates work in the same way.
Therefore, the induction hypothesis also holds in this case for $R_j = S'_j$.

\underline{Case 2.6} (\textit{Certificate-Abort}):
\begin{figure}
\begin{minipage}[b]{.48\textwidth}
    \centering
    \scalebox{0.75}{
        \begin{forest}
            for tree ={edge={very thick,-Stealth},s sep=0.5cm}
            [{$\CB{x_i,x_{b_1}}$},innernode,s sep = 3cm
                [{$0$-cert.},endnodes, child anchor=,edge label={node[midway,above left] {$\,\,x_i = 0\,\,$}}
                ]
                [{},endnodef, child anchor=,edge label={node[midway,above right] {$\,\,x_i = 1\,\,$}}
                ]
            ]
        \end{forest}
    }
\end{minipage}%
\hfill\vline\hfill
\begin{minipage}[b]{.48\textwidth}
    \centering
    \scalebox{0.75}{
        \begin{forest}
            for tree ={edge={very thick,-Stealth},s sep=0.5cm}
            [{$\CB{x_{a_1},x_{b_1}}$},innernode,s sep = 3cm
                [{$0$-cert.},endnodes, child anchor=,edge label={node[midway,above left] {$\,\,x_{a_1} = 0\,\,$}}
                ]
                [{},endnodef, child anchor=,edge label={node[midway,above right] {$\,\,x_{a_1} = 1\,\,$}}
                ]
            ]
        \end{forest}
    }

\end{minipage}

\caption{Case 2.6 (\textit{Certificate-Abort}). $S_j$ on the left, $S'_j$ on the right.}
\label{tab:tree26}
\end{figure}
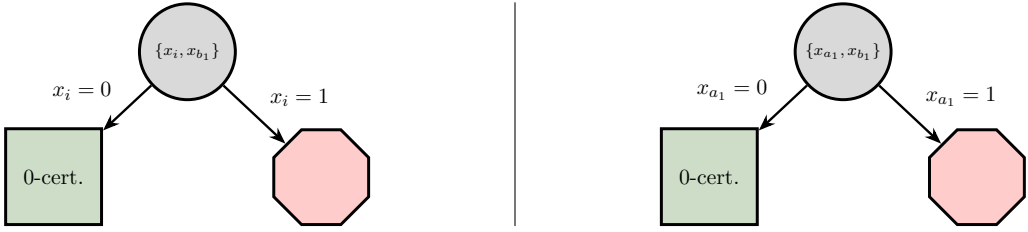

The structure (as depicted in~\Cref{tab:tree26}) implies that exactly one 0 is missing towards a certificate, after considering the outcome of the test on $x_{b_1}$.
We obtain $S'_j$ by replacing element $x_i$ with $x_{a_1}$, with 
$\Ex{\Utilk{\kk}{S'_j}} = \qq[a_1] \geq \qq[i] = \Ex{\Utilk{\kk}{S_j}}$.
The non-increase in amortized cost $\Ex{\Acost{S'_j}} \leq \Ex{\Acost{S_j}}$ easily follows in analogy to Case~2.5.
Since $S'_j$ is a pruning of IO, this concludes Case~2.6 with setting $R_j = S'_j$.

\underline{Case 2.7} (\textit{Certificate-Certificate}):
In this case, we stop with a certificate after the first round for both outcomes of testing $x_i$ (cf.\ \Cref{tab:tree17}).
As in Case 1.7, we are restricted to $\kk \geq 1$, therefore this case can only occur if $x_{b_1} = 0$ (and thus $\ell \geq 1$) and $\kk=1$.
$S_j$ is not a pruning of IO, so we replace it with $S'_j$, which starts with testing $x_{a_1}$ and $x_{b_1}$ in the first round, and whicl will also find a certificate for both realizations of $x_{a_1}$. Note that $S'_j$ is a pruning of IO.
The utility is always $1$ in both cases.
Also, with the same calculation as in Case~2.5, we obtain $\Ex{\Acost{S'_j}} \leq \Ex{\Acost{S_j}}$.
Again, the induction hypothesis holds setting $R_j = S'_j$.
The bounds also work for $\ell = 1$ (similarly to Case~1.7, Case~2.7 cannot occur if $\ell = 0$). 

Having covered all possibilities for Case 2, this concludes the case analysis.

\subsection{Calculations for proof of \Cref{lem:IOtreesurgery}}
\label{app:surgerynurse}
Auxiliary calculations for Case 1.2:
Recall that $\alpha = \frac{\qq[i] - \qq[a_1]}{\qq[i] \cdot \RB{1-\qq[a_1]}}$.

The first of the two bounds for which the calculations are deferred to this section holds with equality:
\begin{align*}
    &\RB{1-\qq[i]} \cdot \Ex{\Acost{a_1}} - \SB{1 - \qq[a_1] - \alpha \cdot \RB{1-\qq[a_1]}} \cdot \Ex{\Acost{i}}\\
    &\quad= \qq[a_1] - \qq[i] \cdot \qq[a_1] - \qq[i] + \qq[a_1] \cdot \qq[i] + \qq[i] \cdot \RB{1-\qq[a_1]} \cdot \alpha\\
    &\quad= \qq[a_1] - \qq[i] + \RB{\qq[i] - \qq[a_1]} = 0\, .
\end{align*}
In the first step, we use that $\Ex{\Acost{a_1}} = \qq[a_1]$ and $\Ex{\Acost{i}} = \qq[i]$ as well as the definition of $\alpha$ in the second step.

The remaining expression bounds the additional cost from recursion and by testing $x_{b_2}$ with higher probability than in $S_j$:
\begin{align*}
    &\SB{\qq[a_1] + \alpha \cdot \RB{1-\qq[a_1]} - \qq[i]} \cdot \Acost{b_2} + \SB{\qq[a_1] + \alpha \cdot \RB{1 - \qq[a_1]}}\\
    &\quad\leq \SB{\alpha \cdot \RB{1-\qq[a_1]} - \alpha \cdot \qq[i] \cdot \RB{1 - \qq[a_1]}} + \SB{1 - \RB{1 - \qq[a_1]} + \alpha \cdot \RB{1 - \qq[a_1]}}\\
    &\quad= 1 + \alpha \cdot \RB{1-\qq[a_1]} \cdot \SB{1-\qq[i]} - \SB{\RB{1-\qq[a_1]} \cdot \RB{1-\alpha}}\\
    &\quad= 1 + \RB{1-\qq[a_1]} \cdot \SB{\RB{1-\qq[i]} \cdot \frac{\qq[i] - \qq[a_1]}{\qq[i] \cdot \RB{1-\qq[a_1]}} - \frac{\qq[i] \cdot \RB{1-\qq[a_1]} - \RB{\qq[i] - \qq[a_1]}}{\qq[i] \cdot \RB{1-\qq[a_1]}}}\\
    &\quad= 1 + \frac{1}{\qq[i]} \cdot \SB{\RB{1-\qq[i]} \cdot \RB{\qq[i] - \qq[a_1]} - \RB{\qq[a_1] - \qq[a_1] \cdot \qq[i]}}\\
    &\quad= 1 + \frac{1-\qq[i]}{\qq[i]} \cdot \RB{\qq[i] - 2 \qq[a_1]}\\
    &\quad\leq 1 \, .
\end{align*}
In the first step, we use $\Acost{b_2} \leq 1$ and $\qq[i] - \qq[a_1] = \alpha \cdot \qq[i] \cdot \RB{1-\qq[a_1]}$. 
In the last step, we use that $\RB{1-\qq[i]} \geq 0$ and, in the case considered here, $2\qq[a_1] \geq 1 \geq \qq[i]$.

\end{document}